\documentclass[a4paper,11pt]{article}
\usepackage{jheppub} 

\usepackage{amssymb,amsfonts,amsmath,bm,bbm}
\usepackage{graphicx}
\usepackage{dcolumn}
\usepackage{xcolor}
\usepackage{epstopdf}

\newcommand{\Msun}{M_\odot}
\newcommand{\Lam}{\Lambda}
\newcommand{\Om}{\Omega}
\newcommand{\Bq}{B_0^{1/4}}
\newcommand{\epsu}{\epsilon_{\mathrm{ud}}}

\newenvironment{ruledtabular}
 {\par\medskip\noindent\begin{center}\begin{minipage}{0.98\textwidth}%
  \hrule height 0.8pt\vspace{5pt}\small\centering}
 {\vspace{5pt}\hrule height 0.8pt\end{minipage}\end{center}\par\medskip}

\begin{document}

\title{Bayesian Multimessenger Constraints on Self-Bound Quark Stars in Revised Perturbative QCD}

\author[a]{Shu-Peng Wang,}\emailAdd{wangshp2026@lzu.edu.cn}
\author[c]{Ju-Xiang Li,}
\author[b]{Zhen-Yan Lu,}\emailAdd{luzhenyan@hnust.edu.cn}
\author[b,d]{Zhi-Jun Ma,}\emailAdd{mazhijun@hnust.edu.cn}
\author[e]{Zhen Zhang,}\emailAdd{zhangzhen@ihep.ac.cn}
\author[f]{Qi Lu}

\affiliation[a]{School of Nuclear Science and Technology,
Lanzhou University, Lanzhou 730000, China}
\affiliation[b]{Hunan Provincial Key Laboratory of Intelligent Sensors and
Advanced Sensor Materials, School of Physics and Electronics,
Hunan University of Science and Technology, Xiangtan 411201, China}
\affiliation[c]{College of Information Engineering,
Suqian University, Suqian 223800, China}
\affiliation[d]{School of Chemistry and Chemical Engineering,
Hunan University of Science and Technology, Xiangtan 411201, China}
\affiliation[e]{State Key Laboratory of Particle Astrophysics,
Institute of High Energy Physics, Chinese Academy of Sciences,
Beijing 100049, China}
\affiliation[f]{School of Physics, Beihang University,
Beijing 100191, China}

\abstract{
We perform a Bayesian multimessenger analysis of self-bound two-flavor $ud$ quark matter described by a thermodynamically consistent revised perturbative-QCD equation of state (EOS), incorporating the GW170817 tidal information, mass--radius measurements, and pulsar mass constraints. The resulting posterior medians and $68\%$ credible intervals are $B_0^{1/4}=137.66^{+1.72}_{-1.86}\,\mathrm{MeV}$, $M_{\max}=2.166^{+0.035}_{-0.036}\,M_\odot$, $R_{1.4}=11.778^{+0.203}_{-0.196}\,\mathrm{km}$, and $\Lambda_{1.4}=768^{+90}_{-81}$. The posterior-predictive support probability for interpreting the $2.59\,M_\odot$ GW190814 secondary as a self-bound $ud$ quark star is of order $10^{-5}$, strongly disfavoring this interpretation within the adopted EOS framework. The reconstructed EOS exhibits characteristic self-bound behavior, with finite energy and baryon number densities at the zero-pressure surface and a median surface energy density of $\epsilon_{\rm surf}\simeq187\,\mathrm{MeV\,fm^{-3}}$. Its pressure is suppressed relative to both the noninteracting $ud+e$ gas and the representative MIT bag-model prediction. Although the running strong coupling decreases with increasing quark chemical potential, its central value in maximum-mass configurations correlates positively with $M_{\max}$, indicating stronger coupling and reduced perturbative reliability toward the high-mass end. Notably, the squared speed of sound remains nearly constant, varying from $c_s^2\simeq0.321$ to $0.324$ between canonical and maximum-mass configurations, whereas the conformal measure $\Delta=1/3-P/\epsilon$ decreases substantially from approximately $0.208$ to $0.076$. This demonstrates that an approximately constant sound speed does not necessarily imply approximate conformality. The relativistic adiabatic index also increases sharply near the zero-pressure surface, reflecting the self-bound nature of the EOS.
}

\keywords{Revised Perturbative QCD, Dense Quark Matter,
Equation of State, Self-Bound Quark Stars,
Tidal Deformability}

\maketitle

\section{Introduction} \label{sec:INTRODUCTION}

The equation of state (EOS) of strongly interacting matter at densities several times the nuclear saturation density remains a key unresolved link between quantum chromodynamics (QCD)~\cite{Gross:2022hyw,MUSES:2023hyz,
Chen:2024aom,Altmann:2024icx} and the observable properties of compact stars
~\cite{Chatziioannou:2024jsr,Lattimer:2010kpa}.
The binary merger GW170817 provided a major constraint on the tidal
response of compact stars~\cite{LIGOScientific:2017vwq,LIGOScientific:2017zic,LIGOScientific:2018cki,LIGOScientific:2018hze,LIGOScientific:2025snk}.
Under the assumptions adopted in the common-EOS, low-spin analysis,
the dimensionless tidal deformability of a $1.4\,\Msun$ star was inferred
to be $\Lam_{1.4}=190^{+390}_{-120}$, favoring relatively compact
configurations~\cite{Chatziioannou:2020pqz,Most:2018hfd,Raithel:2019uzi,Capano:2019eae}.
Complementary mass--radius information has been obtained with the
Neutron Star Interior Composition Explorer (NICER) for
PSR J0030$+$0451~\cite{Riley:2019yda,Miller:2019cac,Bogdanov:2019ixe,Raaijmakers:2019qny,Vinciguerra:2023qxq},
PSR J0740$+$6620~\cite{Riley:2021pdl,Miller:2021qha,Fonseca:2021wxt,Salmi:2022cgy,Raaijmakers:2021uju},
and PSR J0437$-$4715~\cite{Choudhury:2024xbk},
with updated analyses further refining the constraints on
PSR J0740$+$6620~\cite{Dittmann:2024mbo,Salmi:2024aum,Legred:2021hdx}.
At the high-mass end, measurements of PSR J2215$+$5135~\cite{Linares:2018ppq} and PSR J0952$-$0607~\cite{Romani:2022jhd,Romani:2026j0952}, whose updated mass estimate
is $2.35\pm0.11\,\Msun$, test the maximum mass supported by the EOS.
The inferred low mass and small radius of the compact object in
HESS~J1731$-$347 provide a complementary probe of the low-mass
regime~\cite{Doroshenko:2022nwp,Char:2024kgo,Ozel:2016oaf}.
Together, these observations test whether an EOS can produce
sufficiently compact stars while retaining enough pressure to support
the most massive pulsars~\cite{Antoniadis:2013pzd,Demorest:2010bx,NANOGrav:2019jur}. 
Quark matter offers one possible description of matter in compact-star
interiors, either as a core within a hybrid star~\cite{Annala:2019puf,Annala:2021gom,Alford:2019oge,Alford:2004pf,Alford:2007xm,Alford:2006wn,Cao:2026dwk} or as the bulk
constituent of a self-bound star~\cite{Alford:2009qj}.
The latter possibility has long been explored in connection with
the proposals of Bodmer~\cite{Bodmer:1971we} and
Witten~\cite{Witten:1984rs}, particularly for strange quark matter~\cite{Farhi:1984qu,Alcock:1986hz,Weissenborn:2011qu,Weber:2004kj,Zhang:2024xod,DiClemente:2024lzi}.
Self-bound two-flavor $ud$ quark stars provide a distinct EOS hypothesis
whose viability can likewise be tested through their masses, radii,
and tidal deformabilities~\cite{Cao:2023rgh}.
Their stellar properties have been investigated using resummed perturbative QCD,
quasiparticle, and Nambu--Jona-Lasinio approaches~\cite{Restrepo:2022wqn,Restrepo:2025qgp,Ma:2023stj,Yuan:2022dxb,Gorda:2023mkk}.
For a specified $ud$ quark-matter EOS, the central observational
question is whether the same parameter choices can account for both
the compactness favored by tidal measurements and the support
required by high-mass pulsars.

Perturbative QCD provides a systematically improvable description
of cold quark matter at asymptotically high densities~\cite{Fraga:2001id,Kurkela:2009gj,Fraga:2013qra,Fraga:2016yxs,Baym:2017whm,Lattimer:2015nhk}.
Its application to compact stars nevertheless requires care, because
the coupling is larger and uncertainties associated with truncation
and renormalization-scale dependence become more important at stellar
densities. High-density perturbative QCD information has therefore also been used
to constrain lower-density EOSs through thermodynamic consistency
and interpolation~\cite{Kurkela:2014vha,Komoltsev:2021jzg,Komoltsev:2023zor,Annala:2023cwx,Gorda:2023usm}.
Here, we instead adopt a specific thermodynamically consistent
revised perturbative QCD model for $ud$ quark matter
~\cite{Xu:2014zea,Wang:2025lwv}.
In this construction, an additional contribution to the thermodynamic
potential enforces consistency between the pressure and the prescribed
number densities, including the chemical-potential dependence of
the running coupling.
The resulting EOS satisfies the thermodynamic requirement that
the minimum energy per baryon on the stable branch occurs at zero
pressure, while recovering the high-density perturbative QCD behavior.
Its two parameters are the dimensionless renormalization-scale
coefficient $C$ and the vacuum constant $B_0$.
Thermodynamic consistency, however, does not by itself establish
perturbative accuracy at stellar densities; this distinction is
particularly relevant to parameter choices with a large running
coupling~\cite{Bauswein:2018bma,Most:2018eaw,Ecker:2024uqv,Fonseca:2025cxs}.

Previous work with this revised EOS examined the stability
conditions and stellar sequences of $ud$ quark matter
and showed that the GW170817 tidal constraint
substantially restricts the supported maximum mass
~\cite{Wang:2025lwv}.
Bayesian methods have also been widely applied to
hadronic, hybrid, and crossover equations of state~\cite{Grundler:2026zus,Biswas:2021pvm,Finch:2025bao},
as well as to quark stars described by interacting
MIT bag models~\cite{Wang:2024xon,Xie:2025sth}.
In gravitational-wave-based EOS inference,
correlations among the binary component masses
and tidal deformabilities provide important
information for reconstructing the properties
of dense matter
~\cite{De:2018TidalGW170817,Landry:2019EOSGW,
Essick:2020KDEGW}.
Different statistical representations of the
GW170817 tidal information have been explored,
including one-dimensional approximations,
multidimensional kernel density estimates,
and methods that retain correlations among
the binary parameters~\cite{Thrane:2019BayesianGW,Lackey:2015prd,
Botev:2010KDE,Landry:2019EOSGW,Essick:2020KDEGW,
Greif2019,McGinn2024,Williams:2021qyt}.
These developments provide an important basis
for multimessenger constraints on dense-matter
equations of state. Nevertheless, it remains
important to assess how different representations
of the GW170817 tidal information affect the
inferred parameters and stellar predictions
within a specific thermodynamically consistent
revised perturbative-QCD model.
In particular, the simultaneous requirements
of relatively small tidal deformability and
sufficiently large maximum mass motivate a
systematic examination of the observational
viability and stability-allowed parameter
space of self-bound two-flavor $ud$ quark stars.

Previous investigations have shown that the tidal response
of self-bound quark stars is sensitive to their internal
structure~\cite{Postnikov:2010yn}. Studies based on
thermodynamically consistent quasiparticle models and
interacting Nambu--Jona-Lasinio approaches have further
explored the stability, thermodynamic properties, and
stellar configurations of cold quark matter
~\cite{Ma:2023stj,Yuan:2022dxb}.
In a broader context, the sound speed has been studied
as an important probe of the stiffness of dense-matter
equations of state~\cite{Bedaque:2014sqa,Tews:2018kmu},
while the behavior of conformal measures has been used
to characterize the microscopic properties of matter
in massive neutron stars~\cite{Annala:2023cwx}.
Perturbative-QCD studies have also emphasized the
importance of renormalization-scale dependence and
higher-order uncertainties when connecting microscopic
calculations to stellar densities~\cite{Gorda:2023usm}.
Motivated by these developments, we extend the
multimessenger analysis of the revised perturbative-QCD
EOS beyond global stellar observables to investigate
the internal structure and microscopic properties of
self-bound two-flavor $ud$ quark stars. We reconstruct
the posterior EOS and examine the radial distributions
of pressure, energy density, and baryon number density,
together with the central-density constraints for
representative stellar configurations. We further
investigate the running strong coupling and its relation
to the EOS parameters and maximum gravitational mass.
The thermodynamic analysis is complemented by a study
of the squared sound speed, conformal deviation, and
relativistic adiabatic index, which characterize the
density dependence of the effective stiffness and
response of quark matter. Together, these analyses
connect multimessenger constraints on the revised
perturbative-QCD EOS with the internal structure and
high-density properties of self-bound quark stars.

The remainder of this paper is organized as follows.
In Sec.~\ref{sec:model}, we introduce the
thermodynamically consistent revised perturbative-QCD
EOS, describe the stellar-structure and
tidal-deformability calculations, and present
the observational likelihoods and Bayesian
methodology.
In Sec.~\ref{sec:results}, we first examine
the posterior constraints on the EOS parameters
and stellar observables, followed by an analysis
of the mass--tidal feasibility frontier,
maximum-mass requirements, and the implications
for the GW190814 secondary.
We then investigate the reconstructed EOS,
the radial structure and central-density
properties of self-bound quark stars,
and the running strong coupling and
thermodynamic properties of dense quark matter,
including the sound speed, conformal behavior,
and adiabatic index.
Finally, in Sec.~\ref{sec:concl}, we summarize
our principal findings and discuss their
physical implications.

\section{Theoretical framework}
\label{sec:model}

\subsection{Revised perturbative-QCD equation of state}

For $ud$ quark matter at zero temperature the thermodynamic potential density
of the revised perturbative QCD model is~\cite{Wang:2025lwv}
\begin{equation}
\begin{split}
\Omega_{\mathrm{pt}} &=
-\frac{\mu_{\mathrm{u}}^4+\mu_{\mathrm{d}}^4}{4\pi^2}
(1-2\alpha)
-\frac{\mu_{\mathrm{e}}^4}{12\pi^2},\\
\alpha &= \frac{1}{\beta_0\ln(\mu_{\rm R}/\Lambda_{\rm QCD})},\\
\mu_{\rm R} &= C\left[
\frac{\mu_{\mathrm{u}}^4+\mu_{\mathrm{d}}^4}{2}
\right]^{1/4} ,
\end{split}
\label{eq:ompt}
\end{equation}
with $\beta_0 = 11/2 - N_{\mathrm{f}}/3$, where $N_{\mathrm{f}}$ is the number of flavors.
The thermodynamic-consistency correction~\cite{Xu:2014zea}, required for the pressure to vanish
at the minimum of the energy per baryon, is
\begin{equation}
\Om = \Om_{\mathrm{pt}} + \Om_0,
\qquad
\Om_0 = \int\!\left(F\,d\mu_{\mathrm{u}} + Q\,d\mu_{\mathrm{d}}\right) + B_0,
\label{eq:om}
\end{equation}
with $F=\mu_u^3/[2\beta_0\pi^2\ln^2(\mu_{\rm R}/\Lambda_{\rm QCD})]$ and
$Q = F(\mu_{\mathrm{u}}\to\mu_{\mathrm{d}})$. The quark and electron number densities follow from
$n_{\mathrm{q}} = \mu_{\mathrm{q}}^3(1-2\alpha)/\pi^2$ and $n_{\mathrm{e}} = \mu_{\mathrm{e}}^3/3\pi^2$, and the energy
density and pressure are
\begin{equation}
E = \Om + \sum_i \mu_i n_i,
\qquad
P = -\Om .
\label{eq:EP}
\end{equation}

\begin{figure}[tbp]
    \centering
    \includegraphics[width=0.563\textwidth]
    {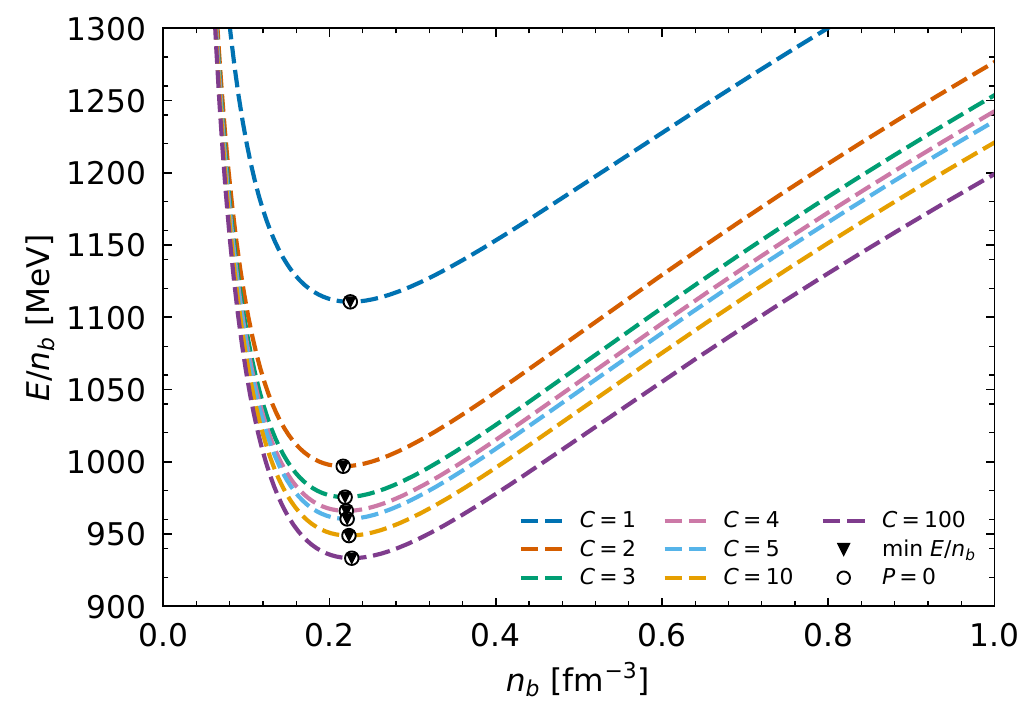}
    \caption{ Density dependence of the energy per baryon for seven values of the scale parameter $C$ at fixed $B_0^{1/4}=142$~MeV. The minimum of each curve and the point where the pressure vanishes fall at the same baryon density.
    }
    \label{fig:consistency}
\end{figure}

This model contains two parameters: the dimensionless scale coefficient $C$ and the
vacuum constant $B_0$, whose fourth root, $\Bq$, is expressed in MeV.
Since $B_0$ enters $\Omega$ only as an additive constant, it does not affect the
$\beta$-equilibrium condition $\mu_{\mathrm{d}} = \mu_{\mathrm{u}} + \mu_{\mathrm{e}}$,
the baryon number density $n_{\mathrm{b}} = \tfrac13(n_{\mathrm{u}}+n_{\mathrm{d}})$,
or the charge-neutrality condition
$\tfrac23 n_{\mathrm{u}} - \tfrac13 n_{\mathrm{d}} - n_{\mathrm{e}} = 0$.
The thermodynamic potential can therefore be written as
\begin{equation}
\Om(n_{\mathrm{b}},C,B_0) = \Om_{\mathrm{M}}(n_{\mathrm{b}},C) + B_0 .
\end{equation}
Thus, for each value of $C$, a single table of $\Om_{\mathrm{M}}$ suffices
for all values of $B_0$.
The physical requirement that $ud$ quark matter not be more bound than nuclear matter is imposed on the energy per baryon of Eq.~\eqref{eq:EP} as $\epsu \equiv \left.E/n_{\mathrm{b}}\right|_{\mathrm{ud}} \geq 930$~MeV, which defines the forbidden region of the stability window
~\cite{Bodmer:1971we,Weissenborn:2011qu,Fattoyev:2017jql,Drischler:2024ebw}. 
Thermodynamic consistency is ensured by the correction term $\Om_0$ in
Eq.~\eqref{eq:om}, constructed by integrating the number densities derived
from the thermodynamic potential.
At zero temperature, along the charge-neutral, $\beta$-equilibrated EOS,
the thermodynamic identity
$P = n_{\mathrm{b}}^2\,\mathrm{d}(E/n_{\mathrm{b}})/\mathrm{d}n_{\mathrm{b}}$
implies that the minimum energy per baryon on the mechanically stable branch
occurs at zero pressure.
This correspondence holds throughout the admissible parameter space.
Figure~\ref{fig:consistency} illustrates it for $C=1$--$100$ at fixed
$B_0^{1/4}=142$~MeV.

\subsection{Stellar structure and tidal deformability}

We adopt units in which $c=1$ and retain $G$ explicitly.
For a given EOS, the structure of a static, spherically symmetric
star is determined by the Tolman--Oppenheimer--Volkoff (TOV)
equations~\cite{Shapiro:1983du}
\begin{equation}
\begin{split}
\frac{dP}{dr} &= -\frac{G m E}{r^2}
\frac{(1+P/E)(1+4\pi r^3 P/m)}{1-2Gm/r},\\
\frac{dm}{dr} &= 4\pi r^2 E,
\end{split}
\end{equation}
where $P(r)$, $E(r)$, and $m(r)$ are the pressure, energy density,
and enclosed gravitational mass, respectively.
For each central pressure $P_{\mathrm{c}}$, these equations are
integrated outward from $m(0)=0$ and $P(0)=P_{\mathrm{c}}$
to the surface, defined by $P(R)=0$.
The stellar mass is then $M=m(R)$, and the compactness is
$\beta=GM/R$.

The quadrupolar tidal response is obtained by simultaneously
integrating the tidal perturbation equation
~\cite{Flanagan:2007ix,Damour:2009vw,Yagi:2013bca,Hinderer:2009ca}
\begin{equation}
r\frac{dy}{dr} + y^2 + yF(r) + r^2 Q(r) = 0,
\qquad y(0)=2,
\label{eq:y}
\end{equation}
where $f(r) = 1-2Gm(r)/r$, $F(r) = [1-4\pi G r^2 (E-P)]/f$, and
$Q(r) = (4\pi/f)\,[G(5E+9P+(E+P)/c_{\mathrm{s}}^2) - 6/(4\pi r^2)] - (\nu')^2$ with
$\nu' = 2G(m+4\pi r^3P)/(r^2 f)$.
The Love number and the dimensionless tidal deformability follow from the
relativistic formulation of the tidal problem used in gravitational-wave
applications~\cite{Hinderer:2009ca, Postnikov:2010yn,Zhao:2018nyf} and are given by
\begin{equation}
\begin{split}
k_2 &= \frac{8\beta^5}{5}(1-2\beta)^2
      \frac{2+2\beta(y_{\mathrm{R}}-1)-y_{\mathrm{R}}}{D},\\
\Lam &= \frac{2}{3}\frac{k_2}{\beta^5}, \qquad  \beta = \frac{GM}{R},
\end{split}
\end{equation}
where $D$ is the usual cubic-plus-logarithmic denominator~\cite{Hinderer:2007mb}.
The $-(\nu')^2$ form of the source term is used in Eq.~\eqref{eq:y}~\cite{Damour:2009vw}. The value fed to the $k_2$ formula is not $y(R)$ obtained by integrating
Eq.~\eqref{eq:y} up to the surface, but the surface-matched value
\begin{equation}
y_{\mathrm{R}} = y(R^{-}) - \frac{4\pi R^{3}\epsilon_{\mathrm{s}}}{M},
\label{eq:yjunc}
\end{equation}
where $y(R^{-})$ and $\epsilon_{\mathrm{s}}$
are evaluated immediately inside the surface.
This correction vanishes when the surface energy density is
zero, but is retained for all self-bound quark-star
configurations considered here.

\subsection{Multimessenger data and observational constraints}

\begin{table}[tbp]
\caption{Measurements used in the baseline likelihood, and the GW190814 mass
used only as a posterior-predictive diagnostic.}
\label{tab:data}
\begin{ruledtabular}
\footnotesize
\setlength\tabcolsep{4pt}
\begin{tabular}{lllc}
Observable & Measurement & Type & Refs. \\
\hline
GW170817 & $\Lam_{1.4}=190^{+390}_{-120}$ & tidal & \cite{LIGOScientific:2018cki} \\
NICER J0030$+$0451 & \begin{tabular}{@{}l@{}}$M=1.44^{+0.15}_{-0.14}\,\Msun$\\$R=13.02^{+1.24}_{-1.06}$~km\end{tabular} & $M$--$R$ & \cite{Riley:2019yda, Miller:2019cac} \\
NICER J0740$+$6620 & \begin{tabular}{@{}l@{}}$M=2.08^{+0.07}_{-0.07}\,\Msun$\\$R=12.39^{+1.30}_{-0.98}$~km\end{tabular} & $M$--$R$ & \cite{Riley:2021pdl, Miller:2021qha} \\
PSR J0437$-$4715 & \begin{tabular}{@{}l@{}}$M=1.418\pm0.037\,\Msun$\\$R=11.36^{+0.95}_{-0.63}$~km\end{tabular} & $M$--$R$ & \cite{Choudhury:2024xbk} \\
4U 1702$-$429 & \begin{tabular}{@{}l@{}}$M=1.9\pm0.3\,\Msun$\\$R=12.4\pm0.4$~km\end{tabular} & $M$--$R$ & \cite{Nattila:2017wtj} \\
HESS J1731$-$347 & \begin{tabular}{@{}l@{}}$M=0.77^{+0.20}_{-0.17}\,\Msun$\\$R=10.4^{+0.86}_{-0.78}$~km\end{tabular} & $M$--$R$ & \cite{Doroshenko:2022nwp} \\
 PSR J0952$-$0607 & $M=2.35\pm0.11\,\Msun$ & mass & \cite{Romani:2026j0952} \\
PSR J2215$+$5135 & $M=2.27^{+0.17}_{-0.15}\,\Msun$ & mass & \cite{Linares:2018ppq} \\
GW190814 sec. & $M=2.59^{+0.08}_{-0.09}\,\Msun$ & mass & \cite{LIGOScientific:2020zkf} \\
\end{tabular}
\end{ruledtabular}
\end{table}

The likelihood combines three classes of observational constraints, listed in Table~\ref{tab:data} together with their primary references. The central values and quoted uncertainties are adopted from the corresponding sources. All observational data sets are treated as statistically independent, with their likelihood contributions constructed using the prescriptions described below. The first class consists of the GW170817 tidal information. For the one-dimensional split-Gaussian prescription, we adopt the constraint on the single-star dimensionless tidal deformability at $1.4,M_\odot$, $\Lambda_{1.4}=190^{+390}_{-120}$~\cite{LIGOScientific:2018cki}. We also consider two-dimensional and four-dimensional kernel
density estimation (KDE) representations of the GW170817 posterior, retaining the correlations among the binary parameters and tidal deformabilities.

Five mass--radius measurements form the second family: the NICER observations
of PSR J0030$+$0451 and PSR J0740$+$6620, the NICER resolution of PSR
J0437$-$4715, and the two broader X-ray constraints 4U~1702$-$429 and
HESS~J1731$-$347~\cite{Raaijmakers:2019qny, Raaijmakers:2019dks, Legred:2021hdx, Capano:2019eae, Choudhury:2024xbk}. PSR J0437$-$4715 is the most informative of the five.
Its radio-timing mass uncertainty is about half that of the next-tightest mass
constraint in the set and roughly four times smaller than that of
PSR J0030$+$0451, and it sits essentially at $1.4\,\Msun$, so it constrains
$\Lam_{1.4}$ directly. The five measurements are adopted with the published uncertainties and treated as mutually independent, so a common cross-source systematic is not propagated into the likelihood; the radio-timing masses are treated as conditionally independent of the NICER targets given the EOS parameters
~\cite{Bogdanov:2019ixe,Salmi:2022cgy,Vinciguerra:2023qxq,Kini:2026rjx,Raaijmakers:2019dks,Raaijmakers:2021uju,Kunert:2021hgm}.

The third family consists of the mass measurements of
PSR J0952$-$0607, the heaviest pulsar known and one of
the most demanding constraints on $M_{\max}$, and PSR
J2215$+$5135~\cite{Romani:2026j0952,Linares:2018ppq}.
Finally, the $2.59\,\Msun$ secondary of GW190814 is listed in the table but is
not part of the baseline likelihood: whether it is a quark star is one of the questions the analysis is meant to answer~\cite{Bombaci:2020vgw, Dexheimer:2020rlp,
Biswas:2020xna,Most:2020bba,Essick:2020ghc}.





The following likelihood factors map the measurements summarized above onto each stable EOS sequence.
For the mass--radius (M--R) likelihood, the one-sided
uncertainties are selected independently according to
the signs of the mass and radius residuals:
\begin{equation}
\begin{aligned}
\sigma_{\mathrm{M}}(M) &=
\begin{cases}
\sigma_{\mathrm{M},-}, & M<M_{\mathrm{obs}},\\
\sigma_{\mathrm{M},+}, & M\geq M_{\mathrm{obs}},
\end{cases}\\[6pt]
\sigma_{\mathrm{R}}(R) &=
\begin{cases}
\sigma_{\mathrm{R},-}, & R<R_{\mathrm{obs}},\\
\sigma_{\mathrm{R},+}, & R\geq R_{\mathrm{obs}}.
\end{cases}
\end{aligned}
\label{eq:asym}
\end{equation}
The normalization widths are defined as the arithmetic
means of the two one-sided uncertainties:
\begin{equation}
\bar\sigma_{\mathrm{M}}=
\frac{\sigma_{\mathrm{M},-}+\sigma_{\mathrm{M},+}}{2},
\qquad
\bar\sigma_{\mathrm{R}}=
\frac{\sigma_{\mathrm{R},-}+\sigma_{\mathrm{R},+}}{2}.
\label{eq:sigbar}
\end{equation}

The M--R likelihood is obtained by marginalizing over
the stellar mass along the stable branch
~\cite{Chatziioannou:2020pqz,Lourenco:2021lpn,Capano:2019eae,Legred:2021hdx,Raithel:2019uzi}:
\begin{equation}
\mathcal{L}_{\mathrm{MR}} =
\frac{1}{2\pi\bar\sigma_{\mathrm{M}}\bar\sigma_{\mathrm{R}}M_{\max}}
\int_0^{M_{\max}}\!\! dM\,
\exp\left[
-\frac{(M-M_{\mathrm{obs}})^2}{2\sigma_{\mathrm{M}}^2(M)}
-\frac{(R(M)-R_{\mathrm{obs}})^2}{2\sigma_{\mathrm{R}}^2(R(M))}
\right],
\label{eq:lmr}
\end{equation}
where $M_{\mathrm{obs}}$ and $R_{\mathrm{obs}}$ denote
the measured mass and radius, respectively. The prefactor
normalizes the two split-Gaussian factors, while
$1/M_{\max}$ corresponds to uniform weighting in stellar
mass along the stable branch. This marginalization treats
the stellar mass as a nuisance parameter and accounts
for the observational uncertainties in both mass and
radius, while naturally penalizing models with
$M_{\max}<M_{\mathrm{obs}}$.
For mass-only measurements, the same asymmetric mass
uncertainty prescription is evaluated at
$M=M_{\max}$. The likelihood represents the probability
that the EOS can support a star at least as massive as
the observed object, accounting for the measurement
uncertainty
~\cite{Rezzolla:2017aly,Tsokaros:2020hli}:
\begin{equation}
\ln \mathcal{L}_{\mathrm{M}} =
\ln\Phi\!\left(
\frac{M_{\max}-M_{\mathrm{obs}}}
{\sigma_{\mathrm{M}}(M_{\max})}
\right),
\label{eq:lmass}
\end{equation}
where $\Phi$ is the standard normal cumulative
distribution function.

\subsection{GW170817 tidal-likelihood prescriptions}
\label{sec:gw_tidal}

To investigate the impact of different treatments
of gravitational-wave tidal information on the inferred
EOS constraints, we consider three alternative
prescriptions for incorporating the GW170817 observation.
These prescriptions differ in the dimensionality of the
gravitational-wave posterior information retained in the
EOS likelihood. They range from a one-dimensional
approximation based on the tidal deformability of a
canonical-mass star to multidimensional density estimates
that preserve correlations among the binary parameters
and the individual tidal deformabilities.


\subsubsection{One-dimensional split-Gaussian prescription}

As a baseline prescription, we approximate the GW170817
tidal constraint by an asymmetric Gaussian distribution
in the dimensionless tidal deformability of a
$1.4\,\Msun$ star, $\Lambda_{1.4}$.
Under the common-EOS and low-spin assumptions,
the GW170817 analysis gives
$\Lambda_{1.4}=190^{+390}_{-120}$ at the
90\% credible level~\cite{De:2018TidalGW170817,
Thrane:2019BayesianGW}.
This one-dimensional representation provides a
computationally efficient way to incorporate the
gravitational-wave tidal information into the
EOS likelihood while accounting for the
asymmetry of the inferred constraint.

For each EOS parameter set, the predicted value of
$\Lambda_{1.4}$ is obtained by interpolating the
tidal deformability along the stable stellar
sequence at $M=1.4\,\Msun$.
The corresponding log-likelihood is written as
\begin{equation}
\ln\mathcal{L}_{\mathrm{GW}}^{(1D)}
= -\frac12
\left(
\frac{\Lambda_{1.4}-\Lambda_{\mathrm{obs}}}
{\sigma_\pm}
\right)^2
-\ln\!\left(\sqrt{2\pi}\,\sigma_\pm\right),
\label{eq:llam}
\end{equation}
where $\Lambda_{\mathrm{obs}}=190$ and
$\sigma_+$ or $\sigma_-$ denotes the adopted
width on the corresponding side of the
observational constraint.
The width is selected according to the sign
of $\Lambda_{1.4}-\Lambda_{\mathrm{obs}}$,
allowing the likelihood to account for
the asymmetric tidal-deformability constraint.

This prescription compresses the gravitational-wave
tidal information into a single canonical-mass
observable. Consequently, it does not explicitly
retain the dependence on the binary mass ratio
or the correlations between the component
masses and their individual tidal deformabilities.
It therefore serves as a simplified reference
for assessing the impact of the more detailed
two-dimensional and four-dimensional KDE
prescriptions introduced below.

\subsubsection{Two-dimensional KDE prescription.}
\label{subsubsec:2d_kde}

The two-dimensional prescription retains the correlation
between the effective tidal deformability $\tilde{\Lambda}$
and the binary mass ratio $q$, which is lost in a one-dimensional approximation. The corresponding joint
posterior is reconstructed using a KDE, preserving the
correlated GW170817 tidal information
~\cite{LIGOScientific:2018hze,De:2018TidalGW170817,
Thrane:2019BayesianGW,Botev:2010KDE,
Landry:2019EOSGW,Essick:2020KDEGW,Greif2019}.
The effective tidal deformability is calculated directly
from the EOS as~\cite{LIGOScientific:2018hze,Landry:2019EOSGW}
\begin{equation}
\tilde{\Lambda}(M_1,M_2)
=
\frac{16\left[
(M_1+12M_2)M_1^4\Lambda_1+
(M_2+12M_1)M_2^4\Lambda_2
\right]}
{13(M_1+M_2)^5}.
\label{eq:lambdatilde}
\end{equation}
where $\Lambda_i=\Lambda(M_i)$ is obtained from the EOS.
Since the chirp mass is measured with a fractional
uncertainty of approximately $0.1\%$, it is
approximated by its posterior median $\mathcal{M}_{\mathrm{obs}}=1.1872\,M_\odot$. 
The component masses are therefore related through
\begin{equation}
\mathcal{M}(M_1,M_2)
=
\frac{(M_1M_2)^{3/5}}
{(M_1+M_2)^{1/5}}
=
\mathcal{M}_{\mathrm{obs}},
\label{eq:chirpmass_fixed}
\end{equation}
so that $M_2$ is determined by $M_1$.



For the baseline construction, a uniform binary-mass
prior is adopted over the triangular domain
$M_{\min}\leq M_2\leq M_1\leq M_{\max}$,
with $M_{\min}=0.5\,M_\odot$. Its normalization is
given by the area of this triangular domain,
\begin{equation}
\mathcal{N}_{M}
=
\frac{1}{2}(M_{\max}-M_{\min})^2.
\label{eq:mass_norm_2d}
\end{equation}
The two-dimensional tidal-likelihood prescription is
then evaluated along the fixed-chirp-mass curve as
\begin{equation}
\mathcal{L}_{\mathrm{GW}}^{(2D)}
=
\frac{1}{\mathcal{N}_{M}}
\int_{\mathcal{I}_2}\!dM_1\,
\hat p_{\mathrm{GW}}^{(2D)}
\left[
\tilde{\Lambda}_{\mathrm{EOS}}
\big(M_1,M_2(M_1)\big),q(M_1)
\right].
\label{eq:lgw2d}
\end{equation}
where $\mathcal{I}_2$ denotes the range of $M_1$ for which
$M_{\min}\leq M_2(M_1)\leq M_1\leq M_{\max}$. This prescription directly evaluates the reconstructed
gravitational-wave density at the EOS-predicted effective
tidal deformability and binary mass ratio. Compared with
the one-dimensional split-Gaussian approximation,
the two-dimensional construction retains the dependence
of the tidal information on the binary mass ratio,
thereby providing a more detailed representation of
the GW170817 tidal constraints.

\subsubsection{Four-dimensional KDE prescription.}

The four-dimensional prescription retains the correlations
among the component masses and individual tidal
deformabilities, with a common EOS requiring
$\Lambda_i=\Lambda(M_i)$ for each component.
This provides a direct connection between the GW posterior
and the EOS-predicted stellar properties
~\cite{LIGOScientific:2018hze,Landry:2019EOSGW,
Essick:2020KDEGW,Ayriyan:2021KDE,Botev:2010KDE,Greif2019}. We retain the component-level mass and tidal information
using a four-dimensional Gaussian kernel density estimate.
The density is constructed in the coordinates
\begin{equation}
\boldsymbol{x}
=
(\mathcal{M},q,\Lambda_1,\Lambda_2),
\label{eq:4d_coordinates}
\end{equation}
where $\mathcal{M}$ is the chirp mass and
$q=M_2/M_1$ is the mass ratio.
The component masses are converted to
the source frame using
$m_i=m_{i,\mathrm{det}}/(1+z)$, where the redshift is
estimated from the luminosity distance using the
low-redshift approximation. The four-dimensional KDE can be expressed schematically as
\begin{equation}
\hat p_{\mathrm{GW}}^{(4D)}(\boldsymbol{x})
=
\frac{1}{N_{\mathrm{train}}}
\sum_{n=1}^{N_{\mathrm{train}}}
\frac{
\exp\!\left[
-\frac12
(\boldsymbol{x}-\boldsymbol{x}_n)^{T}
H^{-1}
(\boldsymbol{x}-\boldsymbol{x}_n)
\right]
}{
(2\pi)^2 |H|^{1/2}
},
\label{eq:kde4d}
\end{equation}
where $\boldsymbol{x}_n$
denotes the training samples, and $H$ is the
bandwidth covariance matrix determined using
Scott's rule~\cite{Scott:1992kde}. For each pair
$(\mathcal{M}_a,M_{2,b})$, the corresponding primary
mass $M_{1,ab}$ is obtained by solving
\begin{equation}
\mathcal{M}(M_{1,ab},M_{2,b})
=
\mathcal{M}_a,
\qquad M_{1,ab}\geq M_{2,b}.
\label{eq:mass_solve_4d}
\end{equation}
Only configurations satisfying
$M_{1,ab}\leq M_{\max}$ are retained. The corresponding
tidal deformabilities are obtained by interpolation
along the EOS sequence,
\begin{equation}
\Lambda_{1,ab}=\Lambda(M_{1,ab}),
\qquad
\Lambda_{2,b}=\Lambda(M_{2,b}).
\end{equation}

The four-dimensional tidal-likelihood prescription
is defined as the mean KDE density over the retained
EOS-dependent evaluation points,
\begin{equation}
\begin{aligned}
\mathcal{L}_{\mathrm{GW}}^{(4D)}
=&~
\frac{1}{N_{\mathrm{eval}}}
\sum_{(a,b)\in\mathcal{S}_{\mathrm{EOS}}}
\hat p_{\mathrm{GW}}^{(4D)}
\big(
\mathcal{M}_a,q_{ab},
\Lambda_{1,ab},\Lambda_{2,b}
\big),
\end{aligned}
\label{eq:lgw4d}
\end{equation}
where $q_{ab}=M_{2,b}/M_{1,ab}$,
$\mathcal{S}_{\mathrm{EOS}}$ denotes the set of
retained grid points, and $N_{\mathrm{eval}}$ is
their total number. The corresponding log-likelihood is
$\ln\mathcal{L}_{\mathrm{GW}}^{(4D)}$. The public GW170817 samples are posterior samples rather
than prior-independent likelihood samples~\cite{LIGOScientific:2018hze,Landry:2019EOSGW}, satisfying
\begin{equation}
p(\boldsymbol{\theta}\mid d_{\mathrm{GW}})
\propto
\mathcal{L}(d_{\mathrm{GW}}\mid\boldsymbol{\theta})
\pi_{\mathrm{GW}}(\boldsymbol{\theta}).
\label{eq:gw_posterior_prior}
\end{equation}
Consequently, KDEs constructed directly from these samples
retain the original parameter-estimation prior.

\subsubsection{Combined multimessenger likelihood.}

The three GW170817 tidal-likelihood prescriptions
are incorporated into the same multimessenger framework.
For each prescription, the M--R and mass-only likelihood
contributions are kept identical, while only the
gravitational-wave tidal-likelihood factor is varied.
The total log-likelihood is therefore written as
\begin{equation}
\ln\mathcal{L}_{\mathrm{tot}}^{(k)}
=
\sum_{i\in\mathrm{MR}}\ln\mathcal{L}_{\mathrm{MR},i}
+
\sum_{j\in\mathrm{M}}\ln\mathcal{L}_{\mathrm{M},j}
+
\ln\mathcal{L}_{\mathrm{GW}}^{(k)},
\qquad k\in\{1D,2D,4D\}.
\label{eq:ltot_three}
\end{equation}
where the M--R and mass-only contributions are common
to all three prescriptions. This construction ensures
that differences in the resulting posterior distributions
can be attributed to the adopted representation of the
GW170817 tidal information.

\subsection{EOS and stellar modeling}
\label{sec:sampling}

Surrogate models based on neural networks and Gaussian
processes are widely used in dense-matter inference
~\cite{Guo:2023mhf,Gorda:2025aiu,Kashyap:2025cpd,Fujimoto:2021zas}.
Here, the stellar observables are tabulated on a
$(C,\Bq)$ grid, including $\min(E/n_{\mathrm{b}})$,
$M_{\max}$, $R(M)$, and $\Lam(M)$, with a mass spacing
of $0.01\,\Msun$. Bilinear interpolation is used for
likelihood evaluation.
The posterior distributions are sampled using
\texttt{emcee}~\cite{Foreman-Mackey:2012any,Hogg:2017akh}
with 96 walkers and 12,000 steps, discarding the first
4,000 steps as burn-in. Independent nested-sampling
calculations are performed with
\texttt{dynesty}~\cite{Speagle:2019ivv,Ashton:2022grj},
using 1,000 live points and a stopping tolerance
of $\Delta\ln Z=0.01$ to cross-check the posterior
and calculate the Bayesian evidence~\cite{Skilling:2006gxv,Feroz:2008xx}.

An equal-weight mixture of the three posterior
distributions is constructed to obtain unified
constraints:
\begin{equation}
p_{\mathrm{mix}}(\boldsymbol{\theta}|d)
=
\frac{1}{3}
\left[
p_{\mathrm{1D}}(\boldsymbol{\theta}|d)
+
p_{\mathrm{2D}}(\boldsymbol{\theta}|d)
+
p_{\mathrm{4D}}(\boldsymbol{\theta}|d)
\right],
\label{eq:posterior_mixture}
\end{equation}
where $\boldsymbol{\theta}=(C,B_0^{1/4})$.
Each posterior sample is assigned a weight
$1/(3N_k)$, where $N_k$ is the number of samples
from prescription $k$. The combined samples are
used to calculate the marginal constraints and
derived stellar properties, with the median and
central $68\%$ credible interval determined from
the 16th, 50th, and 84th percentiles
~\cite{Lewis:2025GetDist}.


\section{Results and discussion}
\label{sec:results}

\subsection{Constraints on the revised pQCD EOS}

\subsubsection{GW170817 tidal constraints and statistical representations}
\label{subsec:gw_posterior}

\begin{figure}[tbp]
    \centering
    \includegraphics[width=0.98\textwidth]
    {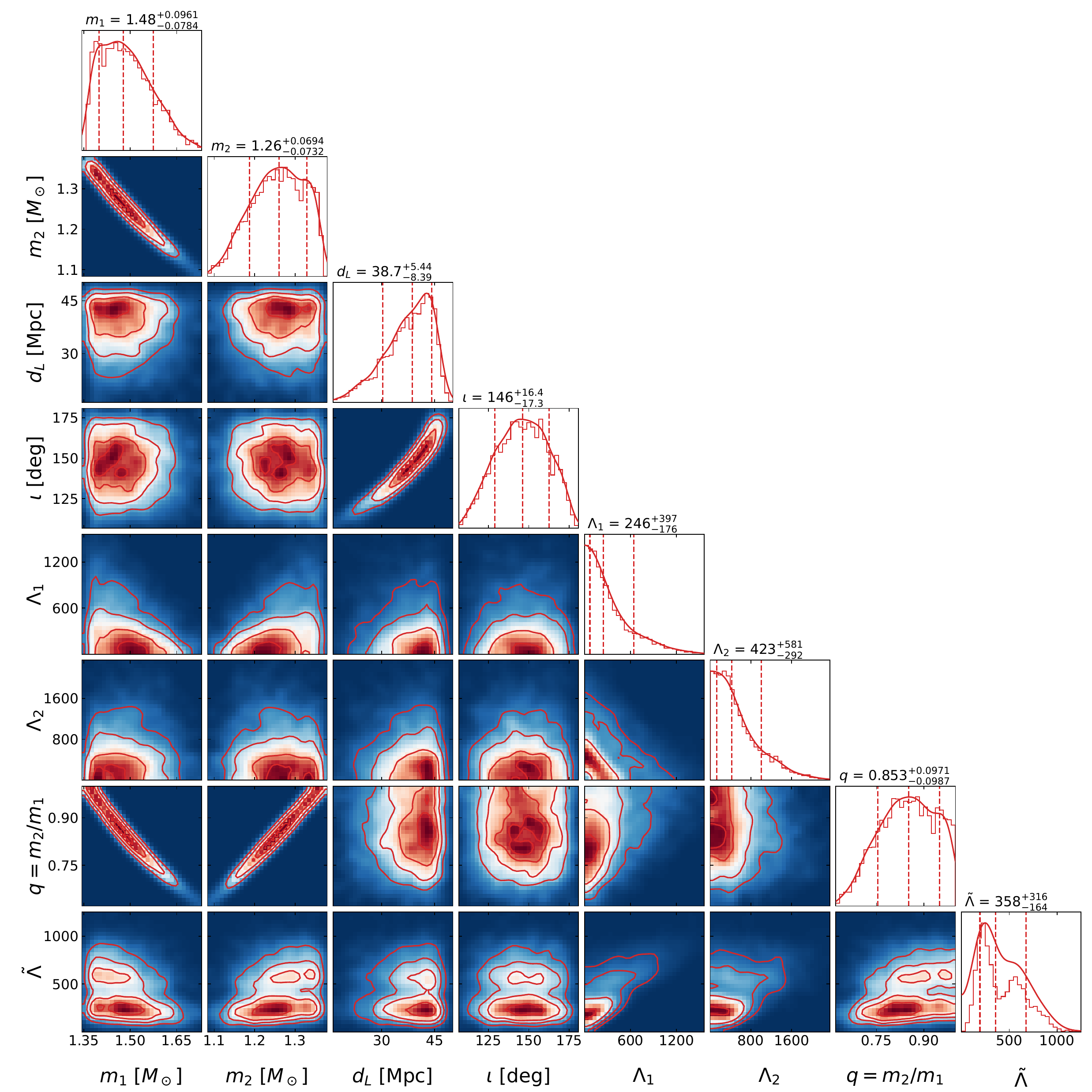}
    \caption{
    Corner plot of the gravitational-wave posterior
    distributions for GW170817, showing the binary
    parameters $m_1$, $m_2$, $d_L$, $\iota$,
    $\Lambda_1$, $\Lambda_2$, $q$, and $\tilde{\Lambda}$.
    The diagonal panels show the marginalized
    one-dimensional posterior distributions, while
    the lower-triangular panels display the corresponding
    two-dimensional joint distributions.
    }
    \label{fig:gw_corner}
\end{figure}

Before discussing the EOS constraints, we first examine
the gravitational-wave posterior distributions of
GW170817, which provide the observational input for
the tidal-likelihood constructions considered in this work
~\cite{LIGOScientific:2018hze}.
Figure~\ref{fig:gw_corner} presents the marginalized
and joint posterior distributions of the binary parameters
obtained from the GW170817 parameter-estimation samples.
The diagonal panels show the marginalized distributions
of the individual parameters, while the lower-triangular
panels illustrate their mutual correlations. In particular,
the component masses $m_1$ and $m_2$ exhibit a pronounced
anticorrelation, reflecting the strong constraint imposed
by the accurately measured chirp mass. The luminosity
distance $d_L$ and inclination angle $\iota$ also show
a correlated posterior distribution. In contrast, the
individual tidal deformabilities $\Lambda_1$ and
$\Lambda_2$ exhibit broad and asymmetric distributions,
indicating substantial uncertainties in the tidal
properties of the two components~\cite{De:2018TidalGW170817,Carson:2019rjx,Ripley:2023lsq}.

The correlations among the component masses, tidal
deformabilities, and derived quantities $q$ and
$\tilde{\Lambda}$ are particularly relevant for the
subsequent EOS inference. Under the common-EOS assumption,
the individual tidal deformabilities are determined by
the corresponding stellar masses through
$\Lambda_i=\Lambda(M_i)$. Consequently, the joint
gravitational-wave posterior contains information beyond
the marginalized distribution of $\tilde{\Lambda}$ alone.
This provides the observational basis for the three
likelihood prescriptions used in the present inference, ranging from
the one-dimensional split-Gaussian approximation to
the two-dimensional KDE in $(\tilde{\Lambda},q)$
and the four-dimensional KDE in
$(\mathcal{M},q,\Lambda_1,\Lambda_2)$.

\subsubsection{EOS parameter constraints and inference robustness}
\label{sec:parameter}

\begin{figure}[tbp]
\centering
\includegraphics[
    width=0.9\textwidth,
    height=0.78\textheight,
    keepaspectratio
]{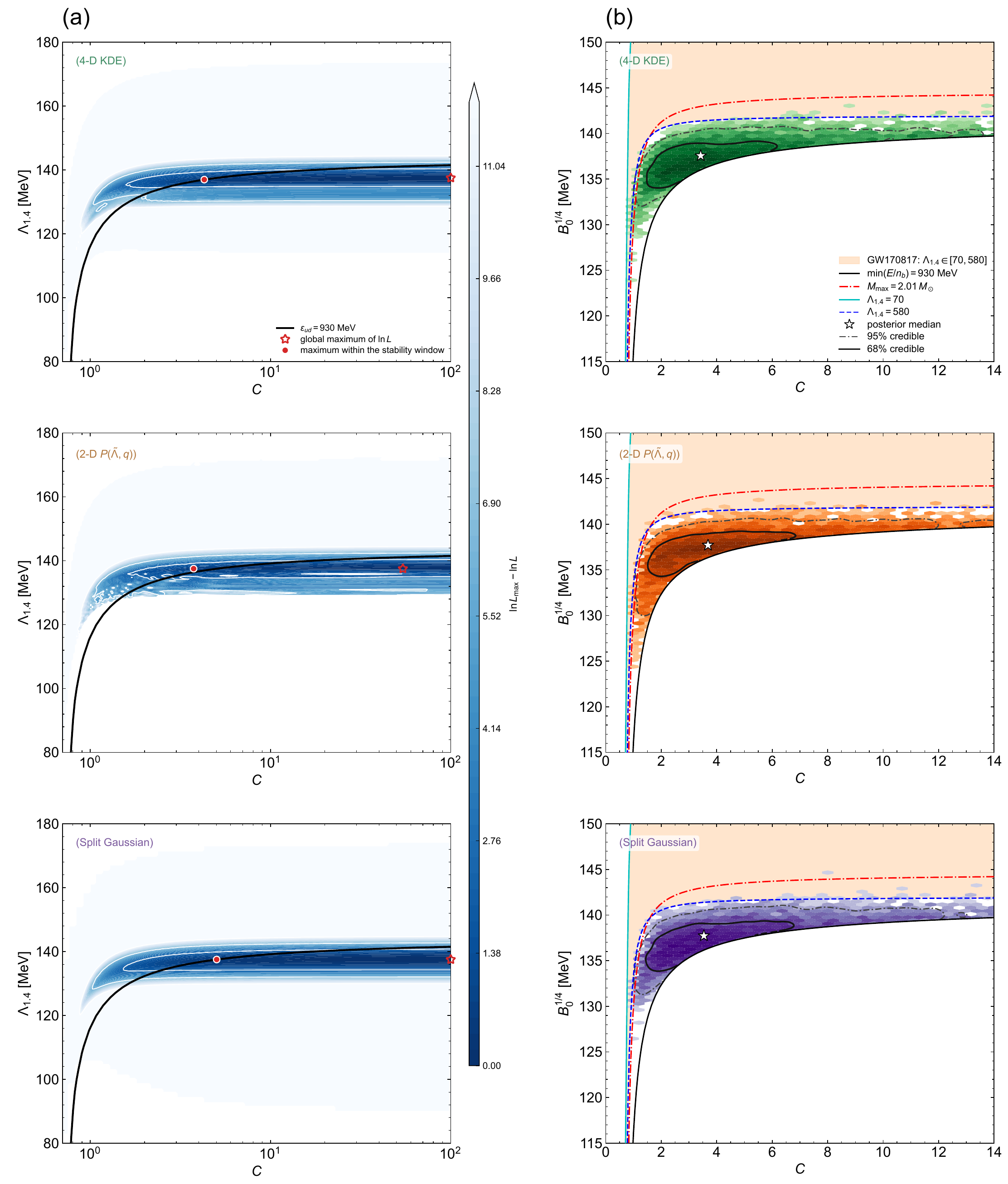}
\caption{
Comparison of the likelihood distributions and posterior
constraints obtained with the three GW170817 tidal-likelihood
prescriptions. Panel (a) shows the likelihood distributions
in the $(C,\Lambda_{1.4})$ plane, while panel (b) presents
the posterior distributions in the $(C,B_0^{1/4})$ plane.
From top to bottom, the rows correspond to the 4-D KDE,
2-D KDE of $p(\tilde{\Lambda},q)$, and split-Gaussian
prescription.
In panel (a), the color scale represents
$\Delta\chi^2=2(\ln\mathcal{L}_{\max}-\ln\mathcal{L})$,
and the white contours indicate the 68.3\%, 95.4\%,
and 99.7\% confidence levels. The black curve denotes
the stability boundary, while the open red star and
filled red circle mark the global and stability-allowed
maximum-likelihood points, respectively.
In panel (b), the orange-shaded region indicates
$70\leq\Lambda_{1.4}\leq580$. The black curve denotes
the stability boundary, and the red dash-dotted curve
marks $M_{\max}=2.01\,M_\odot$. The cyan and blue dashed
curves indicate $\Lambda_{1.4}=70$ and $580$,
respectively. The white stars mark the posterior medians,
and the contours enclose the 68\% and 95\% credible regions.
}
\label{fig:like_param}
\end{figure}

\begin{figure}[tbp]
\includegraphics[width=0.95\textwidth]{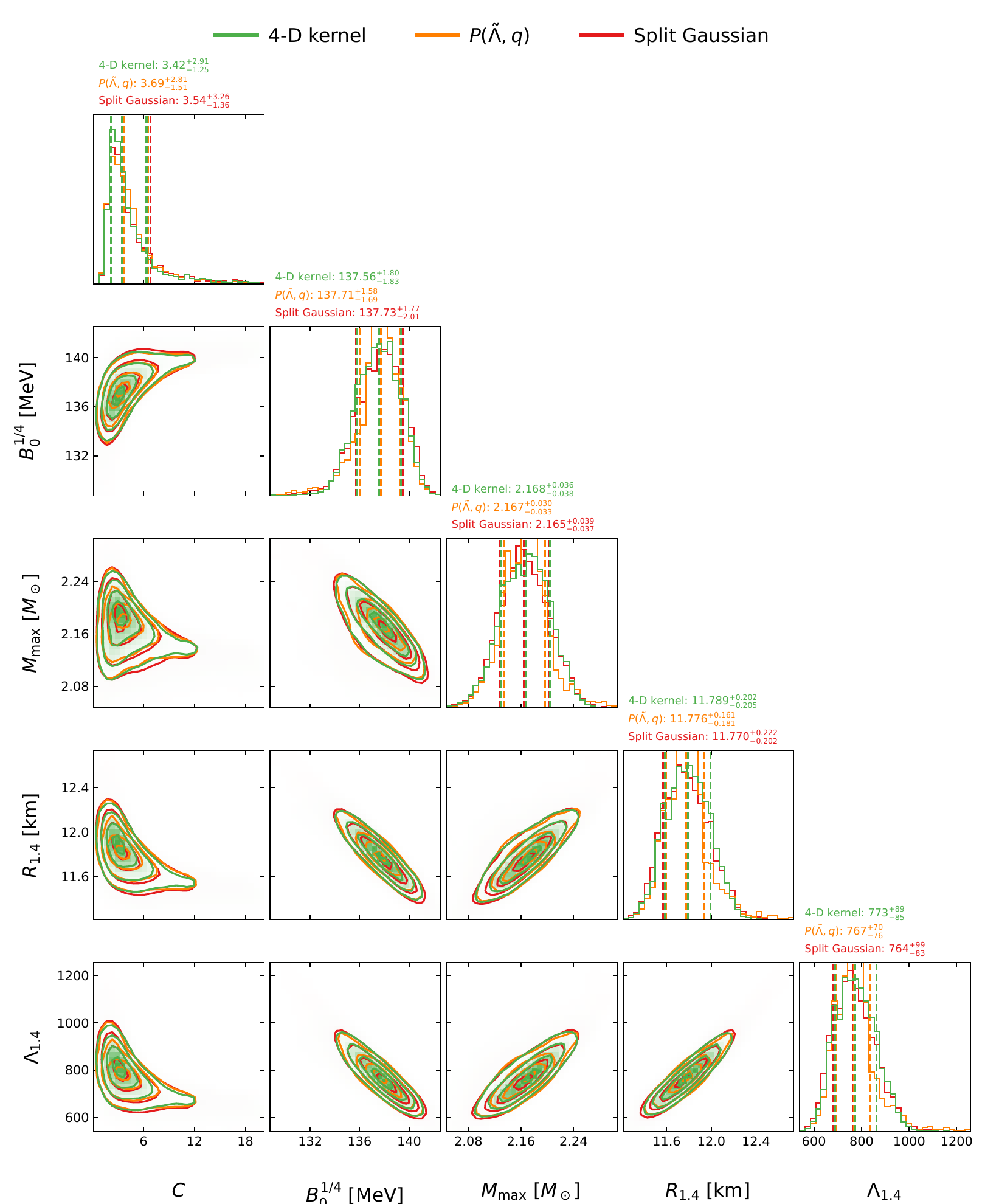}
\caption{Comparison of the joint and marginal posterior
distributions of the EOS parameters
$(C,B_0^{1/4})$ and the derived stellar observables
$(M_{\max},R_{1.4},\Lambda_{1.4})$ obtained using
the three GW170817 tidal-likelihood prescriptions.
The green, orange, and red curves represent the
four-dimensional KDE, two-dimensional KDE of
$p(\tilde{\Lambda},q)$, and one-dimensional
split-Gaussian approximation, respectively.
The diagonal panels show the one-dimensional
marginal posterior distributions, while the
off-diagonal panels display the corresponding
two-dimensional joint posterior distributions.
The vertical dashed lines indicate the
16th, 50th, and 84th percentiles.
The numerical annotations give the posterior
medians and central $68\%$ credible intervals
for each likelihood prescription.}
\label{fig:corner}
\end{figure}

The multimessenger likelihood and the stability requirement
jointly determine the physically admissible parameter region.
Figure~\ref{fig:like_param} compares the likelihood distributions
and posterior constraints obtained using the three GW170817
tidal-likelihood prescriptions.
Figure~\ref{fig:like_param}(a) shows the likelihood distributions
in the $(C,\Lambda_{1.4})$ plane, while
Fig.~\ref{fig:like_param}(b) presents the corresponding posterior
distributions in the $(C,\Bq)$ plane.
From top to bottom, the results correspond to the
four-dimensional KDE, two-dimensional KDE of
$P(\tilde{\Lambda},q)$, and one-dimensional split-Gaussian
approximation.
As shown in Fig.~\ref{fig:like_param}(a), all three prescriptions
exhibit a similar curved likelihood structure, with
$\Lambda_{1.4}$ increasing rapidly at small $C$ and gradually
approaching a plateau at larger $C$. The black curve represents
the stability boundary $\epsu=930$~MeV.
The open red star marks the global maximum-likelihood
point in the extended scan, which occurs at the upper
boundary $C=100$, whereas the filled red circle indicates
the maximum within the stability-allowed region.
The latter lies on the stability boundary, indicating
that the unconstrained likelihood maximum is excluded
by the stability requirement.
The posterior distributions in Fig.~\ref{fig:like_param}(b)
show strong overlap among the three prescriptions.
The posterior medians of $C$ are $3.42$, $3.69$, and $3.54$
for the four-dimensional KDE, two-dimensional KDE, and
split-Gaussian prescriptions, respectively.
The corresponding medians of $B_0^{1/4}$ are
$137.56$, $137.71$, and $137.73$~MeV.
Despite some differences in the posterior widths and shapes,
the inferred central values remain highly consistent.
The orange-shaded region in Fig.~\ref{fig:like_param}(b)
corresponds to the parameter combinations satisfying
the adopted tidal-deformability interval
$70\leq\Lambda_{1.4}\leq580$.
Its intersection with the stability-allowed region
illustrates the additional restrictions imposed by
the combined observational and stability requirements.

Figure~\ref{fig:corner} presents the marginalized and
joint posterior distributions of the EOS parameters
$C$ and $B_0^{1/4}$, together with the derived stellar
observables $M_{\max}$, $R_{1.4}$, and $\Lambda_{1.4}$.
The diagonal panels show the one-dimensional marginal
distributions, while the lower-triangular panels display
the corresponding two-dimensional joint distributions.
Different colors represent the four-dimensional KDE,
two-dimensional KDE of $p(\tilde{\Lambda},q)$, and
split-Gaussian prescriptions, respectively.
The marginal distributions of $B_0^{1/4}$, $M_{\max}$,
$R_{1.4}$, and $\Lambda_{1.4}$ show substantial overlap
among the three prescriptions, indicating consistent
constraints on the EOS parameters and derived stellar
properties.
In contrast, $C$ exhibits a broader and strongly
right-skewed posterior distribution, indicating that
it is less tightly constrained by the multimessenger
observations.
The joint distributions reveal several correlations
among the EOS parameters and stellar observables.
In particular, $B_0^{1/4}$ is anticorrelated with
$M_{\max}$, $R_{1.4}$, and $\Lambda_{1.4}$, reflecting
the influence of the vacuum contribution on the
stellar mass-supporting capability, radius, and
tidal response.
Meanwhile, $M_{\max}$ is positively correlated
with both $R_{1.4}$ and $\Lambda_{1.4}$.
A particularly strong positive correlation is
observed between $R_{1.4}$ and $\Lambda_{1.4}$,
consistent with the strong dependence of tidal
deformability on stellar compactness.
The distributions involving $C$ also exhibit
nonlinear and asymmetric structures.
These correlations illustrate how the EOS
parameters jointly influence the maximum mass,
stellar radius, and tidal response of self-bound
quark stars.
In particular, the positive correlation between
$M_{\max}$ and $\Lambda_{1.4}$ suggests a tension
between achieving a large maximum mass and a
small tidal deformability within the inferred
EOS family.
Overall, the three prescriptions yield consistent
constraints on the stellar properties, despite
some differences in the detailed shapes of
their joint posterior distributions.


\subsubsection{Comparison with a normalizing-flow reconstruction.}

\begin{figure}[tbp]
    \centering
    \includegraphics[width=0.563\textwidth]
    {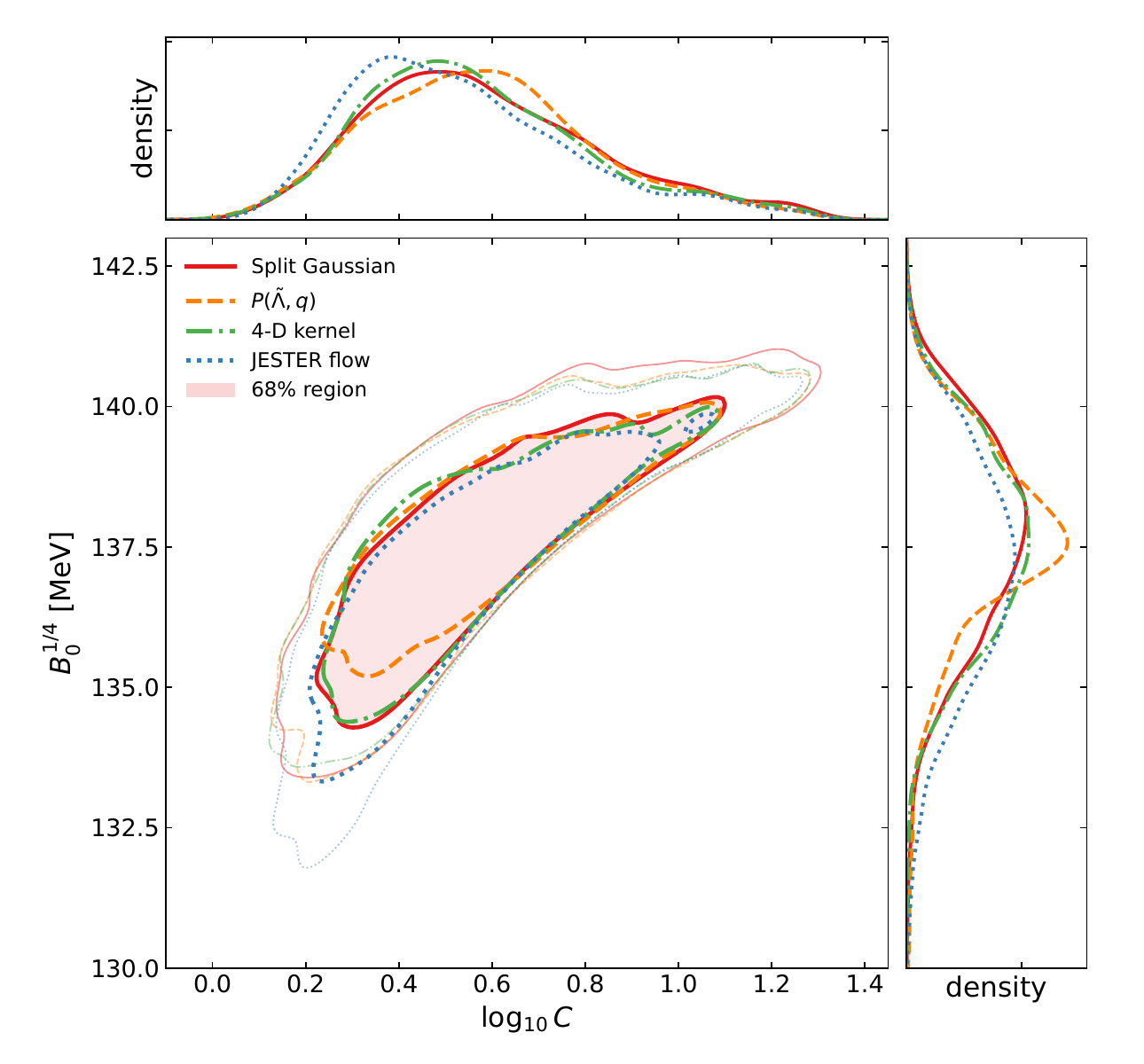}
    \caption{Joint and marginal posterior constraints on the revised
perturbative-QCD EOS parameters from the three baseline GW170817
tidal-likelihood prescriptions and the JESTER normalizing-flow
comparison. The central panel shows the posterior structure in
$(\log_{10} C,B_0^{1/4})$, and the adjacent panels display the
corresponding marginal distributions. The JESTER-flow result is
included as a robustness comparison and is not part of the
three-prescription equal-weight mixture.}
    \label{fig:jester_posterior}
\end{figure}

To further assess the robustness of the inferred EOS
parameters, we compare the three baseline GW170817
tidal-likelihood prescriptions with the JESTER
normalizing-flow approach~\cite{Wouters:2025zju,McGinn2024}.
Figure~\ref{fig:jester_posterior} presents the
one-dimensional marginalized and two-dimensional
joint posterior distributions of the revised
perturbative-QCD EOS parameters
$(\log_{10}C,B_0^{1/4})$.
The upper and right panels show the marginalized
posterior densities of $\log_{10}C$ and
$B_0^{1/4}$, respectively, while the central
panel displays their joint posterior distributions.
The four prescriptions exhibit substantial
overlap in their main regions of posterior
support, although differences are visible
in the peak locations, distribution widths,
and posterior tails.
The marginal distribution of $B_0^{1/4}$
is concentrated within a relatively narrow
range, whereas $\log_{10}C$ exhibits a broader
and more asymmetric distribution.
The joint posterior contours show an elongated
and curved structure, indicating a nonlinear
positive correlation between the two EOS
parameters.
This structure reflects a parameter degeneracy
in which different combinations of $C$ and
$B_0^{1/4}$ can provide comparable agreement
with the multimessenger observations.
Despite differences in the detailed posterior
shapes, the JESTER-flow contours broadly
overlap with those obtained from the three
baseline prescriptions~\cite{McGinn2024,Williams:2021qyt,Wouters:2025csq}.
This agreement indicates that the preferred
EOS parameter region is not substantially
shifted by the alternative gravitational-wave
posterior representation considered here.
The JESTER-flow result therefore provides
an additional consistency check on the
parameter inference, rather than an independent
observational constraint.

\begin{figure}[tbp]
    \centering
    \includegraphics[width=0.9\textwidth]
    {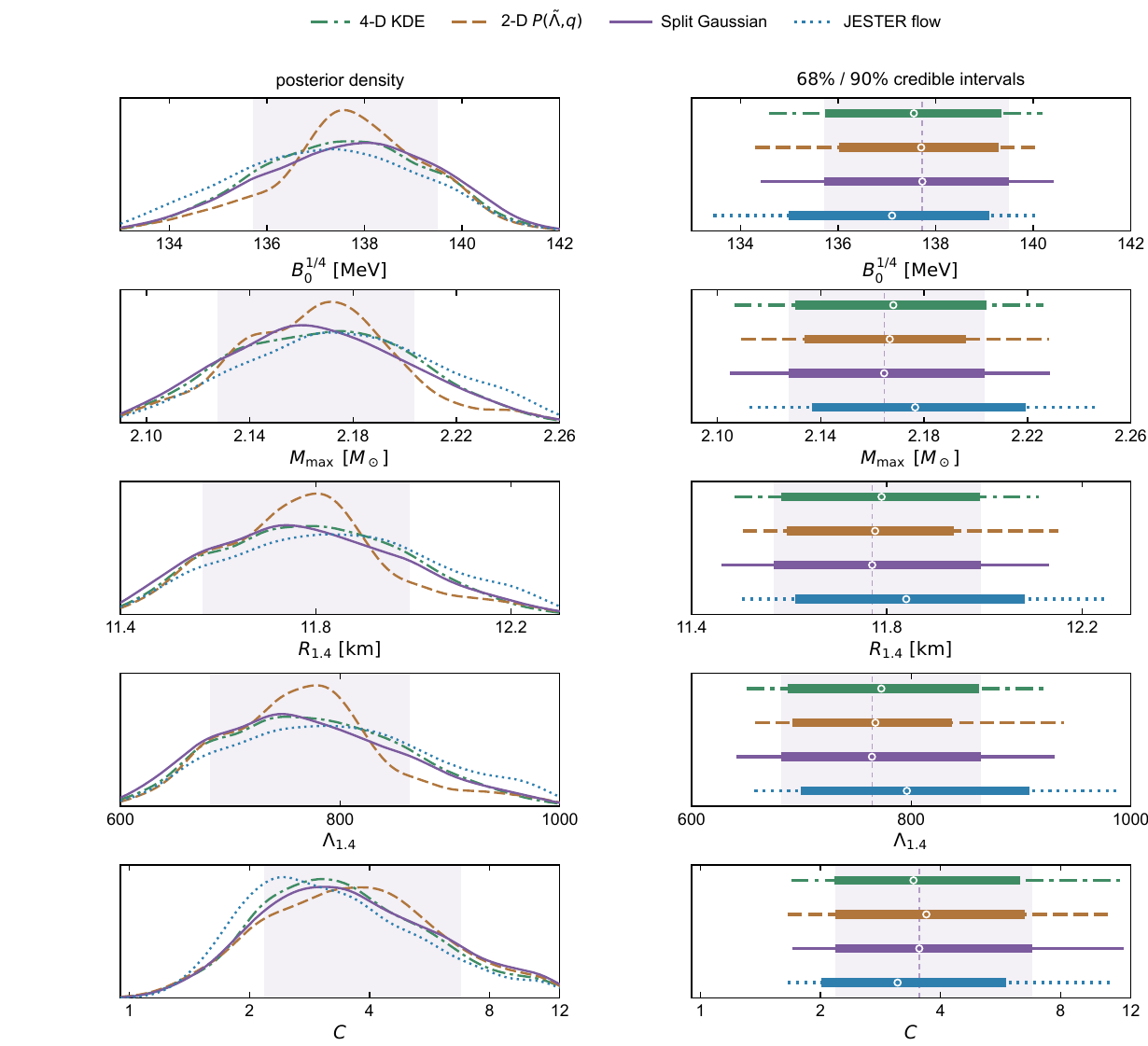}
    \caption{Comparison of the marginal posterior distributions and credible intervals of the EOS parameters and derived stellar observables obtained from four posterior constructions based on the GW170817 tidal information.
The left-hand panels show the one-dimensional marginal posterior distributions, with the green, orange, purple, and blue curves representing the four-dimensional KDE, two-dimensional KDE of $P(\tilde{\Lambda},q)$, split-Gaussian (SG), and JESTER-flow posteriors, respectively.
The right-hand panels summarize the corresponding posterior medians and 68\% and 90\% credible intervals.
The thick horizontal bars and extended thin lines indicate the 68\% and 90\% credible intervals, respectively, while the circular markers denote the posterior medians.
The light-purple shaded bands indicate the 68\% credible intervals of the equal-weight mixture of the three baseline posteriors (4D KDE, 2D KDE, and SG), excluding JESTER-flow; the vertical dashed lines in the right-hand panels mark the corresponding mixture medians.
    }
    \label{fig:posterior_comparison}
\end{figure}

Figure~\ref{fig:posterior_comparison} provides
a more quantitative comparison of the marginal
posterior constraints on the EOS parameters
$C$ and $B_0^{1/4}$ and the derived stellar
observables $M_{\max}$, $R_{1.4}$, and
$\Lambda_{1.4}$.
The left panels show the one-dimensional
marginal posterior distributions, while
the right panels compare the corresponding
posterior medians and $68\%$ and $90\%$
credible intervals.
The four-dimensional KDE, two-dimensional
KDE of $p(\tilde{\Lambda},q)$, and
one-dimensional split-Gaussian approximation
constitute the three baseline prescriptions,
whereas the JESTER normalizing-flow result
is included as an additional comparison.
The marginal posterior distributions of
$B_0^{1/4}$, $M_{\max}$, $R_{1.4}$, and
$\Lambda_{1.4}$ exhibit substantial overlap
among the three baseline prescriptions.
The two-dimensional KDE generally produces
somewhat narrower distributions, while
the other prescriptions show modest
differences in their widths and tails.
Nevertheless, their posterior medians and
credible intervals remain broadly consistent,
indicating that the main stellar predictions
are relatively insensitive to the choice
of tidal-likelihood representation.
In contrast, the posterior distribution of
$C$ is broader and asymmetric, with an
extended tail toward larger values.
This behavior indicates that $C$ is less
tightly constrained by the multimessenger
data; its dependence on the adopted
parameter prior is examined separately.
The JESTER-flow results also overlap
substantially with the baseline constraints,
although some differences remain in the
tails and credible-interval widths.

\begin{table}[t]
    \centering
    \caption{Posterior medians and central $68\%$
    credible intervals of the EOS parameters and
    derived stellar observables obtained from the
    equal-weight mixture of the three baseline
    posterior distributions.}
    \label{tab:combined_posterior}
    \begin{tabular}{lc}
        \hline
        Parameter & Combined posterior \\
        \hline
        $C$ & $3.56^{+2.97}_{-1.38}$ \\
        $B_0^{1/4}$ [$\mathrm{MeV}$]
        & $137.66^{+1.72}_{-1.86}$ \\
        $M_{\max}$ [$M_\odot$]
        & $2.166^{+0.035}_{-0.036}$ \\
        $R_{1.4}$ [$\mathrm{km}$]
        & $11.778^{+0.203}_{-0.196}$ \\
        $\Lambda_{1.4}$ & $768^{+90}_{-81}$ \\
        \hline
    \end{tabular}
\end{table}

The equal-weight mixture of the three
baseline posterior distributions provides
a unified summary of the inferred EOS
parameters and stellar observables.
The resulting posterior medians and
central $68\%$ credible intervals are
summarized in table~\ref{tab:combined_posterior}.
The combined posterior favors a
$1.4\,\Msun$ tidal deformability
$\Lambda_{1.4}=768^{+90}_{-81}$,
whose central $68\%$ credible interval
lies entirely above the adopted
GW170817-motivated upper benchmark of $580$.
This result is consistent with the tension
between the relatively small tidal
deformability favored by GW170817 and
the large maximum mass required by
PSR J0952$-$0607.
Taken together, Figs.~\ref{fig:jester_posterior}
and~\ref{fig:posterior_comparison} demonstrate
that the main posterior constraints remain
broadly consistent across the examined
representations of the gravitational-wave
tidal information.


\subsection{Stellar observables and multimessenger consistency}

\subsubsection{Mass--radius relations and tidal deformability}
\label{subsec:eos_likelihood}

\begin{figure}[tbp]
\centering
\includegraphics[width=0.8\textwidth]{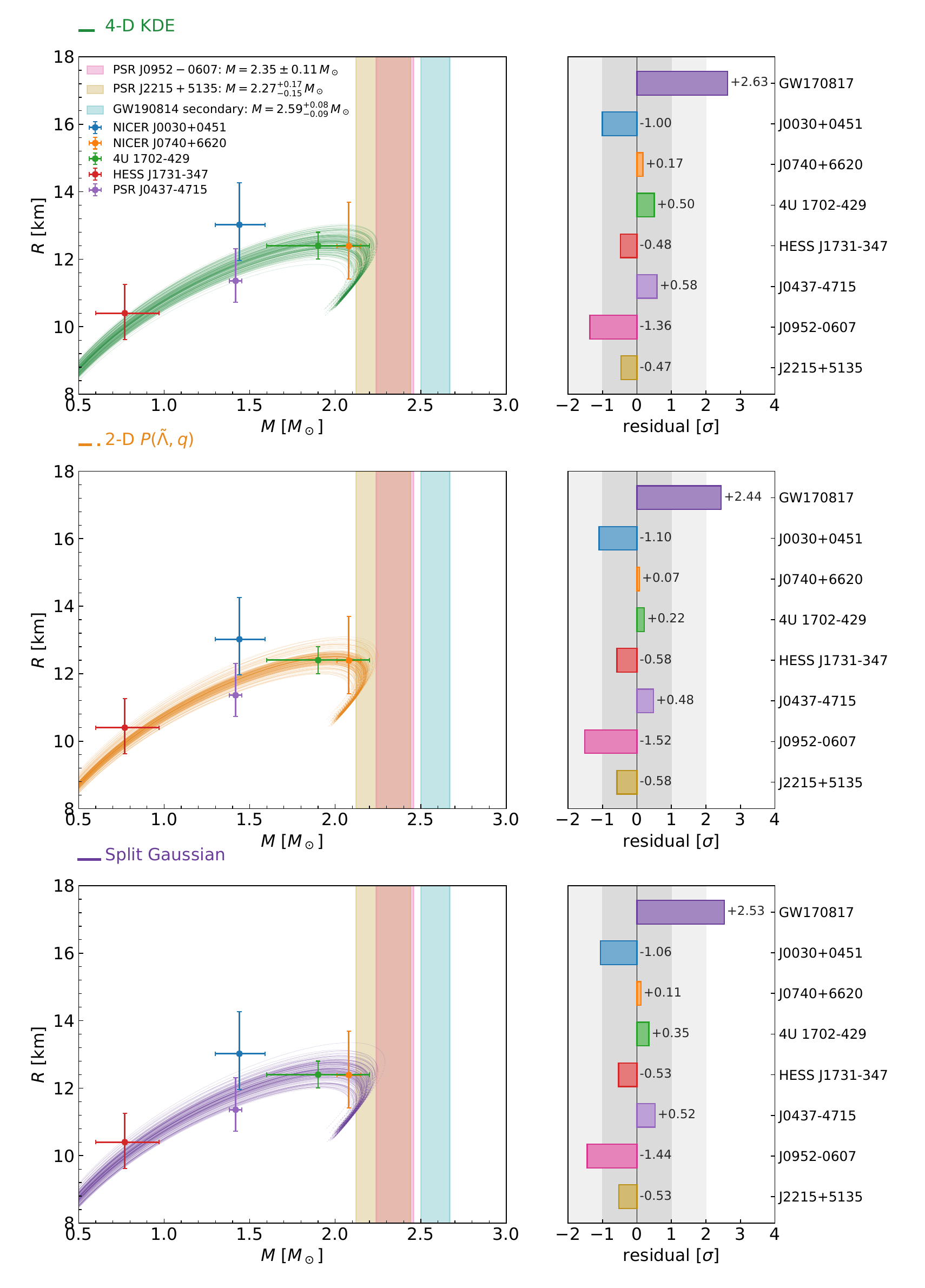}
\caption{
Posterior mass--radius relations and residuals of the
observational constraints obtained using the three
GW170817 tidal-likelihood prescriptions.
The upper, middle, and lower rows correspond to the
four-dimensional KDE, two-dimensional KDE of
$p(\tilde{\Lambda},q)$, and one-dimensional
split-Gaussian approximation, respectively.
The left panels show the posterior mass--radius
relations, with individual curves representing stable
branches from posterior EOS samples. The observational
points with error bars indicate the five adopted
mass--radius measurements. The vertical shaded bands
represent the mass constraints from PSR J0952$-$0607
and PSR J2215$+$5135, together with the mass of the
GW190814 secondary component.
The right panels show the residuals of the eight
likelihood contributions at the maximum-likelihood
point of each prescription, expressed in units of
the corresponding one-sided observational uncertainties.
}
\label{fig:mr}
\end{figure}

Figure~\ref{fig:mr} compares the posterior mass--radius
relations and the residuals of the observational
constraints obtained using the three GW170817
tidal-likelihood prescriptions.
The upper, middle, and lower rows correspond to
the four-dimensional KDE, two-dimensional KDE
of $p(\tilde{\Lambda},q)$, and one-dimensional
split-Gaussian prescriptions, respectively.
In the left panels, each colored curve represents
the stable branch of an individual posterior
EOS sample, with the spread of the curves
reflecting the uncertainties in the predicted
mass--radius relations.
The observational points with error bars show
the five adopted M--R constraints, including
the NICER measurements of PSR J0030$+$0451,
PSR J0740$+$6620, and PSR J0437$-$4715,
as well as the X-ray constraints from
4U~1702$-$429 and HESS~J1731$-$347.
The vertical shaded bands indicate the
mass constraints from PSR J0952$-$0607
and PSR J2215$+$5135, together with
the mass of the GW190814 secondary
component for reference.
The three tidal-likelihood prescriptions yield
very similar posterior mass--radius relations,
indicating that the predicted stellar structure
is relatively insensitive to the adopted
representation of the GW170817 tidal information.
In particular, the inferred radii of
$1.4\,\Msun$ stars are
$R_{1.4}=11.789^{+0.202}_{-0.205}$~km,
$11.776^{+0.161}_{-0.181}$~km, and
$11.770^{+0.222}_{-0.202}$~km for the
four-dimensional KDE, two-dimensional KDE,
and split-Gaussian prescriptions, respectively.
The corresponding maximum masses are also
closely clustered around $2.17\,\Msun$.
The posterior stellar sequences pass through
or near several of the adopted mass--radius
measurements, indicating that the revised
perturbative-QCD EOS can accommodate much
of the available mass--radius information.
Nevertheless, the predicted maximum masses
remain below the central mass estimate
of PSR J0952$-$0607, indicating that
high-mass pulsar observations place stronger
demands on the mass-supporting capability
of the inferred EOS.
The right panels of Fig.~\ref{fig:mr} provide
a complementary assessment of the agreement
with the individual observations.
They show the residuals of the eight likelihood
contributions evaluated at the maximum-likelihood
point of each prescription, expressed in units
of the corresponding one-sided observational
uncertainties.
The residuals associated with GW170817 are
$+2.63\sigma$, $+2.44\sigma$, and
$+2.53\sigma$ for the four-dimensional KDE,
two-dimensional KDE, and split-Gaussian
prescriptions, respectively.
In contrast, the remaining seven observational
constraints have residuals with absolute values
of approximately $1.5\sigma$ or smaller,
indicating that the mass--radius and mass-only
measurements are reasonably well accommodated
by the model.
The similar residual patterns across the
three prescriptions further indicate that
the discrepancy associated with the GW170817
tidal constraint is not substantially reduced
by changing its statistical representation.

\begin{figure}[tbp]
    \centering
    \includegraphics[width=0.92\textwidth]
    {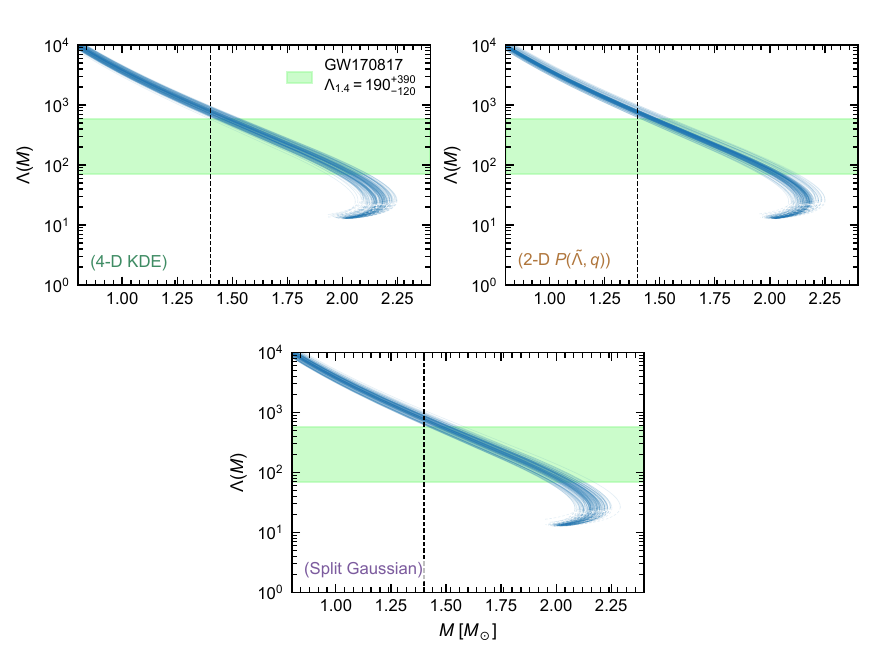}
    \caption{Tidal deformability $\Lambda(M)$ along the stable branches
of the posterior EOS samples obtained using the three
GW170817 tidal-likelihood prescriptions.
The upper, middle, and lower panels correspond to
the four-dimensional KDE, two-dimensional KDE of
$p(\tilde{\Lambda},q)$, and one-dimensional
split-Gaussian approximation, respectively.
Each blue curve represents an individual posterior
EOS sample. The green shaded band indicates the
GW170817 $90\%$ credible interval for the canonical
tidal deformability,
$\Lambda_{1.4}=190^{+390}_{-120}$, corresponding to
$70\leq\Lambda_{1.4}\leq580$.
The vertical dashed line marks $M=1.4\,M_\odot$.
    }
    \label{fig:lam}
\end{figure}

Although the posterior mass--radius relations
are broadly consistent with much of the
adopted observational information, tidal
deformability provides an additional and
complementary test of the EOS.
Figure~\ref{fig:lam} compares the tidal
deformability $\Lam(M)$ along the stable
branches of the posterior EOS samples
obtained using the three tidal-likelihood
prescriptions.
The upper, middle, and lower panels correspond
to the four-dimensional KDE, two-dimensional
KDE of $p(\tilde{\Lambda},q)$, and
one-dimensional split-Gaussian prescriptions,
respectively.
Each blue curve represents an individual
posterior EOS sample.
Since $\Lam$ varies by several orders of
magnitude along the stable branch, the
vertical axis is logarithmic.
The green shaded band indicates the GW170817
$90\%$ credible interval for the canonical
tidal deformability,
$\Lam_{1.4}=190^{+390}_{-120}$, corresponding
to $70\leq\Lam_{1.4}\leq580$,
while the vertical dashed line marks
$M=1.4\,\Msun$.
The three prescriptions yield very similar
predictions for the canonical tidal
deformability.
The posterior medians are
$\Lam_{1.4}=773^{+89}_{-85}$,
$767^{+70}_{-76}$, and
$764^{+99}_{-83}$ for the four-dimensional
KDE, two-dimensional KDE, and split-Gaussian
prescriptions, respectively.
In all three cases, the lower endpoints
of the central $68\%$ credible intervals
lie above the GW170817 upper benchmark
of $580$.
Specifically, these lower endpoints are
approximately $688$, $691$, and $681$,
respectively.
The posterior branches typically cross
the upper boundary of the GW170817
tidal-deformability interval near
$M\simeq1.47\,\Msun$ and remain above
this boundary at lower masses.
These results show that the preference
for relatively large tidal deformabilities
is a common feature of the inferred
EOS configurations rather than an effect
associated with a single tidal-likelihood
prescription.
This behavior highlights the complementary
roles of mass--radius and tidal-deformability
measurements in constraining the EOS.
Although the posterior stellar sequences
accommodate much of the mass--radius
information, their predicted canonical
tidal deformabilities remain systematically
larger than the adopted GW170817 benchmark.
Together with the maximum-mass constraints,
this suggests that reducing the tidal
deformability while maintaining sufficient
mass support is particularly challenging
within the inferred revised perturbative-QCD
EOS family.

To provide a unified summary of the constraints
from the three prescriptions, we also consider
their equal-weight posterior mixture.
The resulting combined canonical tidal
deformability is
$\Lam_{1.4}=768^{+90}_{-81}$,
whose central $68\%$ credible interval
also lies entirely above the GW170817
$90\%$ upper endpoint.
The consistency of the individual posterior
predictions and their mixture indicates
that the relatively large canonical tidal
deformability is a robust feature of
the present analysis with respect to
the three tidal-likelihood representations.
Thus, the tension with the GW170817
tidal-deformability benchmark persists
despite the different treatments of the
gravitational-wave information.
The inferred bulk stellar properties are
broadly consistent with results from other
quark-matter and hybrid-star models
~\cite{Annala:2021gom,Cao:2023rgh,
Oikonomou:2023otn,Cao:2020zxi}.
An independent Bayesian analysis of bag-model
stars~\cite{Miao:2021nuq,Xie:2025sth}
obtains
$M_{\mathrm{TOV}}=2.15^{+0.16}_{-0.14}\,\Msun$,
$R_{1.4}=11.50^{+0.52}_{-0.55}$~km, and
$\Lam_{1.4}=650^{+230}_{-190}$.
These results overlap with the corresponding
constraints obtained from our equal-weight
posterior mixture.
The inferred radius also falls within the
$11.3$--$11.9$~km range reported for
interacting MIT-bag stars~\cite{Xie:2025sth}.
Similarly, a hybrid-star model with a
color-flavor-locked core described using
a nonlocal Nambu--Jona-Lasinio framework
yields $R_{1.4}=11$--$12$~km and
$M_{\max}=2.0$--$2.2\,\Msun$
~\cite{Ayriyan:2025rub},
consistent with the corresponding ranges
found here.

Previous studies have demonstrated that the
compatibility of quark-star models with
astrophysical observations depends sensitively
on the underlying EOS and the treatment of
strong interactions.
Investigations of strange and hybrid stars
have shown that strong-interaction effects
and color superconductivity can play important
roles in supporting massive compact-star
configurations~\cite{Weissenborn:2011qu}. Bayesian analyses of strange quark stars
in normal and color-flavor-locked phases
have used the tidal information from
GW170817 and GW190425 to constrain their
maximum masses, radii, and tidal
deformabilities~\cite{Miao:2021nuq}.
Within an interacting Nambu--Jona-Lasinio
description of color-flavor-locked strange
matter, parameter regions compatible with
GW170817 tidal measurements, massive-pulsar
constraints, and NICER mass--radius
observations have also been identified
~\cite{Lourenco:2021lpn}.
More generally, multimessenger observations
have been used to constrain dense-matter
equations of state and investigate the
possible presence of quark cores in massive
neutron stars~\cite{Annala:2021gom},
while potential observational signatures
of quark matter in neutron-star interiors
have been discussed~\cite{Alford:2019oge}.
For the thermodynamically consistent revised
perturbative-QCD EOS considered here,
previous investigations have shown that
imposing the GW170817 tidal constraint
restricts the maximum mass of self-bound
$ud$ quark stars to approximately
$2.08\,\Msun$~\cite{Wang:2025lwv}.
The present results place the inferred
bulk stellar properties within the ranges
predicted by several independent descriptions
of quark matter.
However, agreement in the mass--radius
properties does not necessarily imply
compatibility with the tidal-deformability
constraint.
For the revised perturbative-QCD EOS
considered here, the systematically larger
$\Lam_{1.4}$ values indicate a persistent
tension with the GW170817 benchmark,
despite the broadly consistent predictions
obtained from the three tidal-likelihood
prescriptions.


\subsubsection{Predictions in the GW170817 component-tidal plane.}

\begin{figure}[tbp]
    \centering
    \includegraphics[width=0.563\textwidth]
    {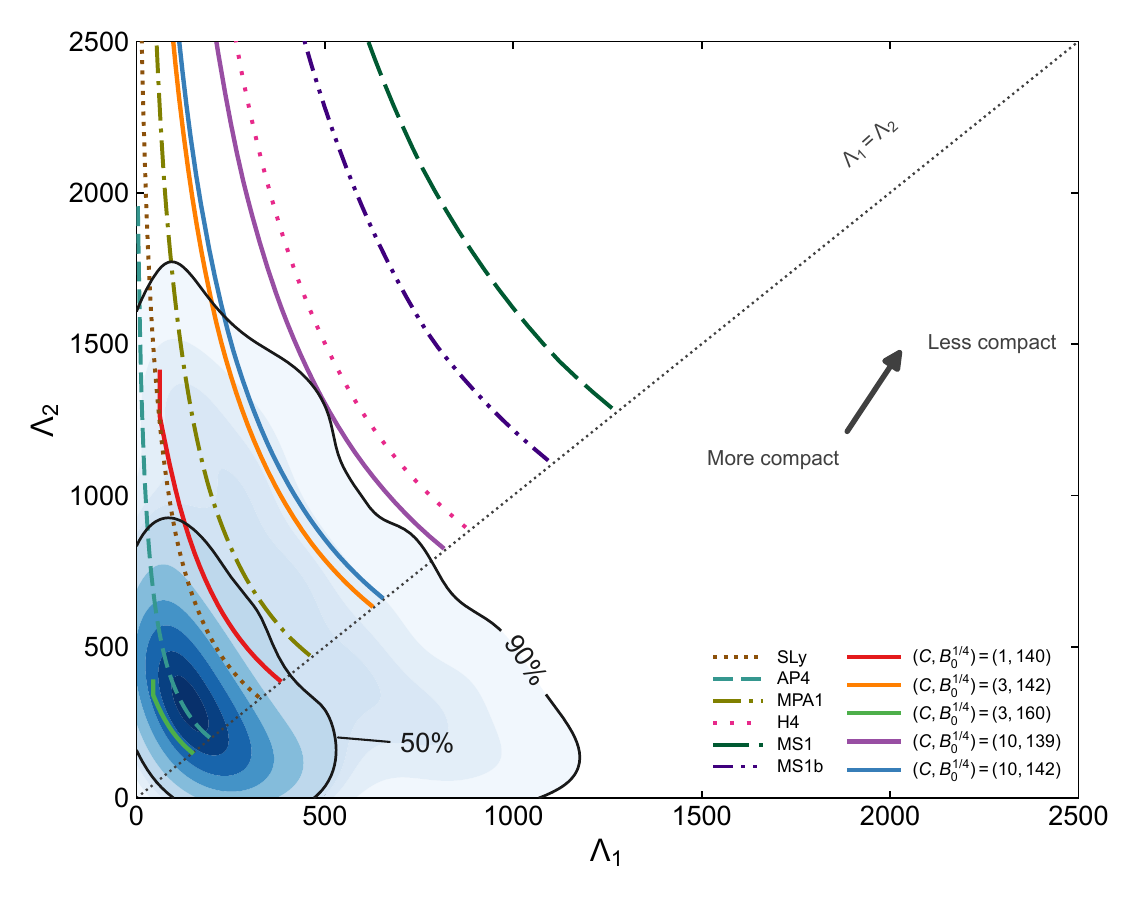}
    \caption{
    Comparison of the GW170817 posterior distribution
    and representative EOS predictions in the
    $(\Lambda_1,\Lambda_2)$ plane.
    The blue shaded contours represent the joint posterior
    distribution of the component tidal deformabilities.
    The dashed curves correspond to six representative
    hadronic EOS models (SLy, AP4, MPA1, H4, MS1, and MS1b),
    taken from the piecewise-polytrope compilation~\cite{Read:2009yp},
    while the solid colored curves show predictions from
    the revised perturbative-QCD EOS for five representative
    parameter combinations $(C,B_0^{1/4})$.
    The gray diagonal indicates $\Lambda_1=\Lambda_2$.
    }
    \label{fig:tidal_plane}
\end{figure}

Figure~\ref{fig:tidal_plane} compares the
GW170817 posterior distribution with the
tidal-deformability predictions of six
representative hadronic EOS models
(SLy, APR4, MPA1, H4, MS1, and MS1b),
and the revised perturbative-QCD EOS in
the $(\Lambda_1,\Lambda_2)$ plane.
The blue shaded contours represent the joint
posterior distribution of the two component
tidal deformabilities, while the dashed curves
correspond to the hadronic EOS models and the
solid colored curves show the predictions of
the revised perturbative-QCD EOS for five
representative parameter combinations
$(C,B_0^{1/4})$.
The gray diagonal line $\Lambda_1=\Lambda_2$
indicates equal tidal deformabilities for
the two components.
For a common EOS, the individual tidal
deformabilities are determined by their
respective masses, $\Lambda_i=\Lambda(M_i)$.
Since the more massive component generally
has a smaller tidal deformability, one
expects $\Lambda_1\leq\Lambda_2$ for
$m_1\geq m_2$.
Accordingly, the EOS predictions exhibit
a characteristic anticorrelated behavior
in the $(\Lambda_1,\Lambda_2)$ plane.
As indicated by the arrow, smaller tidal
deformabilities correspond to more compact
stellar configurations, whereas larger
values indicate less compact stars.
The different EOS models occupy distinct
regions of the component-tidal plane,
reflecting differences in their predicted
stellar compactness and tidal response.
The representative hadronic EOS curves
span a broad range of tidal deformabilities,
with some extending well beyond the main
GW170817 posterior region.
Similarly, the revised perturbative-QCD
predictions exhibit a clear dependence
on the parameters $C$ and $B_0^{1/4}$.
Some parameter combinations yield tidal
deformabilities that overlap with regions
of appreciable GW170817 posterior support,
whereas others predict larger values
and lie farther from the preferred region.
This demonstrates that the component-tidal
observations provide a direct means of
distinguishing among different EOS
parameter choices.

\subsubsection{Maximum mass and tidal deformability}
\label{sec:tension}

\begin{figure}[tbp]
    \centering
    \includegraphics[width=0.920\textwidth]
    {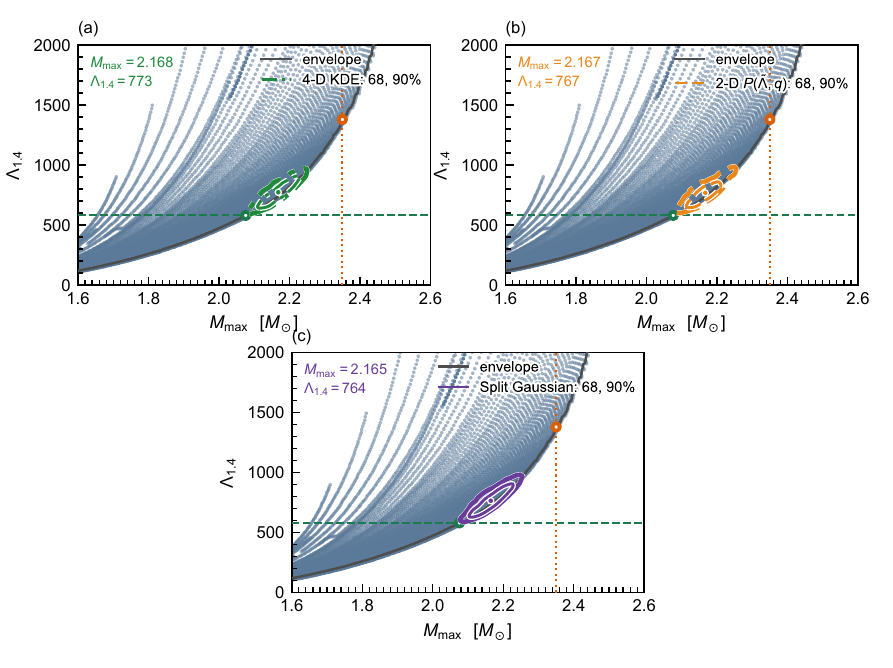}
    \caption{Relation between the maximum stellar mass
    $M_{\max}$ and the canonical tidal deformability
    $\Lambda_{1.4}$ for the three GW170817
    tidal-likelihood prescriptions.
    The upper, middle, and lower panels correspond
    to the four-dimensional KDE, two-dimensional KDE
    of $p(\tilde{\Lambda},q)$, and one-dimensional
    split-Gaussian approximation, respectively.
    The light-blue points represent stable EOS
    configurations evaluated on the surrogate grid,
    and the dark-gray solid curve indicates the
    minimum tidal deformability attainable at each
    maximum mass.
    The colored contours show the $68\%$ and
    $90\%$ posterior credible regions for each
    prescription. The green horizontal dashed line
    marks the adopted GW170817 $90\%$ upper-limit
    benchmark, $\Lambda_{1.4}=580$, while the orange
    vertical dotted line indicates the central
    measured mass of PSR J0952$-$0607,
    $M_{\mathrm{J0952}}=2.35\,M_\odot$.
    The open circles mark the intersections of
    these two benchmark lines with the lower
    boundary. The posterior medians of
    $M_{\max}$ and $\Lambda_{1.4}$ are also
    indicated in each panel.}
    \label{fig:feas}
\end{figure}

The surrogate EOS is evaluated over the stability-allowed
parameter region to examine the relation between the canonical
tidal deformability $\Lam_{1.4}$ and the maximum stellar mass
$M_{\max}$. Figure~\ref{fig:feas} shows the results of this
scan. Each light-blue point represents a stable EOS
configuration evaluated on the surrogate grid, while the
dark-gray solid curve indicates the minimum tidal
deformability attainable at each value of $M_{\max}$.
Stable configurations below this curve are not found within
the parameter region considered. The colored contours show
the $68\%$ and $90\%$ posterior credible regions obtained
using the three GW170817 tidal-likelihood prescriptions.

The green horizontal dashed line marks the adopted
GW170817 $90\%$ upper-limit benchmark,
$\Lam_{1.4}=580$. It intersects the lower boundary at
approximately $M_{\max}\simeq2.077\,\Msun$.
The orange vertical dotted line indicates the central
measured mass of PSR J0952$-$0607,
$M_{\mathrm{J0952}}=2.35\,\Msun$. At this mass, the
minimum attainable tidal deformability is approximately
$\Lam_{1.4}\simeq1.34\times10^3$.
A grid-based scan gives
$\Lam_{1.4}^{\min}\simeq1358$ for configurations with
$M_{\max}\geq2.35\,\Msun$, while a direct scan along the
stability boundary, followed by interpolation to
$M_{\max}=2.35\,\Msun$, gives approximately
$1.337\times10^3$. Given the small difference between
these two estimates, the corresponding lower limit is
quoted as an approximate result,
$\Lam_{1.4}\sim1.34\times10^3$.
These results show that the maximum mass and tidal
deformability cannot simultaneously satisfy the two
adopted observational benchmarks within the scanned
stability-allowed EOS region. In particular, requiring
a smaller tidal deformability limits the maximum mass,
whereas supporting a more massive star requires a
larger tidal deformability. The high-mass end of the
boundary corresponds to smaller values of $C$ and
relatively stronger running coupling.

\begin{table}[tbp]
\caption{Maximum-mass and tidal-deformability bounds
associated with different observational benchmarks,
together with the corresponding posterior probabilities
for the three GW170817 tidal-likelihood prescriptions.}
\label{tab:feasibility3}
\begin{ruledtabular}
\resizebox{0.98\textwidth}{!}{%
  \begin{tabular}{llcccc}
 & & & \multicolumn{3}{c}{Posterior probability of the condition} \\
Benchmark condition & Consequence & GW170817 residual & 4-D KDE & 2-D $P(\tilde\Lambda,q)$ & Split-Gaussian \\
\hline
$M_{\max}\geq 2.08\,\Msun$ & $\Lam_{1.4}\geq 590$ & $\geq+1.69\,\sigma$ & 0.990 & 0.992 & 0.989 \\
$M_{\max}\geq 2.27\,\Msun$ & $\Lam_{1.4}\geq 1029$ & $\geq+3.54\,\sigma$ & 0.001 & 0.012 & 0.003 \\
$M_{\max}\geq 2.35\,\Msun$ & $\Lam_{1.4}\sim1.34\times10^3$ & $\sim+4.8\,\sigma$ & $<10^{-4}$ & $<10^{-4}$ & $<10^{-4}$ \\
\hline
$\Lam_{1.4}\leq 580$ & $M_{\max}\leq 2.077\,\Msun$ & $\leq+1.64\,\sigma$ & 0.002 & 0.002 & 0.003 \\
$\Lam_{1.4}\leq 800$ & $M_{\max}\leq 2.195\,\Msun$ & $\leq+2.57\,\sigma$ & 0.617 & 0.700 & 0.646 \\
$\Lam_{1.4}\leq 1000$ & $M_{\max}\leq 2.263\,\Msun$ & $\leq+3.42\,\sigma$ & 0.995 & 0.978 & 0.989 \\
\end{tabular}
}
\end{ruledtabular}
\end{table}

Table~\ref{tab:feasibility3} provides a more quantitative
comparison of the feasibility bounds and posterior
probabilities for the three tidal-likelihood prescriptions.
The feasibility bounds are identical for all three
prescriptions because they are determined by the
stability-allowed EOS parameter region, rather than by
the posterior construction method.
Requiring $\Lam_{1.4}\leq580$ limits the maximum
attainable mass to approximately $2.077\,\Msun$.
Conversely, supporting a maximum mass of at least
$2.35\,\Msun$ requires
$\Lam_{1.4}\sim1.34\times10^3$.
Thus, these two observational benchmarks cannot be
satisfied simultaneously within the scanned parameter
region. The posterior probabilities lead to the same
conclusion: only $0.2\%$--$0.3\%$ of the posterior
samples satisfy $\Lam_{1.4}\leq580$, while the
probability of obtaining
$M_{\max}\geq2.35\,\Msun$ is below $10^{-4}$
for all three prescriptions.

At the best-fit point
$(C,\Bq)=(5.07,137.55\,\mathrm{MeV})$,
the normalized residuals of the tidal and mass
constraints are
\begin{equation}
\left\{
\begin{aligned}
\Lambda_{1.4} &: \quad +2.53\,\sigma,\\
M_{\max}-M_{\mathrm{J0952}} &: \quad -1.44\,\sigma,\\
M_{\max}-M_{\mathrm{J2215}} &: \quad -0.52\,\sigma.
\end{aligned}
\right.
\label{eq:tension_residuals}
\end{equation}
These results indicate that the tension is mainly
associated with the GW170817 tidal constraint and
the high-mass measurement of PSR J0952$-$0607.
When the model is restricted to
$\Lam_{1.4}=580$, the total log-likelihood decreases
by approximately $4.6$, and the residual associated
with PSR J0952$-$0607 becomes more negative,
changing from $-1.44\,\sigma$ to $-2.5\,\sigma$.
This shows that satisfying the tidal-deformability
benchmark makes it more difficult for the model
to support the high mass of PSR J0952$-$0607.
The feasibility bounds should not be interpreted
as strict observational exclusions, since both
tidal-deformability and mass measurements carry
observational uncertainties. For comparison, the
resulting mass ceiling is comparable to the
$M_{\mathrm{TOV}}\lesssim2.16^{+0.17}_{-0.15}\,\Msun$
bound inferred for nucleonic EOSs using GW170817
and the kilonova of GRB~170817A
~\cite{Rezzolla:2017aly,Nathanail:2021tay,Weih:2019rzo,Capano:2019eae,Essick:2020ghc,Chatziioannou:2020pqz}.

The Bayesian evidence provides an additional way
to examine how the different observational data
affect the overall agreement with the EOS model.
Table~\ref{tab:evidence3} summarizes the evidence
for the three tidal-likelihood prescriptions under
different combinations of observational data.
For each prescription, the evidence differences
are calculated relative to its own baseline.
Removing GW170817 increases $\ln Z$ by
$9.43$--$11.42$, while removing the five
mass--radius measurements or the two mass-only
constraints increases it by $9.33$--$9.41$
or $5.06$--$5.16$, respectively.
In contrast, adding GW190814 decreases $\ln Z$
by $11.39$--$11.98$.
These results show that the Bayesian evidence
depends strongly on the observational data
included in the analysis. However, evidence
differences obtained from different data
combinations should not be interpreted as a
direct ranking of the tensions associated with
individual observations
~\cite{Biswas:2021pvm,Thrane:2019BayesianGW}.

\begin{table}[tbp]
\caption{Bayesian evidence for the three
tidal-likelihood prescriptions.
The evidence differences are defined as
$\Delta\ln Z\equiv\ln Z-\ln Z_{\rm all}$,
where $\ln Z_{\rm all}$ is the evidence obtained
using the full baseline data set for each
prescription.}
\label{tab:evidence3}
\begin{ruledtabular}
\resizebox{0.98\textwidth}{!}{%
  \begin{tabular}{lcccccc}
 & \multicolumn{3}{c}{$\ln Z$} & \multicolumn{3}{c}{$\Delta\ln Z$} \\
Data set & 4-D KDE & 2-D $P(\tilde\Lambda,q)$ & Split-Gaussian & 4-D KDE & 2-D $P(\tilde\Lambda,q)$ & Split-Gaussian \\
\hline
Baseline constraints & $-28.14$ & $-26.14$ & $-26.93$ & $0$ & $0$ & $0$ \\
Without GW170817 & $-16.71$ & $-16.71$ & $-16.71$ & $+11.42$ & $+9.43$ & $+10.21$ \\
Without the five $M$--$R$ data & $-18.81$ & $-16.75$ & $-17.52$ & $+9.33$ & $+9.39$ & $+9.41$ \\
Without the two mass-only data & $-23.04$ & $-21.08$ & $-21.77$ & $+5.09$ & $+5.06$ & $+5.16$ \\
With GW190814 added & $-40.12$ & $-37.54$ & $-38.81$ & $-11.98$ & $-11.39$ & $-11.89$ \\
\end{tabular}
}
\end{ruledtabular}
\end{table}

\subsubsection{Maximum-mass constraints and GW190814}
\label{sec:mmax}

The maximum mass supported by the EOS provides an important
test of its compatibility with massive neutron-star
observations and the secondary component of GW190814.
The posterior medians and central $68\%$ credible intervals
obtained from the three tidal-likelihood prescriptions are
\begin{equation}
\left\{
\begin{aligned}
M_{\max}^{\rm 4D}
&=2.168^{+0.036}_{-0.038}\,\Msun,\\
M_{\max}^{\rm 2D}
&=2.167^{+0.029}_{-0.033}\,\Msun,\\
M_{\max}^{\rm SG}
&=2.165^{+0.039}_{-0.037}\,\Msun.
\end{aligned}
\right.
\label{eq:mmax_three}
\end{equation}

\begin{figure}[tbp]
\centering
\includegraphics[width=0.563\textwidth]
{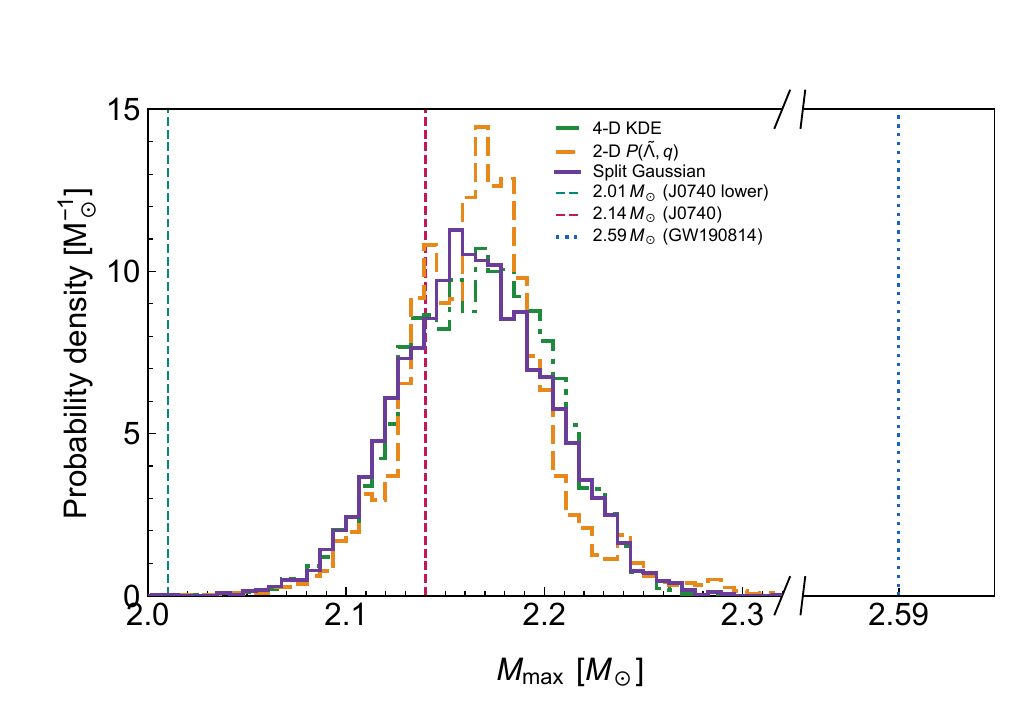}
\caption{Posterior probability densities of the maximum mass
$M_{\max}$ obtained from the nested-sampling runs using
three different GW170817 tidal-likelihood prescriptions.
The green dash-dotted, orange dashed, and purple solid
curves represent the four-dimensional KDE, the
two-dimensional KDE of $P(\tilde{\Lambda},q)$, and the
split-Gaussian approximation, respectively.
The vertical reference lines indicate the conservative
$2.01\,\Msun$ lower-mass benchmark for PSR J0740$+$6620, the earlier $2.14\,M_\odot$ radio-timing mass estimate, and the $2.59\,\Msun$ secondary of
GW190814.
}
\label{fig:mmax}
\end{figure}

To examine whether the reconstructed EOS can support the
massive compact stars indicated by observations, we compare
the inferred maximum-mass distributions with the corresponding
observational mass constraints. As shown in Fig.~\ref{fig:mmax},
the three tidal-likelihood prescriptions yield very similar
results, with the maximum mass concentrated around
$2.17\,\Msun$. This indicates that the inferred maximum mass
is largely insensitive to the adopted likelihood prescription.
For PSR J0740$+$6620, we consider two observational mass
benchmarks, $2.01\,\Msun$ and $2.14\,\Msun$. The probabilities
that the EOS supports a maximum mass exceeding these values are
\begin{equation}
\begin{array}{c|ccc}
 & \mathrm{4D} & \mathrm{2D} & \mathrm{SG} \\
\hline
P(M_{\max}>2.01\,\Msun)
& 1.0000 & 0.9999 & 0.9998 \\
P(M_{\max}>2.14\,\Msun)
& 0.7548 & 0.7817 & 0.7474
\end{array}
\label{eq:pmax}
\end{equation}
All three prescriptions readily satisfy the $2.01\,\Msun$
benchmark, while the probabilities of exceeding
$2.14\,\Msun$ range from approximately $75\%$ to $78\%$.
We further examine whether the EOS can support the secondary
of GW190814, whose central mass is $2.59\,\Msun$.
None of the posterior samples reaches this value.
The largest inferred maximum mass among the three posterior
sample sets is $2.337\,\Msun$, below the lower edge of the secondary's $68\%$ mass interval
$[2.50,2.67]\,\Msun$. Although the absence of samples above
this threshold does not imply an exactly vanishing posterior
probability, it indicates that the present EOS provides very
little support for a stable quark star with a mass as large
as that of the GW190814 secondary.

It is important to distinguish the posterior tail probability $P(M_{\max}>2.59\,\Msun)$ from the posterior-predictive support for the GW190814 mass measurement itself. The former quantifies the fraction of EOS configurations capable of supporting a star more massive than the central observed value, whereas the latter additionally accounts for the observational uncertainty through the mass-only likelihood. We therefore quantify the compatibility of GW190814 with the baseline analysis using the posterior-predictive support probability
\begin{equation}
\mathcal{P}_{\rm qs}^{(k)}
\equiv
\left\langle
\Phi\!\left(
\frac{M_{\max}-2.59\,\Msun}
{\sigma_{\mathrm M}}
\right)
\right\rangle_{p_k(\boldsymbol{\theta}|d_{\rm base})},
\label{eq:bayes}
\end{equation}
where $k$ labels the tidal-likelihood prescription, $\Phi$ is the standard normal cumulative distribution function, and the average is taken over the posterior obtained from the baseline data excluding GW190814.

The posterior-predictive support probability is directly related to the Bayesian evidence ratio:
\begin{equation}
\mathcal{P}_{\rm qs}^{(k)}
=
\frac{
\mathcal{Z}_{\rm base+GW190814}^{(k)}
}{
\mathcal{Z}_{\rm base}^{(k)}
}
=
\exp\!\left(\Delta\ln Z_k\right),
\label{eq:bayes_evidence}
\end{equation}
where $\Delta\ln Z_k$ denotes the change in the logarithmic evidence after including the GW190814 mass constraint. Using the evidence values reported in Table~\ref{tab:evidence3}, we obtain
\begin{equation}
\left\{
\begin{aligned}
\mathcal{P}_{\rm qs}^{\rm 4D}
&=\exp(-11.98)
\simeq 6.3\times10^{-6},\\
\mathcal{P}_{\rm qs}^{\rm 2D}
&=\exp(-11.39)
\simeq 1.13\times10^{-5},\\
\mathcal{P}_{\rm qs}^{\rm SG}
&=\exp(-11.89)
\simeq 6.9\times10^{-6}.
\end{aligned}
\right.
\label{eq:pqs}
\end{equation}
As an independent consistency check, we also evaluate the posterior-predictive average directly using the equal-weight posterior samples in Eq.~\eqref{eq:bayes}. This gives
\begin{equation}
\left\{
\begin{aligned}
\mathcal{P}_{\rm qs,\,sample}^{\rm 4D}
&=6.00\times10^{-6},\\
\mathcal{P}_{\rm qs,\,sample}^{\rm 2D}
&=1.05\times10^{-5},\\
\mathcal{P}_{\rm qs,\,sample}^{\rm SG}
&=6.85\times10^{-6}.
\end{aligned}
\right.
\label{eq:pqs_sample}
\end{equation}
This agreement provides an independent consistency check between the nested-sampling posteriors and the numerical evidence calculations.
For the equal-weight posterior mixture, the direct sample-based estimate is $7.78\times10^{-6}$.
All three tidal-likelihood prescriptions therefore yield extremely small posterior-predictive support for the GW190814 mass measurement. Although the two-dimensional KDE gives a slightly larger support probability than the other two prescriptions, all three estimates remain at the order of $10^{-5}$.

The interpretation of the GW190814 secondary remains model-dependent. Several multimessenger analyses favor a black-hole companion, consistent with arguments based on the observed mass scale and causal constraints~\cite{Tews:2020ylw, Most:2020bba, Nathanail:2021tay, Godzieba:2020tjn}. Nevertheless, other studies have shown that sufficiently stiff hadronic EOSs, rapid rotation, or exotic neutron-star matter may also allow the support of unusually massive compact stars~\cite{Fattoyev:2020cws, Tsokaros:2020hli, Lim:2020zvx, Tan:2020ics, Dexheimer:2020rlp, Zhang:2020zsc,Godzieba:2020tjn}. Alternative quark-star scenarios, including two-flavor and strange quark matter as well as color-superconducting configurations, have also been investigated~\cite{Cao:2020zxi, Roupas:2020nua, Oikonomou:2023otn}. Other analyses emphasize the dependence of the inferred maximum mass on the adopted EOS assumptions and priors~\cite{Lopes:2021yga, Xie:2025sth,Grundler:2026zus,Gupta:2022qgg,Walker:2024loo}.
The present result should therefore not be interpreted as a model-independent identification of the GW190814 secondary. Rather, it addresses the more specific question of whether the adopted revised perturbative-QCD two-flavor self-bound EOS can support an object with the measured mass of the GW190814 secondary. The consistently small posterior-predictive support obtained from all three tidal-likelihood prescriptions indicates that such an interpretation is strongly disfavored within the present EOS framework.

\begin{figure}[tbp]
    \centering
    \includegraphics[
        width=0.563\textwidth,
        keepaspectratio
    ]{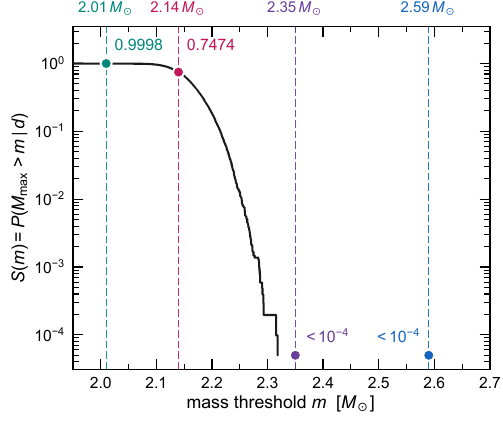}

    \caption{
    Posterior survival probability of the maximum
    gravitational mass as a function of the mass
    threshold $m$.
    The black solid curve represents
    $S(m)=P(M_{\max}>m\mid d)$, evaluated from
    the posterior distribution of the maximum
    gravitational mass.
    The vertical dashed lines indicate four
    representative mass thresholds:
    $2.01\,M_\odot$,
    $2.14\,M_\odot$,
    $2.35\,M_\odot$, and
    $2.59\,M_\odot$.
    }
    \label{fig:mass_survival}
\end{figure}

To quantify the posterior support for different
maximum-mass thresholds, we introduce the
survival function of the maximum gravitational
mass,
\begin{equation}
S(m)
=
P(M_{\max}>m\mid d)
=
\int_m^\infty
p(M_{\max}\mid d)\,\mathrm{d}M_{\max},
\label{eq:mass_survival_function}
\end{equation}
where $p(M_{\max}\mid d)$ is the posterior
probability density of the maximum mass
conditioned on the observational data $d$.
The survival function measures the posterior
probability that an EOS can support a stable
stellar configuration with a maximum mass
exceeding a specified threshold $m$.
Figure~\ref{fig:mass_survival} shows the
resulting survival probability as a function
of the mass threshold.
The probability decreases rapidly with
increasing $m$, particularly between
approximately $2.1\,\Msun$ and $2.3\,\Msun$,
consistent with the maximum-mass posterior
being concentrated near $2.17\,\Msun$.
For the two representative mass benchmarks
of PSR J0740$+$6620, the survival probabilities
are $S(2.01\,\Msun)=0.9998$ and
$S(2.14\,\Msun)=0.7474$.
These values indicate that nearly all
posterior EOS configurations can support
a star above $2.01\,\Msun$, while
approximately $75\%$ can support a
maximum mass exceeding $2.14\,\Msun$.
In contrast, the survival probabilities
at $2.35\,\Msun$ and $2.59\,\Msun$
are both below $10^{-4}$.
These thresholds correspond to the
central-mass benchmarks of
PSR J0952$-$0607 and the GW190814
secondary, respectively.
The sharp decline in the survival
probability therefore indicates that
the inferred revised perturbative-QCD
EOS can readily support stars near
$2\,\Msun$, whereas configurations
capable of supporting substantially
larger masses receive very little
posterior support.
This behavior is consistent with the
maximum-mass limitations identified
in the mass--tidal feasibility analysis.


\subsection{Reconstructed equation of state and quark-star structure}

\subsubsection{Posterior equation of state}
\label{sec:eos}

\begin{figure}[tbp]
\centering
\includegraphics[width=0.563\textwidth]
{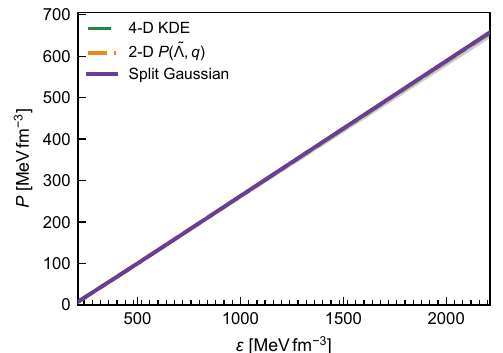}
\caption{
Posterior pressure--energy-density relations of the two-flavor
$ud$ quark-matter EOS obtained using three different
GW170817 tidal-likelihood prescriptions.
The green dash-dotted, orange dashed, and purple solid curves
represent the four-dimensional KDE, the two-dimensional KDE
of $P(\tilde{\Lambda},q)$, and the Split-Gaussian (SG)
approximation, respectively.
The shaded regions indicate the corresponding central
$16$--$84\%$ posterior intervals.
}
\label{fig:eos}
\end{figure}

Figure~\ref{fig:eos} compares the posterior EOS profiles
obtained from the three GW170817 tidal-likelihood
prescriptions. Their median pressure--energy-density
relations are nearly identical, with a maximum pressure
difference of $0.94\,\mathrm{MeV\,fm^{-3}}$
(less than $0.18\%$ for
$\epsilon>1000\,\mathrm{MeV\,fm^{-3}}$).
The relative ordering of the curves varies with energy
density, showing no systematic preference for a stiffer
EOS. The posterior uncertainties are mainly driven by
$C$, while $\Bq$ contributes less.

At $\epsilon_0=800\,\mathrm{MeV\,fm^{-3}}$, we quantify
the width of the central $68\%$ pressure interval as
\begin{equation}
w_{16\text{--}84}
=
P_{84}(\epsilon_0)-P_{16}(\epsilon_0),
\label{eq:wband}
\end{equation}
where $P_{16}$ and $P_{84}$ denote the corresponding
posterior percentiles. At the nearest grid point,
$\epsilon\simeq797.3\,\mathrm{MeV\,fm^{-3}}$,
the widths are $6.06$, $4.51$, and
$5.99\,\mathrm{MeV\,fm^{-3}}$ for the 4-D KDE,
2-D KDE, and SG prescriptions, respectively.
Thus, the 2-D KDE gives a slightly narrower posterior
band, while the other two yield comparable uncertainties.

The reconstructed EOS also exhibits the characteristic
self-bound nature of the adopted two-flavor quark matter.
Each posterior realization terminates at its own surface
energy density, where the pressure vanishes, rather than
extending continuously to zero energy density.
The corresponding surface energy densities are
$\epsilon_{\rm surf}=188^{+9}_{-10}\,\mathrm{MeV\,fm^{-3}}$
(central $68\%$ of the posterior; individual realizations
span roughly $160$--$210\,\mathrm{MeV\,fm^{-3}}$), slightly
below the lower end of the displayed energy-density
range. Unlike conventional neutron-star EOS constructions
that generally include a low-density hadronic crust,
the present model describes the stellar surface directly
through the zero-pressure condition of quark matter,
without an additional hadronic branch. Overall, the close agreement among the three posterior
EOS profiles demonstrates that the inferred thermodynamic
properties are robust against the choice of GW170817
tidal-likelihood prescription.


\subsubsection{Radial profiles of self-bound stars.}

\label{sec:radial_structure}

\begin{figure}[tbp]
    \centering
    \includegraphics[
        width=\textwidth,
        keepaspectratio
    ]{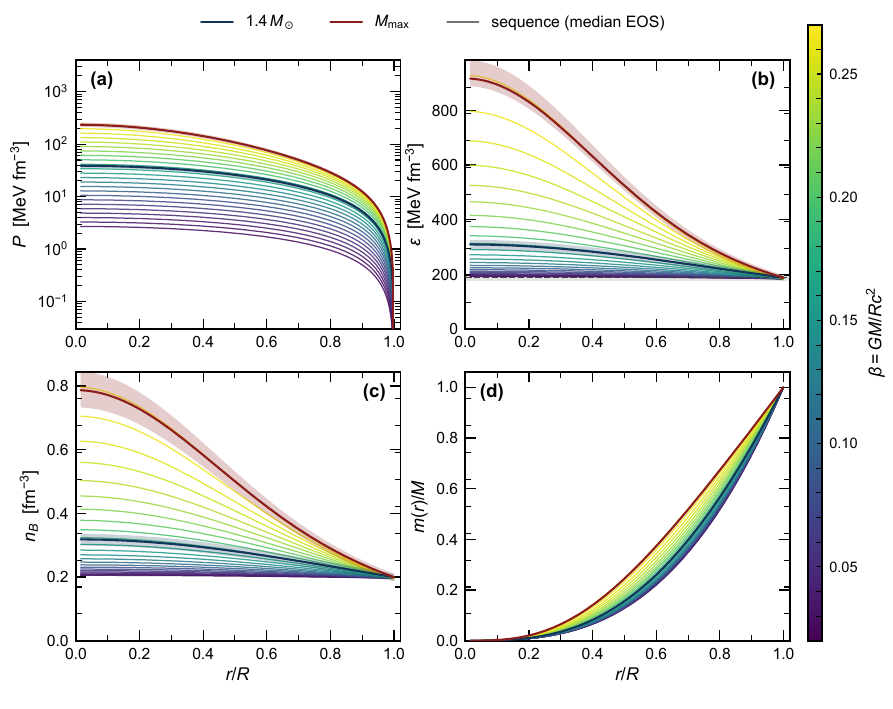}

    \caption{
    Radial profiles of representative self-bound
    two-flavor $ud$ quark-star configurations.
    Panels (a), (b), and (c) show the pressure,
    energy density, and baryon number density as
    functions of the normalized radial coordinate
    $r/R$, respectively. Panel (d) shows the enclosed
    gravitational mass normalized by the total
    stellar mass, $m(r)/M$.
    The thick blue and red curves represent the
    $1.4\,M_\odot$ and maximum-mass configurations,
    respectively, while the thin colored curves show
    the stellar sequence obtained using the median
    posterior EOS. The color scale indicates the
    stellar compactness
    $\beta=GM/(Rc^2)$.
    The shaded regions indicate the corresponding
    posterior uncertainties.
    }
    \label{fig:radial_structure}
\end{figure}

To investigate the internal structure of the
inferred self-bound quark stars, we examine
the radial profiles of representative stellar
configurations obtained from the median-posterior
EOS.
Figure~\ref{fig:radial_structure} shows the
pressure, energy density, baryon number density,
and enclosed gravitational mass as functions
of the normalized radial coordinate $r/R$.
The thick blue and red curves correspond to
the canonical $1.4\,\Msun$ and maximum-mass
configurations, respectively, while the thin
colored curves represent the stellar sequence
constructed from the median EOS.
Their colors indicate the stellar compactness
$\beta=GM/(Rc^2)$, and the shaded regions
show the corresponding posterior uncertainties.

As shown in Figs.~\ref{fig:radial_structure}(a)--(c),
the pressure, energy density, and baryon number
density decrease monotonically from the stellar
center toward the surface.
The maximum-mass configuration has substantially
higher central values of these quantities than
the $1.4\,\Msun$ configuration.
This trend reflects the higher central pressures
and densities required to support more massive
quark stars against gravitational compression.
The profiles also vary systematically with
compactness, indicating that the internal
thermodynamic conditions are closely connected
to the macroscopic stellar properties.
A particularly important feature is observed
at the stellar surface.
Although the pressure vanishes at $r=R$,
both the energy density and baryon number
density remain finite.
For the median-posterior EOS, the surface
energy density is approximately
\begin{equation}
\epsilon_{\rm surf}
\simeq 187\,\mathrm{MeV\,fm^{-3}},
\label{eq:surface_energy_density}
\end{equation}
consistent with the central $68\%$
posterior interval of
$178$--$197\,\mathrm{MeV\,fm^{-3}}$
obtained from the reconstructed EOS ensemble.
The baryon number density likewise remains
nonzero at the zero-pressure surface.
This behavior is a characteristic feature
of self-bound quark matter, in which the
stellar boundary is reached at zero pressure
without the matter density vanishing.
It therefore provides a direct manifestation
of the self-bound nature of the revised
perturbative-QCD EOS.

Figure~\ref{fig:radial_structure}(d) shows
the enclosed gravitational mass normalized
by the total stellar mass, $m(r)/M$.
The enclosed mass increases from zero at
the center to unity at the stellar surface.
At the same normalized radius, the
maximum-mass configuration encloses a
larger fraction of its total mass than
the canonical $1.4\,\Msun$ configuration,
indicating a more centrally concentrated
normalized mass distribution.
This difference further illustrates how
the internal matter distribution changes
along the stellar sequence as the mass
and compactness increase.
Together, these radial profiles show that
the reconstructed revised perturbative-QCD
EOS predicts finite surface densities
and systematic changes in the internal
structure with increasing stellar mass.
In particular, the higher central pressures
and densities of massive configurations
highlight the connection between the
high-density properties of quark matter
and the stellar mass-supporting capability.



\subsubsection{Central densities and stable sequences.}

\label{sec:central_density}

\begin{figure}[tbp]
    \centering
    \includegraphics[width=0.563\textwidth]
    {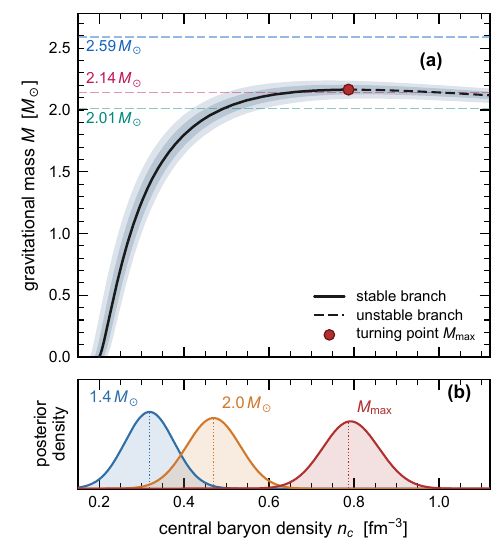}
    \caption{
    Gravitational mass as a function of the central
    baryon number density.
    Panel (a) shows the mass--central-density
    relation obtained from the TOV sequences.
    The solid black curve represents the stable
    branch, whereas the dashed black curve indicates
    the unstable branch. The red circle marks the
    turning point corresponding to the maximum
    gravitational mass, $M_{\max}$. The shaded
    regions represent the uncertainty associated
    with the posterior EOS reconstruction.
    The horizontal dashed lines indicate reference
    gravitational masses of $2.01\,M_\odot$,
    $2.14\,M_\odot$, and $2.59\,M_\odot$.
    Panel (b) displays the posterior distributions
    of the central baryon number density for
    $1.4\,M_\odot$, $2.0\,M_\odot$, and
    maximum-mass configurations, shown in blue,
    orange, and red, respectively. The vertical
    dotted lines indicate the corresponding
    representative central-density estimates.
     }
    \label{fig:mass_central_density}
\end{figure}

To further investigate the internal properties
and stability of the inferred self-bound $ud$
quark stars, Fig.~\ref{fig:mass_central_density}
shows the gravitational mass as a function
of the central baryon number density $n_c$,
together with the corresponding central-density
posterior distributions for representative
stellar configurations.
As shown in Fig.~\ref{fig:mass_central_density}(a),
the gravitational mass initially increases
with central baryon density, but its growth
gradually slows as the stellar sequence
approaches the maximum-mass configuration.
For the representative EOS, the maximum
gravitational mass is approximately
\begin{equation}
M_{\max}\simeq2.16\,\Msun,
\label{eq:maximum_mass_nc}
\end{equation}
at a central baryon number density of
\begin{equation}
n_{c,\max}\simeq0.79\,\mathrm{fm}^{-3}.
\label{eq:maximum_central_density}
\end{equation}
Beyond this turning point, increasing the
central density leads to a decrease in
gravitational mass along the unstable branch.
According to the standard turning-point
criterion for nonrotating stellar sequences,
the maximum-mass point marks the onset of
radial instability.
This behavior shows that increasing the
central density does not indefinitely
increase the mass that the EOS can
support in a stable configuration.
The horizontal reference lines indicate
the mass benchmarks of $2.01$, $2.14$,
and $2.59\,\Msun$.
The representative stable sequence
extends beyond $2.01\,\Msun$ and reaches
a maximum mass close to $2.14\,\Msun$,
whereas the $2.59\,\Msun$ benchmark lies
well above its maximum mass.
This comparison is consistent with the
maximum-mass constraints obtained from
the multimessenger posterior analysis.

Figure~\ref{fig:mass_central_density}(b)
shows that the posterior distributions
of central baryon density shift
systematically toward higher values
as the stellar mass increases.
The distributions peak at approximately
$0.32\,\mathrm{fm}^{-3}$ for
$1.4\,\Msun$ configurations,
$0.47\,\mathrm{fm}^{-3}$ for
$2.0\,\Msun$ configurations, and
$0.79\,\mathrm{fm}^{-3}$ for
maximum-mass configurations.
Their finite widths reflect the
uncertainties in the reconstructed EOS.
The systematic shift indicates that
more massive quark stars probe
progressively denser regions of
quark matter.
In particular, the maximum-mass
configurations reach substantially
higher central densities than
canonical-mass stars, even though
the increase in gravitational mass
becomes relatively small near
the stability limit.
Together, the mass--central-density
relation and the central-density
posterior distributions connect
the internal properties of quark
matter with the maximum mass and
stability of the resulting stars.
The turning point identifies the
limit of stable mass support,
while the posterior distributions
quantify the central densities
associated with different stellar
masses.
These results complement the radial
structure analysis and show how
the high-density behavior of the
revised perturbative-QCD EOS
influences the properties of
self-bound $ud$ quark stars.



\subsection{Strong-interaction and thermodynamic properties}

\subsubsection{Running coupling and perturbative reliability}
\label{sec:strong_coupling}

\begin{figure}[tbp]
    \centering
    \includegraphics[
        height=0.46\textheight,
        keepaspectratio
    ]{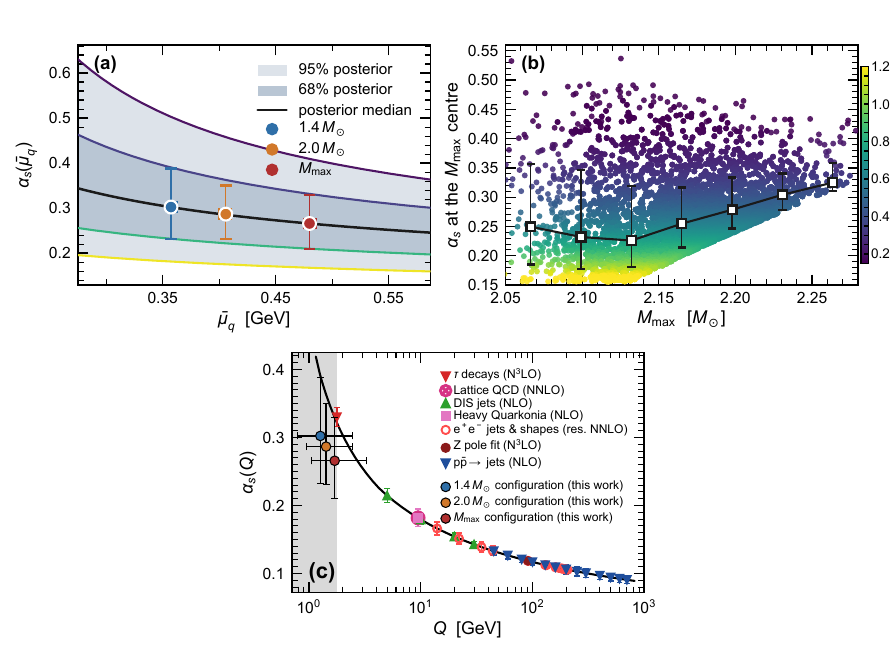}
    \caption{
    Posterior constraints on the strong coupling constant
    and its connection to quark-star properties.
    (a) Dependence of $\alpha_s$ on the averaged quark
    chemical potential $\bar{\mu}_q$.
    The light and dark shaded regions indicate the
    95\% and 68\% posterior credible intervals,
    respectively, while the black solid curve represents
    the posterior median. The blue, orange, and red markers
    represent the $1.4\,M_\odot$, $2.0\,M_\odot$, and
    maximum-mass configurations, respectively.
    (b) Posterior samples of the central strong coupling
    $\alpha_s$ at the maximum-mass configuration as a
    function of $M_{\max}$. The color scale represents
    $\log_{10} C$, with darker colors corresponding to
    smaller values of $C$. Black squares show the binned
    posterior summary, and the error bars indicate the
    corresponding statistical uncertainties.
    (c) Comparison of the strong coupling inferred in
    this work with independent QCD determinations from
    $\tau$ decays, lattice QCD, DIS jets, heavy quarkonia,
    $e^+e^-$ event shapes, $Z$-pole observables, and
    $p\bar{p}$ jets. The black solid curve denotes the
    reference running-coupling behavior. The colored
    circular markers represent the model predictions
    for the three representative quark-star configurations.
    }
    \label{fig:alpha_s_constraints}
\end{figure}

To examine the strong-interaction properties
of the inferred quark matter, we investigate
the running coupling constant $\alpha_s$
and its relation to the stellar observables.
Figure~\ref{fig:alpha_s_constraints} presents
the posterior constraints on $\alpha_s$
as a function of the averaged quark
chemical potential, its correlation
with the maximum stellar mass, and a
comparison with independent QCD
determinations at different energy scales.

As shown in Fig.~\ref{fig:alpha_s_constraints}(a),
the strong coupling constant decreases
with increasing averaged quark chemical
potential $\bar{\mu}_q$.
This behavior is consistent with the
asymptotic freedom of QCD.
The representative central configurations
of $1.4\,\Msun$, $2.0\,\Msun$, and
maximum-mass stars shift toward higher
chemical potentials as the stellar mass
increases, accompanied by a gradual
decrease in the coupling strength.
However, the posterior credible bands
remain relatively broad, reflecting
uncertainties in the inferred EOS parameters.
In particular, because the renormalization
scale is defined by
$\mu_R=C\bar{\mu}_q$, the coupling
depends on both the chemical potential
and the scale coefficient $C$.
Figure~\ref{fig:alpha_s_constraints}(b) shows the relationship between the
central coupling constant of the
maximum-mass configuration and
$M_{\max}$.
The posterior samples exhibit an
overall positive correlation,
with a correlation coefficient
$\rho=+0.39$, although considerable
scatter remains.
The color distribution further shows
that smaller values of $C$ are generally
associated with stronger coupling.
These results indicate that the
maximum-mass predictions are linked
to the renormalization-scale dependence
of the revised perturbative-QCD EOS.

Figure~\ref{fig:alpha_s_constraints}(c)
compares the strong coupling constants
obtained in the present work with
independent QCD results at different
energy scales $Q$.
The external results are obtained from
various sources, including $\tau$-lepton
decays, lattice QCD, deep-inelastic
scattering, heavy-quarkonium production,
$e^+e^-$ annihilation, Z-pole measurements,
and hadron-collider jet production.
As indicated in the legend, these results
are extracted using different perturbative
approximations, ranging from NLO to N$^3$LO.
The external results show a clear trend:
$\alpha_s$ decreases as the energy scale
increases. This is the expected running
behavior of the strong coupling constant
in QCD. The black solid curve illustrates
this behavior and provides a reference
for comparison with the results of
the present work.



\subsubsection{Thermodynamic properties of dense quark matter}
\label{sec:thermodynamic_eos}

The thermodynamic properties of quark matter provide
a direct connection between the microscopic description
of strongly interacting matter and the macroscopic
structure of compact stars.
In particular, the baryon number density and pressure
as functions of the quark chemical potential
characterize the behavior of the reconstructed
equation of state (EOS) over the density range
relevant to self-bound quark stars.
Here, we examine these thermodynamic quantities
and compare the reconstructed pressure with
the predictions of a free $ud+e$ gas and
the MIT bag model.

\begin{figure}[tbp]
    \centering
    \includegraphics[
        width=0.92\textwidth,
        keepaspectratio
    ]{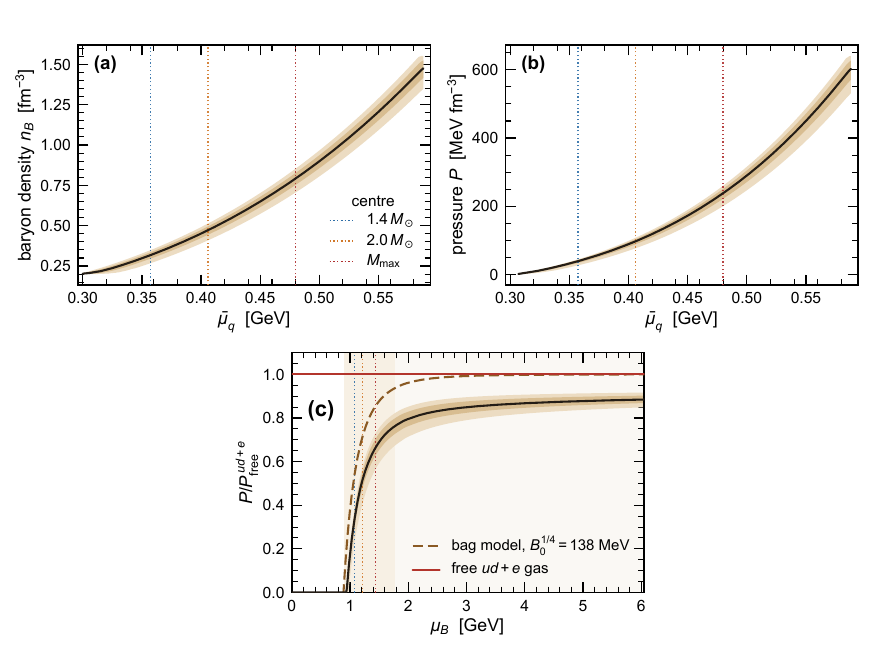}
\caption[Thermodynamic properties of the reconstructed quark-matter EOS.]{Thermodynamic properties of the reconstructed quark-matter equation of state (EOS). Panel (a) shows the baryon number density $n_B$ as a function of the averaged quark chemical potential $\bar{\mu}_q$. Panel (b) shows the pressure $P$ as a function of the averaged quark chemical potential $\bar{\mu}_q$. Panel (c) shows the normalized pressure $P/P_{\mathrm{free}}^{ud+e}$ as a function of the baryon chemical potential $\mu_B$. The brown dashed curve represents the prediction of the MIT bag model with $B_0^{1/4}=138\,\mathrm{MeV}$, while the red horizontal line indicates the reference value for a noninteracting $ud+e$ gas. In all three panels, the black solid curves represent the posterior median predictions, and the shaded regions indicate the corresponding posterior uncertainties. The blue, orange, and red vertical dotted lines indicate the central chemical potentials corresponding to the $1.4\,M_\odot$, $2.0\,M_\odot$, and maximum-mass configurations, respectively.
}
    \label{fig:thermodynamic_eos}
\end{figure}

Figure~\ref{fig:thermodynamic_eos} presents
the posterior predictions for the thermodynamic
properties of quark matter.
The averaged quark chemical potential is defined as
\begin{equation}
\bar{\mu}_q =
\left[
\frac{\mu_u^4+\mu_d^4}{2}
\right]^{1/4},
\label{eq:averaged_quark_chemical_potential}
\end{equation}
where $\mu_u$ and $\mu_d$ are the up- and
down-quark chemical potentials, respectively.
As shown in Figs.~\ref{fig:thermodynamic_eos}(a)
and (b), both the baryon number density $n_B$
and pressure $P$ increase monotonically
with the averaged quark chemical potential
$\bar{\mu}_q$.
The pressure rises more rapidly at higher
chemical potentials, while the baryon number
density also increases continuously.
The posterior uncertainty bands become more
pronounced toward higher chemical potentials,
reflecting the uncertainties in the reconstructed
EOS at higher densities.
The vertical markers further show that
the central chemical potential increases
with stellar mass.
Consequently, more massive quark stars
probe regions of higher baryon density
and pressure.
In particular, the maximum-mass configuration
requires a substantially higher central
pressure than the canonical $1.4\,\Msun$ star.
These results illustrate the connection
between the thermodynamic properties of
dense quark matter and the pressure support
required for massive stellar configurations.

To examine the effects of interactions
on the thermodynamic pressure, we introduce
the normalized pressure
\begin{equation}
\mathcal{R}_P(\mu_B)
=
\frac{P(\mu_B)}
{P_{\mathrm{free}}^{ud+e}(\mu_B)},
\label{eq:normalized_pressure_ratio}
\end{equation}
where $P_{\mathrm{free}}^{ud+e}$ denotes
the pressure of a noninteracting
up-down quark and electron gas.
For the chemical-potential convention
used in the present comparison, the
horizontal-axis variable is defined by
\begin{equation}
\mu_B=3\bar{\mu}_q.
\label{eq:baryon_quark_chemical_potential}
\end{equation}
Here, this relation specifies the effective
chemical-potential scale used for plotting
rather than a general thermodynamic identity
for unequal quark chemical potentials.
Figure~\ref{fig:thermodynamic_eos}(c)
compares the normalized pressure of the
reconstructed EOS with the free-gas
and MIT bag-model predictions.
The black solid curve and its shaded band
represent the posterior median and
uncertainty of the revised perturbative-QCD
EOS, respectively.
The horizontal red line corresponds to
the free $ud+e$ gas, for which
$\mathcal{R}_P=1$.
The brown dashed curve shows the
MIT bag-model prediction with
$B_0^{1/4}=138\,\mathrm{MeV}$.
The reconstructed EOS exhibits a
normalized pressure substantially below
the free-gas value at relatively low
chemical potentials.
As $\mu_B$ increases, the pressure ratio
rises gradually but remains below unity
throughout the displayed range.
The MIT bag-model pressure also approaches
the free-gas reference with increasing
chemical potential, but it does so
more rapidly than the reconstructed EOS.
This difference shows that the revised
perturbative-QCD EOS predicts a stronger
suppression of the pressure relative to
the free-gas reference than the chosen
MIT bag-model parametrization.
The behavior reflects the combined
influence of strong-interaction corrections
and the vacuum contribution included
in the reconstructed EOS.
Thus, the pressure of interacting
quark matter cannot be described
solely by the free-gas limit or
the chosen constant-bag model.
Overall, the reconstructed revised
perturbative-QCD EOS predicts increasing
baryon density and pressure with
chemical potential, while the pressure
remains suppressed relative to the
noninteracting reference.
The comparison with the MIT bag model
further illustrates the effects of
the interaction corrections on the
thermodynamic behavior of self-bound
$ud$ quark matter.
These properties help explain the
internal conditions and mass-supporting
capability of the inferred quark stars.



\subsection{Sound speed, conformal behavior, and adiabatic index}

\subsubsection{Sound speed and conformal properties}

\label{sec:conformal_properties}

The sound speed and the
associated conformal measures provide important
information about the microscopic behavior of dense
quark matter. In particular, the density dependence
of the sound speed, the adiabatic index, and the
deviation from conformal symmetry can be used to
characterize the stiffness of the equation of state
(EOS) and its evolution toward the high-density regime.
These quantities also establish a connection between
the properties of quark matter and the internal
conditions of compact stars.
In this subsection, we investigate the posterior
behavior of several EOS-related quantities, including
the squared speed of sound $c_s^2$, the adiabatic
index $\gamma$, and the conformal measures
$\Delta$ and $\Delta'$. We further examine the
correlations between these quantities for three
representative stellar configurations:
$1.4\,M_\odot$, $2.0\,M_\odot$, and the maximum
Tolman--Oppenheimer--Volkoff mass $M_{\rm TOV}$.

The squared speed of sound is defined as
\begin{equation}
c_s^2 =
\frac{dP}{d\epsilon},
\label{eq:sound_speed_definition}
\end{equation}
where $P$ and $\epsilon$ denote the pressure and
energy density, respectively. 
The adiabatic index is defined as
\begin{equation}
\gamma =
\frac{d\ln P}{d\ln n_B}
=
\frac{\epsilon+P}{P}c_s^2,
\label{eq:adiabatic_index_definition}
\end{equation}
where the second equality follows from the
zero-temperature thermodynamic relation
$d\epsilon=(\epsilon+P)dn_B/n_B$.
To quantify the deviation of the EOS from conformal
behavior, we introduce the conformal measure
\begin{equation}
\Delta =
\frac{1}{3}-\frac{P}{\epsilon}.
\label{eq:conformal_delta_definition}
\end{equation}
Its logarithmic derivative with respect to the
energy density is defined as
\begin{equation}
\Delta' =
\frac{d\Delta}{d\ln\epsilon},
\label{eq:conformal_delta_prime_definition}
\end{equation}
and, for convenience, we also introduce
\begin{equation}
\bar{\Delta}
=
-\Delta'
=
-\frac{d\Delta}{d\ln\epsilon}.
\label{eq:conformal_delta_bar_definition}
\end{equation}
For a conformal system, the corresponding limiting
values are
\begin{equation}
c_s^2=\frac{1}{3},
\qquad
\Delta=0,
\qquad
\Delta'=0,
\qquad
\bar{\Delta}=0.
\label{eq:conformal_limits}
\end{equation}
Therefore, the density dependence of these quantities
provides a useful way to investigate how the
reconstructed quark matter EOS approaches or deviates
from the conformal limit.

\begin{figure}[tbp]
    \centering
    \includegraphics[
        width=\textwidth,
        keepaspectratio
    ]{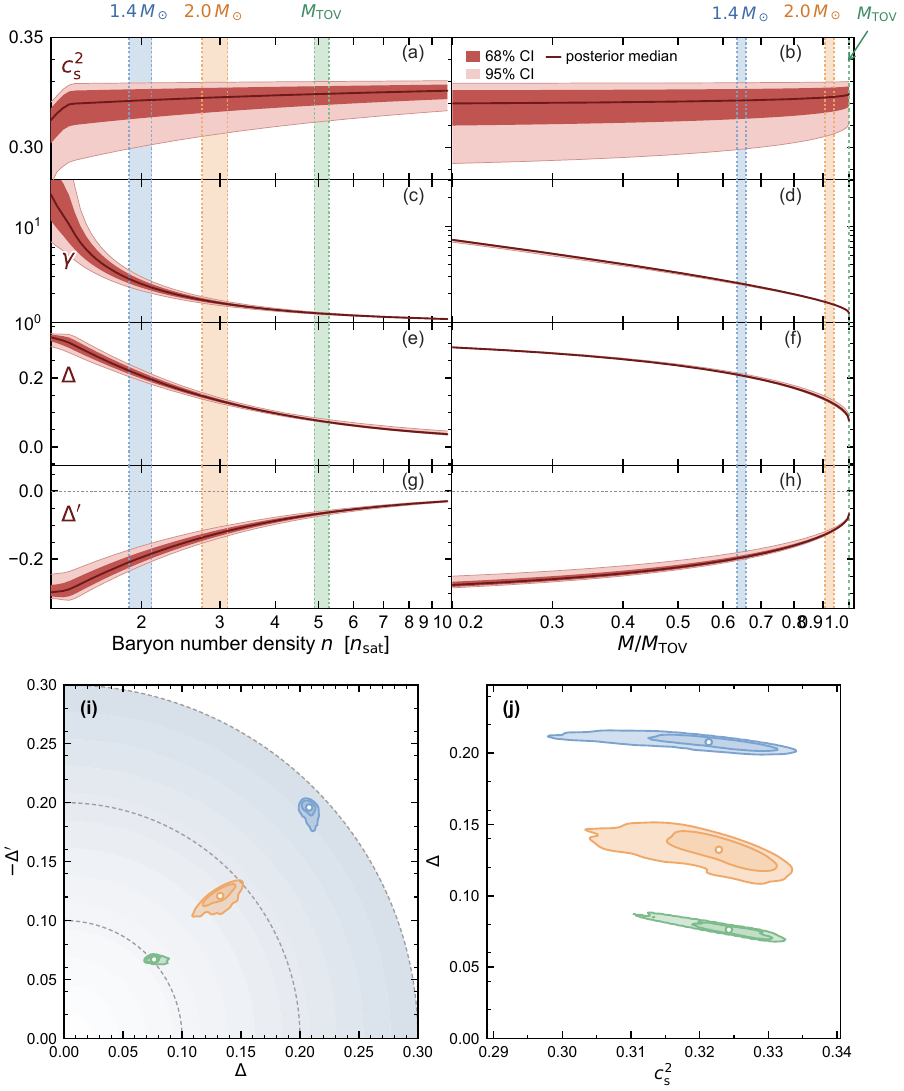}

    \caption[Sound-speed behavior and conformal properties
    of quark matter]{
    The upper panels show the density dependence
    of the squared speed of sound $c_s^2$,
    the adiabatic index $\gamma$, the conformal
    measure $\Delta$, and its logarithmic derivative
    $\Delta'$. The left and right columns display
    these quantities as functions of the baryon
    number density $n/n_{\rm sat}$ and the normalized
    gravitational mass $M/M_{\rm TOV}$, respectively.
    The red solid curves represent the posterior
    medians, while the darker and lighter shaded
    regions indicate the 68\% and 95\% credible
    intervals, respectively.
    The blue, orange, and green vertical bands
    correspond to the central-density regions of
    the $1.4\,M_\odot$, $2.0\,M_\odot$, and
    maximum-mass ($M_{\rm TOV}$) configurations,
    respectively.
    Panel (f) displays the joint posterior
    distributions of $\Delta$ and
    $\bar{\Delta}=-d\Delta/d\ln\epsilon$.
    The dashed curves indicate the reference
    values $d_c=0.1$, $0.2$, and $0.3$.
    Panel (g) shows the joint posterior distributions
    of $\Delta$ and $c_s^2$ for the three
    representative stellar configurations.
    }
    \label{fig:conformal_properties}
\end{figure}

Figure~\ref{fig:conformal_properties} presents the
posterior predictions for these quantities as functions
of the baryon number density and the normalized
stellar mass. The upper eight panels show the density
dependence of $c_s^2$, $\gamma$, $\Delta$, and
$\Delta'$, with the left and right columns corresponding
to the baryon number density $n/n_{\rm sat}$ and the
normalized mass $M/M_{\rm TOV}$, respectively.
Here, $n_{\rm sat}$ denotes the nuclear saturation
density. The red solid curves represent the posterior
median predictions, while the darker and lighter
shaded regions indicate the $68\%$ and $95\%$
credible intervals, respectively.
The blue, orange, and green vertical bands identify
the locations associated with the central regions
of the $1.4\,M_\odot$, $2.0\,M_\odot$, and
maximum-mass configurations, respectively.
In the left panels, these bands represent the
corresponding central-density intervals, whereas
in the right panels they mark the associated
normalized mass intervals. They allow us to compare
the microscopic properties of the EOS at the central
densities sampled by stars with different
gravitational masses.

We first consider the behavior of the squared speed
of sound. As shown in Figs.~\ref{fig:conformal_properties}(a) and (b), $c_s^2$
remains relatively stable over the density range
considered in this work. The posterior median
exhibits only a mild variation, remaining around
$0.32$ over most of the displayed interval.
This value is close to, but slightly below, the
conformal limit $c_s^2=1/3$.
The posterior credible intervals quantify the
uncertainty in this prediction and indicate the
range of sound-speed values allowed by the
reconstructed EOS.
The behavior of the adiabatic index $\gamma$
provides complementary information about the
density dependence of the pressure.
As illustrated in Figs.~\ref{fig:conformal_properties}(c) and (d), $\gamma$
decreases with increasing baryon number density
and gradually approaches a more slowly varying
regime at higher densities.
This behavior reflects the changing relationship
between pressure and baryon number density as
the system evolves from lower to higher densities.
Since $\gamma$ depends on both the squared speed
of sound and the ratio $(\epsilon+P)/P$, it provides
a complementary characterization of the pressure
response of the EOS.

We next examine the conformal measure $\Delta$.
As shown in Figs.~\ref{fig:conformal_properties}(e) and (f), $\Delta$ decreases
systematically with increasing baryon number
density. This behavior indicates that the ratio
$P/\epsilon$ increases toward its conformal value
of $1/3$ as the density increases.
The corresponding derivative $\Delta'$ is shown
in Figs.~\ref{fig:conformal_properties}(g) and (h).
Its posterior median remains negative over most
of the density interval and gradually approaches
zero at higher densities.
Equivalently, the quantity $\bar{\Delta}$
decreases toward zero as the energy density
increases. This behavior indicates that the
conformal deviation becomes less sensitive to
further increases in energy density in the
high-density regime.
An important aspect of this analysis is the
comparison between different stellar configurations.
The vertical bands show that the $1.4\,M_\odot$
configuration probes a relatively low-density region,
whereas the $2.0\,M_\odot$ and maximum-mass configurations probe progressively higher densities. Consequently, the corresponding values of the
sound speed and conformal measures provide
information about different density regimes
of the same reconstructed EOS.

To further investigate the correlations among
the conformal properties and the sound speed,
we examine the joint posterior distributions. Fig.~\ref{fig:conformal_properties}(i) displays the joint posterior distributions
of $\Delta$ and $\bar{\Delta}$ for the three
representative stellar configurations.
The dashed curves indicate the reference values
$d_c=0.1$, $0.2$, and $0.3$.
The posterior distributions occupy distinct
regions in this plane, with both $\Delta$
and $\bar{\Delta}$ decreasing from the
$1.4\,M_\odot$ configuration toward the
maximum-mass configuration.
This behavior is consistent with the density
dependence observed in the upper panels and
further illustrates the systematic reduction
of the conformal deviation in more massive
stellar configurations.
Fig.~\ref{fig:conformal_properties}(j) shows the joint posterior distributions
of the conformal measure $\Delta$ and the squared
speed of sound $c_s^2$.
A particularly interesting feature is the
different degree of variation exhibited by
these two quantities.
While $\Delta$ decreases substantially with
increasing stellar mass, $c_s^2$ remains
confined to a relatively narrow range.
As indicated in the figure, $\Delta$ decreases
from approximately $0.208$ to $0.076$ when
moving from the $1.4\,M_\odot$ configuration
to the maximum-mass configuration.
In contrast, the corresponding squared speed
of sound changes only from approximately
$0.321$ to $0.324$.

This comparison demonstrates that a relatively
weak variation in the sound speed does not
necessarily imply an equally weak variation
in the conformal measure.
Although $c_s^2$ remains close to a nearly
constant value over the relevant density range,
the conformal measure exhibits a more pronounced
variation between the stellar configurations.
Therefore, the sound speed alone may not be
sufficient to fully characterize the evolution
of the EOS toward conformal behavior.
The combined analysis of $c_s^2$, $\Delta$,
and $\bar{\Delta}$ provides a more complete
characterization of the high-density behavior
of the reconstructed EOS.

Overall, these results indicate that the
reconstructed EOS exhibits a relatively stable
sound-speed behavior while its conformal
deviation decreases with increasing density.
The corresponding logarithmic derivative also
approaches zero, indicating a progressively
weaker density dependence of the conformal
measure. The joint posterior distributions
further demonstrate that different stellar
configurations probe distinct regions in the
space of conformal properties.
These findings provide additional insight into
the microscopic behavior of dense quark matter
and complement the thermodynamic analysis
presented in the previous subsection.



\subsubsection{Density dependence of the adiabatic index.}
\label{sec:adiabatic_index}

\begin{figure}[tbp]
    \centering
    \includegraphics[
        height=0.46\textheight,
        keepaspectratio
    ]{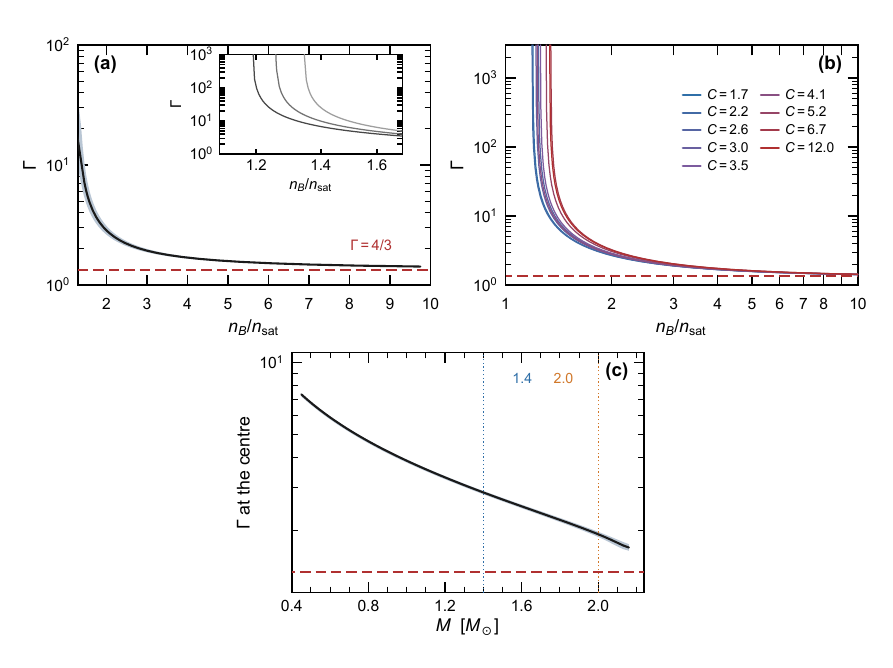}

    \caption[
    Density dependence of the adiabatic index
    and its relation to quark-star structure
    ]{
    Density dependence of the relativistic adiabatic
    index and its implications for quark-star structure.
    (a) Posterior median of the adiabatic index
    $\Gamma$ as a function of the normalized baryon
    number density $n_B/n_{\rm sat}$.
    The shaded region represents the posterior
    uncertainty. The inset provides an enlarged view
    of the low-density region, where the adiabatic
    index exhibits a rapid variation.
    The horizontal red dashed line indicates the
    reference value $\Gamma=4/3$.
    (b) Density dependence of $\Gamma$ for nine
    representative equations of state corresponding
    to different values of the model parameter $C$.
    The values of $C$ range from $1.7$ to $12.0$.
    The red dashed line again denotes $\Gamma=4/3$.
    The different EOS predictions become increasingly
    similar at high baryon number densities.
    (c) Adiabatic index at the stellar centre,
    $\Gamma_{\rm centre}$, as a function of the
    gravitational mass $M$.
    The blue and orange vertical dotted lines indicate
    the $1.4\,M_\odot$ and $2.0\,M_\odot$
    configurations, respectively.
    The horizontal red dashed line represents
    the reference value $\Gamma=4/3$.
    }
    \label{fig:adiabatic_index}
\end{figure}

Following the sound-speed and conformal analyses, we
further examine the density dependence of the relativistic
adiabatic index and its relation to the internal structure
of self-bound $ud$ quark stars.
The adiabatic index describes the relative response of
pressure to changes in baryon number density and provides
a complementary characterization of the reconstructed
equation of state (EOS).

For a cold barotropic EOS, the relativistic adiabatic
index introduced as $\gamma$ in
Eq.~\eqref{eq:adiabatic_index_definition} can equivalently
be written using the notation $\Gamma$:
\begin{equation}
\Gamma =
\frac{\epsilon+P}{P}
\frac{dP}{d\epsilon},
\label{eq:adiabatic_index}
\end{equation}
where $\epsilon$ and $P$ denote the energy density and
pressure, respectively.
Although $\Gamma$ is related to the squared sound speed
through $dP/d\epsilon=c_s^2$, its density dependence
also involves the ratio $(\epsilon+P)/P$.
It therefore provides additional information about
the thermodynamic response of quark matter.

Figure~\ref{fig:adiabatic_index} presents the density
dependence of the adiabatic index, its sensitivity
to the renormalization-scale parameter $C$, and
its variation with stellar mass.
As shown in Fig.~\ref{fig:adiabatic_index}(a),
the adiabatic index decreases rapidly with increasing
baryon number density in the low-density region.
Its decrease becomes progressively slower at higher
densities, where the posterior median gradually
approaches the reference value
\begin{equation}
\Gamma_{\rm ref}=\frac{4}{3}.
\label{eq:gamma_conformal_reference}
\end{equation}
The inset highlights the pronounced increase of
$\Gamma$ toward the low-density boundary of the
self-bound EOS.
This behavior can be understood from
$\Gamma=(\epsilon+P)c_s^2/P$.
For self-bound quark matter, the pressure vanishes
at the stellar surface while the energy density
remains finite.
Consequently, $\Gamma$ increases sharply as
$P$ approaches zero, provided that $c_s^2$
remains finite and nonzero.
The large adiabatic index near the stellar surface
therefore arises primarily from the vanishing
pressure rather than from a corresponding
increase in the squared sound speed.
This provides a direct connection between the
density dependence of $\Gamma$ and the
self-bound nature of the quark matter EOS.

Figure~\ref{fig:adiabatic_index}(b) compares
the adiabatic index for nine representative
EOSs with renormalization-scale parameters
ranging from $C=1.7$ to $C=12.0$.
The predicted curves exhibit noticeable differences
at relatively low baryon densities, indicating
that the pressure response in this region is
sensitive to the model parameter $C$.
However, these differences gradually decrease
as the density increases, and the curves become
increasingly similar in the high-density region.
The corresponding values also approach the
reference line $\Gamma=4/3$.
This behavior indicates that the dependence
of the adiabatic index on $C$ becomes weaker
at higher densities within the explored model
range, although the curves do not coincide
exactly in the displayed interval.
The connection between the adiabatic index
and stellar structure is illustrated in
Fig.~\ref{fig:adiabatic_index}(c).
The central adiabatic index
$\Gamma_{\rm centre}$ decreases systematically
with increasing gravitational mass.
The vertical dotted lines indicate the
representative masses of $1.4\,\Msun$
and $2.0\,\Msun$.
As established by the mass--central-density
relations discussed above, more massive
quark stars generally have higher central
baryon densities.
Their smaller central adiabatic indices
are therefore consistent with the density
dependence shown in panels (a) and (b).
This result connects the local thermodynamic
response of dense quark matter with the
central conditions of stellar configurations
of different masses.

Although $\Gamma_{\rm centre}$ decreases
with increasing stellar mass, it remains
above the reference value $4/3$ throughout
the displayed mass range.
It is important to emphasize that
$\Gamma=4/3$ is used here as a reference
for comparing the density dependence of
the EOS, rather than as a standalone
stability criterion for relativistic
stellar configurations.
In particular, the stability of the
nonrotating quark-star sequence is
determined by its equilibrium structure
and the onset of radial instability,
as discussed in connection with the
maximum-mass turning point.

Overall, these results show that the
relativistic adiabatic index has a strong
density dependence near the self-bound
surface, while its variation becomes
weaker at higher densities.
The comparison among different values
of $C$ indicates that the low-density
behavior is more sensitive to the EOS
parameters, whereas the predictions
become increasingly similar in the
high-density region.
The decreasing central adiabatic index
with stellar mass further connects
these thermodynamic properties to
the internal conditions of self-bound
$ud$ quark stars.
Together with the sound-speed and
conformal analyses, these results
provide a more complete description
of the density-dependent response
of the reconstructed revised
perturbative-QCD EOS.




\section{Conclusions}
\label{sec:concl}

We have performed a Bayesian multimessenger analysis of self-bound two-flavor $ud$ quark stars described by a thermodynamically consistent revised perturbative-QCD EOS, incorporating the GW170817 tidal information, five mass--radius measurements, and two pulsar-mass constraints. Three representations of the GW170817 tidal information, namely four-dimensional KDE, two-dimensional KDE, and one-dimensional split-Gaussian approximations, yield consistent posterior constraints, which are further validated through nested sampling, direct numerical integration, the full EOS solver, and comparison with the JESTER normalizing-flow representation. The equal-weight posterior mixture gives $C=3.56^{+2.97}{-1.38}$, $B_0^{1/4}=137.66^{+1.72}{-1.86}$~MeV, $M_{\max}=2.166^{+0.035}{-0.036},\Msun$, $R{1.4}=11.778^{+0.203}{-0.196}$~km, and $\Lambda{1.4}=768^{+90}_{-81}$. The inferred maximum mass is concentrated around $2.17,\Msun$, while the posterior-predictive support probability for interpreting the $2.59,\Msun$ GW190814 secondary as a self-bound $ud$ quark star is only of order $10^{-5}$. This interpretation is therefore strongly disfavored within the present EOS framework, although no model-independent conclusion about the nature of the secondary can be drawn.

The reconstructed stellar profiles exhibit the characteristic self-bound behavior, with vanishing pressure but finite energy and baryon number densities at the surface. For the median-posterior EOS, the surface energy density is $\epsilon_{\rm surf}\simeq187,\mathrm{MeV,fm^{-3}}$. The normalized pressure remains below the noninteracting $ud+e$ gas reference, reflecting the effects of strong interactions and the vacuum contribution. The running strong coupling decreases with increasing chemical potential, while the high-mass end of the feasibility frontier corresponds to relatively stronger coupling and consequently less reliable perturbative control. Furthermore, the squared speed of sound remains nearly constant, varying from $c_s^2\simeq0.321$ to $0.324$, whereas the conformal measure $\Delta=1/3-P/\epsilon$ decreases substantially from approximately $0.208$ to $0.076$ between canonical and maximum-mass configurations. This demonstrates that an approximately constant sound speed does not necessarily imply approximate conformality. The relativistic adiabatic index also exhibits a pronounced density dependence, including an enhancement near the zero-pressure surface, further reflecting the self-bound nature of the EOS.

The consistency among the different tidal-information representations and the comparison with the JESTER reconstruction indicate that the inferred mass--tidal tension is not primarily an artifact of the statistical treatment of the GW170817 data, but rather reflects an intrinsic feature of the adopted revised perturbative-QCD EOS: configurations supporting larger maximum masses tend to predict larger tidal deformabilities. These findings emphasize the importance of simultaneously considering maximum-mass, radius, and tidal-deformability constraints when assessing self-bound quark-matter models. They also demonstrate that the sound speed alone is insufficient to characterize the conformal behavior of dense quark matter. Future investigations should incorporate higher-order perturbative corrections, renormalization-scale uncertainties, and nonperturbative QCD effects, and extend the Bayesian framework to three-flavor, strange, color-superconducting, and hybrid EOS models. Improved gravitational-wave and pulsar observations, together with extensions to finite-temperature and rapidly rotating configurations, will further test the viability of these models and clarify whether the mass--tidal tension identified here is specific to the adopted two-flavor EOS or represents a more general constraint on self-bound quark matter.

\section*{Acknowledgments}

We thank Hao-Chen Tian, Xun Chen, and the members of the Hunan University of Science and Technology compact-star group for helpful discussions on Bayesian analysis. This work was supported by the
National Natural Science Foundation of China (Grant No.~12205093), the Natural Science Foundation of Hunan Province (Grant
No.~2026JJ50352), and the Scientific Research Foundation of Hunan Provincial
Education Department (Grants No.~25B0432 and No.~25C0302). Z.Z. was also supported by the National Program on Key Research and Development Project of China (Grants No. 2021YFA0718500 and No. 2025YFF0511102), and the Chinese Academy of Sciences (Grant No. E32983U810).

\section*{Data availability}

The observational data used in this work are taken from the publications cited in the manuscript. The derived EOS grid, posterior samples, and analysis code are available from the corresponding author.

\bibliographystyle{JHEP}  
\bibliography{references}

@article{MUSES:2023hyz,
    author = "Kumar, Rajesh and others",
    collaboration = "MUSES",
    title = "{Theoretical and experimental constraints for the equation of state of dense and hot matter}",
    doi = "10.1007/s41114-024-00049-6",
    journal = "Living Rev. Rel.",
    volume = "27",
    number = "1",
    pages = "3",
    year = "2024"
}

@article{Chen:2024aom,
    author = "Chen, Jinhui and others",
    title = "{Properties of the QCD matter: review of selected results from the relativistic heavy ion collider beam energy scan (RHIC BES) program}",
    doi = "10.1007/s41365-024-01591-2",
    journal = "Nucl. Sci. Tech.",
    volume = "35",
    number = "12",
    pages = "214",
    year = "2024"
}

@article{Altmann:2024icx,
    author = "Altmann, Javira and others",
    title = "{QCD challenges from pp to AA collisions: 4th edition}",
    doi = "10.1140/epjc/s10052-024-12650-8",
    journal = "Eur. Phys. J. C",
    volume = "84",
    number = "4",
    pages = "421",
    year = "2024"
}

@article{Gross:2022hyw,
    author = "Gross, Franz and others",
    title = "{50 Years of Quantum Chromodynamics}",
    doi = "10.1140/epjc/s10052-023-11949-2",
    journal = "Eur. Phys. J. C",
    volume = "83",
    pages = "1125",
    year = "2023"
}

@article{LIGOScientific:2018hze,
  author        = {Abbott, B. P. and others},
  collaboration = {LIGO Scientific Collaboration and Virgo Collaboration},
  title         = {Properties of the binary neutron star merger {GW170817}},
  journal       = {Phys. Rev. X},
  volume        = {9},
  pages         = {011001},
  year          = {2019},
  doi           = {10.1103/PhysRevX.9.011001}
}

@article{Chatziioannou:2020pqz,
  author  = {Chatziioannou, Katerina},
  title   = {Neutron-star tidal deformability and equation-of-state constraints},
  journal = {Gen. Relativ. Gravit.},
  volume  = {52},
  pages   = {109},
  year    = {2020},
  doi     = {10.1007/s10714-020-02754-3}
}

@article{Bedaque:2014sqa,
  author  = {Bedaque, Paulo F. and Steiner, Andrew W.},
  title   = {Sound velocity bound and neutron stars},
  journal = {Phys. Rev. Lett.},
  volume  = {114},
  pages   = {031103},
  year    = {2015},
  doi     = {10.1103/PhysRevLett.114.031103}
}

@article{Annala:2019puf,
  author  = {Annala, Eemeli and Gorda, Tyler and Kurkela, Aleksi and Vuorinen, Aleksi},
  title   = {Evidence for quark-matter cores in massive neutron stars},
  journal = {Nat. Phys.},
  volume  = {16},
  pages   = {907},
  year    = {2020},
  doi     = {10.1038/s41567-020-0914-9}
}

@article{Salmi:2024aum,
  author  = {Salmi, Tuomo and others},
  title   = {The radius of the high-mass pulsar {PSR J0740+6620} with 3.6 yr of {NICER} data},
  journal = {Astrophys. J.},
  volume  = {974},
  pages   = {294},
  year    = {2024},
  doi     = {10.3847/1538-4357/ad5f1f}
}

@article{LIGOScientific:2017vwq,
  author        = {Abbott, B. P. and others},
  collaboration = {LIGO Scientific Collaboration and Virgo Collaboration},
  title         = {{GW170817}: Observation of gravitational waves from a binary neutron star inspiral},
  journal       = {Phys. Rev. Lett.},
  volume        = {119},
  pages         = {161101},
  year          = {2017},
  doi           = {10.1103/PhysRevLett.119.161101}
}

@article{LIGOScientific:2018cki,
  author        = {Abbott, B. P. and others},
  collaboration = {LIGO Scientific Collaboration and Virgo Collaboration},
  title         = {{GW170817}: Measurements of neutron star radii and equation of state},
  journal       = {Phys. Rev. Lett.},
  volume        = {121},
  pages         = {161101},
  year          = {2018},
  doi           = {10.1103/PhysRevLett.121.161101}
}

@article{Riley:2019yda,
  author  = {Riley, Thomas E. and others},
  title   = {A {NICER} view of {PSR J0030+0451}: Mass and radius from {NICER} data and implications for the properties of neutron star matter},
  journal = {Astrophys. J. Lett.},
  volume  = {887},
  pages   = {L21},
  year    = {2019},
  doi     = {10.3847/2041-8213/ab481c}
}

@article{Miller:2019cac,
  author  = {Miller, M. C. and others},
  title   = {{PSR J0030+0451} mass and radius from {NICER} data and implications for the properties of neutron star matter},
  journal = {Astrophys. J. Lett.},
  volume  = {887},
  pages   = {L24},
  year    = {2019},
  doi     = {10.3847/2041-8213/ab50c5}
}

@article{Riley:2021pdl,
  author  = {Riley, Thomas E. and others},
  title   = {A {NICER} view of the massive pulsar {PSR J0740+6620} informed by radio timing and {XMM-Newton} spectroscopy},
  journal = {Astrophys. J. Lett.},
  volume  = {918},
  pages   = {L27},
  year    = {2021},
  doi     = {10.3847/2041-8213/ac0a81}
}

@article{Miller:2021qha,
  author  = {Miller, M. C. and others},
  title   = {The radius of {PSR J0740+6620} from a joint {NICER} and {XMM-Newton} analysis},
  journal = {Astrophys. J. Lett.},
  volume  = {918},
  pages   = {L28},
  year    = {2021},
  doi     = {10.3847/2041-8213/ac089b}
}

@article{Dittmann:2024mbo,
  author  = {Dittmann, Alexander J. and others},
  title   = {A More Precise Measurement of the Radius of {PSR J0740+6620} Using Updated {NICER} Data},
  journal = {Astrophys. J.},
  volume  = {974},
  pages   = {295},
  year    = {2024},
  doi     = {10.3847/1538-4357/ad5f1e}
}

@article{Romani:2022jhd,
  author  = {Romani, Roger W. and Kandel, D. and Filippenko, Alexei V. and Brink, Thomas G. and Zheng, WeiKang},
  title   = {{PSR J0952-0607}: The fastest and heaviest known galactic neutron star},
  journal = {Astrophys. J. Lett.},
  volume  = {934},
  pages   = {L17},
  year    = {2022},
  doi     = {10.3847/2041-8213/ac8007}
}

@article{Linares:2018ppq,
  author  = {Linares, M. and Shahbaz, T. and Casares, J.},
  title   = {Peering into the dark side: Magnesium lines establish a massive neutron star in {PSR J2215+5135}},
  journal = {Astrophys. J.},
  volume  = {859},
  pages   = {54},
  year    = {2018},
  doi     = {10.3847/1538-4357/aabde6}
}

@article{Doroshenko:2022nwp,
  author  = {Doroshenko, Victor and Suleimanov, Valery and P{\"u}hlhofer, Gerd and Santangelo, Andrea},
  title   = {A strangely light neutron star within a supernova remnant},
  journal = {Nature Astronomy},
  volume  = {6},
  pages   = {1444},
  year    = {2022},
  doi     = {10.1038/s41550-022-01800-1}
}

@article{LIGOScientific:2025snk,
  author        = {Abac, A. G. and others},
  collaboration = {LIGO Scientific Collaboration and Virgo Collaboration and KAGRA Collaboration},
  title         = {Open data from {LIGO}, {Virgo}, and {KAGRA} through the first part of the fourth observing run},
  journal       = {Astrophys. J.},
  volume        = {1004},
  pages         = {232},
  year          = {2026},
  doi           = {10.3847/1538-4357/ae211e}
}

@article{Choudhury:2024xbk,
  author  = {Choudhury, Devarshi and others},
  title   = {A {NICER} view of the nearest and brightest millisecond pulsar: {PSR J0437-4715}},
  journal = {Astrophys. J. Lett.},
  volume  = {971},
  pages   = {L20},
  year    = {2024},
  doi     = {10.3847/2041-8213/ad5a6f}
}

@article{Bodmer:1971we,
  author  = {Bodmer, A. R.},
  title   = {Collapsed nuclei},
  journal = {Phys. Rev. D},
  volume  = {4},
  pages   = {1601},
  year    = {1971},
  doi     = {10.1103/PhysRevD.4.1601}
}

@article{Witten:1984rs,
  author  = {Witten, Edward},
  title   = {Cosmic separation of phases},
  journal = {Phys. Rev. D},
  volume  = {30},
  pages   = {272},
  year    = {1984},
  doi     = {10.1103/PhysRevD.30.272}
}

@article{Farhi:1984qu,
  author  = {Farhi, Edward and Jaffe, R. L.},
  title   = {Strange matter},
  journal = {Phys. Rev. D},
  volume  = {30},
  pages   = {2379},
  year    = {1984},
  doi     = {10.1103/PhysRevD.30.2379}
}

@article{Alcock:1986hz,
  author  = {Alcock, Charles and Farhi, Edward and Olinto, Angela},
  title   = {Strange stars},
  journal = {Astrophys. J.},
  volume  = {310},
  pages   = {261},
  year    = {1986},
  doi     = {10.1086/164679}
}

@article{Weissenborn:2011qu,
  author  = {Weissenborn, S. and Sagert, I. and Pagliara, G. and Hempel, M. and Schaffner-Bielich, J.},
  title   = {Quark Matter in Massive Neutron Stars},
  journal = {Astrophys. J. Lett.},
  volume  = {740},
  pages   = {L14},
  year    = {2011},
  doi     = {10.1088/2041-8205/740/1/L14}
}

@article{Zhang:2024xod,
  author  = {Zhang, Xiao-Li and Huang, Yong-Feng and Zou, Ze-Cheng},
  title   = {Recent progresses in strange quark stars},
  journal = {Front. Astron. Space Sci.},
  volume  = {11},
  pages   = {1409463},
  year    = {2024},
  doi     = {10.3389/fspas.2024.1409463}
}

@article{Alford:2007xm,
  author  = {Alford, Mark G. and Schmitt, Andreas and Rajagopal, Krishna and Sch{\"a}fer, Thomas},
  title   = {Color superconductivity in dense quark matter},
  journal = {Rev. Mod. Phys.},
  volume  = {80},
  pages   = {1455},
  year    = {2008},
  doi     = {10.1103/RevModPhys.80.1455}
}

@article{DiClemente:2024lzi,
  author  = {Di Clemente, Francesco and Casolino, Marco and Drago, Alessandro and Lattanzi, Massimiliano and Ratti, Claudia},
  title   = {Strange quark matter as dark matter: 40 years later, a reappraisal},
  journal = {Mon. Not. R. Astron. Soc.},
  volume  = {537},
  pages   = {1056},
  year    = {2025},
  doi     = {10.1093/mnras/staf087}
}

@article{Char:2024kgo,
  author  = {Char, Prasanta and Biswas, Bhaskar},
  title   = {Compact object of {HESS J1731-347} and its implication on neutron star matter},
  journal = {Phys. Rev. D},
  volume  = {113},
  pages   = {044002},
  year    = {2026},
  doi     = {10.1103/f8dq-t7ky}
}

@article{Fonseca:2025cxs,
  author  = {Fonseca, {\'U}rsula and Fraga, Eduardo S.},
  title   = {Cooling of quark stars from perturbative {QCD}},
  journal = {Phys. Rev. D},
  volume  = {113},
  pages   = {083015},
  year    = {2026},
  doi     = {10.1103/3nyr-rcc3}
}

@article{Fraga:2001id,
  author  = {Fraga, Eduardo S. and Pisarski, Robert D. and Schaffner-Bielich, J{\"u}rgen},
  title   = {Small, dense quark stars from perturbative {QCD}},
  journal = {Phys. Rev. D},
  volume  = {63},
  pages   = {121702},
  year    = {2001},
  doi     = {10.1103/PhysRevD.63.121702}
}

@article{Kurkela:2014vha,
  author  = {Kurkela, Aleksi and Fraga, Eduardo S. and Schaffner-Bielich, J{\"u}rgen and Vuorinen, Aleksi},
  title   = {Constraining neutron star matter with quantum chromodynamics},
  journal = {Astrophys. J.},
  volume  = {789},
  pages   = {127},
  year    = {2014},
  doi     = {10.1088/0004-637X/789/2/127}
}

@article{Komoltsev:2021jzg,
  author  = {Komoltsev, Oleg and Kurkela, Aleksi},
  title   = {How perturbative {QCD} constrains the equation of state at neutron-star densities},
  journal = {Phys. Rev. Lett.},
  volume  = {128},
  pages   = {202701},
  year    = {2022},
  doi     = {10.1103/PhysRevLett.128.202701}
}

@article{Komoltsev:2023zor,
  author  = {Komoltsev, Oleg and Somasundaram, Rahul and Gorda, Tyler and Kurkela, Aleksi and Margueron, J{\'e}r{\^o}me and Tews, Ingo},
  title   = {Equation of state at neutron-star densities and beyond from perturbative {QCD}},
  journal = {Phys. Rev. D},
  volume  = {109},
  pages   = {094030},
  year    = {2024},
  doi     = {10.1103/PhysRevD.109.094030}
}

@article{Annala:2023cwx,
  author  = {Annala, Eemeli and Gorda, Tyler and Hirvonen, Joonas and Komoltsev, Oleg and Kurkela, Aleksi and N{\"a}ttil{\"a}, Joonas and Vuorinen, Aleksi},
  title   = {Strongly interacting matter exhibits deconfined behavior in massive neutron stars},
  journal = {Nat. Commun.},
  volume  = {14},
  pages   = {8451},
  year    = {2023},
  doi     = {10.1038/s41467-023-44051-y}
}

@article{Ecker:2024uqv,
  author  = {Ecker, Christian and Gorda, Tyler and Kurkela, Aleksi and Rezzolla, Luciano},
  title   = {Constraining the equation of state in neutron-star cores via the long-ringdown signal},
  journal = {Nat. Commun.},
  volume  = {16},
  pages   = {1320},
  year    = {2025},
  doi     = {10.1038/s41467-025-56500-x}
}

@article{Bauswein:2018bma,
  author  = {Bauswein, Andreas and Bastian, Niels-Uwe F. and Blaschke, David B. and Chatziioannou, Katerina and Clark, James A. and Fischer, Tobias and Oertel, Micaela and Stergioulas, Nikolaos and Zagkouris, K.},
  title   = {Identifying a first-order phase transition in neutron-star mergers through gravitational waves},
  journal = {Phys. Rev. Lett.},
  volume  = {122},
  pages   = {061102},
  year    = {2019},
  doi     = {10.1103/PhysRevLett.122.061102}
}

@article{Most:2018eaw,
  author  = {Most, Elias R. and Papenfort, L. Jens and Dexheimer, Veronica and Hanauske, Matthias and St{\"o}cker, Horst and Rezzolla, Luciano},
  title   = {Signatures of quark-hadron phase transitions in general-relativistic neutron-star mergers},
  journal = {Phys. Rev. Lett.},
  volume  = {122},
  pages   = {061101},
  year    = {2019},
  doi     = {10.1103/PhysRevLett.122.061101}
}

@article{Foreman-Mackey:2012any,
  author  = {Foreman-Mackey, Daniel and Hogg, David W. and Lang, Dustin and Goodman, Jonathan},
  title   = {emcee: The {MCMC} hammer},
  journal = {Publ. Astron. Soc. Pac.},
  volume  = {125},
  pages   = {306},
  year    = {2013},
  doi     = {10.1086/670067}
}

@article{Grundler:2026zus,
  author  = {Grundler, Xavier and Li, Bao-An},
  title   = {Bayesian constraints on the neutron star equation of state with a smooth hadron-quark crossover},
  journal = {Phys. Rev. D},
  volume  = {113},
  pages   = {103012},
  year    = {2026},
  doi     = {10.1103/pn2b-vls3}
}

@article{Wang:2024xon,
  author  = {Wang, Ziming and Gao, Yong and Liang, Dicong and Zhao, Junjie and Shao, Lijing},
  title   = {Vetting quark-star models with gravitational waves in the hierarchical {B}ayesian framework},
  journal = {J. Cosmol. Astropart. Phys.},
  volume  = {11},
  number  = {2024},
  pages   = {038},
  year    = {2024},
  doi     = {10.1088/1475-7516/2024/11/038}
}

@article{Xie:2025sth,
  author  = {Xie, Wen-Jie and Xia, Cheng-Jun and Zhang, Chen and Xu, Renxin},
  title   = {Bayesian constraints on quark stars from multimessenger observations},
  journal = {Phys. Rev. D},
  volume  = {114},
  pages   = {023016},
  year    = {2026},
  doi     = {10.1103/66yl-pg5k}
}

@article{Skilling:2006gxv,
  author  = {Skilling, John},
  title   = {Nested sampling for general {B}ayesian computation},
  journal = {Bayesian Anal.},
  volume  = {1},
  pages   = {833},
  year    = {2006},
  doi     = {10.1214/06-BA127}
}

@article{Speagle:2019ivv,
  author  = {Speagle, Joshua S.},
  title   = {dynesty: A dynamic nested sampling package for estimating {B}ayesian posteriors and evidences},
  journal = {Mon. Not. R. Astron. Soc.},
  volume  = {493},
  pages   = {3132},
  year    = {2020},
  doi     = {10.1093/mnras/staa278}
}

@article{Biswas:2021pvm,
  author  = {Biswas, Bhaskar},
  title   = {Bayesian Model Selection of Neutron Star Equations of State Using Multi-messenger Observations},
  journal = {Astrophys. J.},
  volume  = {926},
  pages   = {75},
  year    = {2022},
  doi     = {10.3847/1538-4357/ac447b}
}

@article{Kashyap:2025cpd,
  author  = {Kashyap, Rahul and Gupta, Ish and Dhani, Arnab and Bapna, Sanjay and Sathyaprakash, B. S.},
  title   = {Optimizing {B}ayesian model selection for equation of state of cold neutron stars},
  journal = {Phys. Rev. D},
  volume  = {113},
  pages   = {063019},
  year    = {2026},
  doi     = {10.1103/kybh-44xf},
  primaryClass  = {gr-qc}
}

@article{Wang:2025lwv,
  author  = {Wang, Shu-Peng and Lu, Zhen-Yan and Ma, Zhi-Jun and Yang, Rong-Yao and Xu, Jia-Feng and Fu, Xiang},
  title   = {Probing the nonstrange quark star equation of state with compact stars and gravitational waves},
  journal = {Phys. Rev. D},
  volume  = {112},
  pages   = {094036},
  year    = {2025},
  doi     = {10.1103/d7pv-1dqt}
}

@article{Flanagan:2007ix,
  author  = {Flanagan, {\'E}anna {\'E}. and Hinderer, Tanja},
  title   = {Constraining neutron-star tidal {L}ove numbers with gravitational-wave detectors},
  journal = {Phys. Rev. D},
  volume  = {77},
  pages   = {021502},
  year    = {2008},
  doi     = {10.1103/PhysRevD.77.021502}
}

@article{Damour:2009vw,
  author  = {Damour, Thibault and Nagar, Alessandro},
  title   = {Relativistic tidal properties of neutron stars},
  journal = {Phys. Rev. D},
  volume  = {80},
  pages   = {084035},
  year    = {2009},
  doi     = {10.1103/PhysRevD.80.084035}
}

@article{Hinderer:2009ca,
  author  = {Hinderer, Tanja and Lackey, Benjamin D. and Lang, Ryan N. and Read, Jocelyn S.},
  title   = {Tidal deformability of neutron stars with realistic equations of state and their gravitational wave signatures in binary inspiral},
  journal = {Phys. Rev. D},
  volume  = {81},
  pages   = {123016},
  year    = {2010},
  doi     = {10.1103/PhysRevD.81.123016}
}

@article{Hinderer:2007mb,
  author  = {Hinderer, Tanja},
  title   = {Tidal {L}ove numbers of neutron stars},
  journal = {Astrophys. J.},
  volume  = {677},
  pages   = {1216},
  year    = {2008},
  doi     = {10.1086/533487}
}

@article{Postnikov:2010yn,
  author  = {Postnikov, Sergey and Prakash, Madappa and Lattimer, James M.},
  title   = {Tidal {L}ove Numbers of Neutron and Self-Bound Quark Stars},
  journal = {Phys. Rev. D},
  volume  = {82},
  pages   = {024016},
  year    = {2010},
  doi     = {10.1103/PhysRevD.82.024016}
}

@book{Shapiro:1983du,
  author    = {Shapiro, Stuart L. and Teukolsky, Saul A.},
  title     = {Black Holes, White Dwarfs, and Neutron Stars: The Physics of Compact Objects},
  publisher = {Wiley},
  address   = {New York},
  year      = {1983},
  doi       = {10.1002/9783527617661}
}

@article{Yagi:2013bca,
  author  = {Yagi, Kent and Yunes, Nicol{\'a}s},
  title   = {{I-Love-Q}: Unexpected Universal Relations for Neutron Stars and Quark Stars},
  journal = {Science},
  volume  = {341},
  pages   = {365},
  year    = {2013},
  doi     = {10.1126/science.1236462}
}

@article{Ozel:2016oaf,
  author  = {{\"O}zel, Feryal and Freire, Paulo},
  title   = {Masses, radii, and the equation of state of neutron stars},
  journal = {Annu. Rev. Astron. Astrophys.},
  volume  = {54},
  pages   = {401},
  year    = {2016},
  doi     = {10.1146/annurev-astro-081915-023322}
}

@article{Nattila:2017wtj,
  author  = {N{\"a}ttil{\"a}, Joonas and Miller, M. Coleman and Steiner, Andrew W. and Kajava, Jari J. E. and Suleimanov, Valery F. and Poutanen, Juri},
  title   = {Neutron star mass and radius measurements from atmospheric model fits to {X}-ray burst cooling tail spectra},
  journal = {Astron. Astrophys.},
  volume  = {608},
  pages   = {A31},
  year    = {2017},
  doi     = {10.1051/0004-6361/201731082}
}

@article{LIGOScientific:2020zkf,
  author        = {Abbott, R. and others},
  collaboration = {LIGO Scientific Collaboration and Virgo Collaboration},
  title         = {{GW190814}: Gravitational waves from the coalescence of a 23 solar mass black hole with a 2.6 solar mass compact object},
  journal       = {Astrophys. J. Lett.},
  volume        = {896},
  pages         = {L44},
  year          = {2020},
  doi           = {10.3847/2041-8213/ab960f}
}

@article{Guo:2023mhf,
  author  = {Guo, Ling-Jun and Xiong, Jia-Ying and Ma, Yao and Ma, Yong-Liang},
  title   = {Insights into neutron star equation of state by machine learning},
  journal = {Astrophys. J.},
  volume  = {965},
  pages   = {47},
  year    = {2024},
  doi     = {10.3847/1538-4357/ad2e8d}
}

@article{Gorda:2025aiu,
  author  = {Gorda, Tyler and Komoltsev, Oleg and Kurkela, Aleksi},
  title   = {Constrained {G}aussian-process-bridge prior for neutron-star equation-of-state inference},
  journal = {Astrophys. J.},
  volume  = {1002},
  pages   = {40},
  year    = {2026},
  doi     = {10.3847/1538-4357/ae552a}
}

@article{Fattoyev:2020cws,
  author  = {Fattoyev, F. J. and Horowitz, C. J. and Piekarewicz, J. and Reed, Brendan},
  title   = {{GW190814}: Impact of a 2.6 solar mass neutron star on the nucleonic equations of state},
  journal = {Phys. Rev. C},
  volume  = {102},
  pages   = {065805},
  year    = {2020},
  doi     = {10.1103/PhysRevC.102.065805}
}

@article{Tsokaros:2020hli,
  author  = {Tsokaros, Antonios and Ruiz, Milton and Shapiro, Stuart L.},
  title   = {{GW190814}: Spin and equation of state of a neutron star companion},
  journal = {Astrophys. J.},
  volume  = {905},
  pages   = {48},
  year    = {2020},
  doi     = {10.3847/1538-4357/abc421}
}

@article{Biswas:2020xna,
  author  = {Biswas, Bhaskar and Nandi, Rana and Char, Prasanta and Bose, Sukanta and Stergioulas, Nikolaos},
  title   = {{GW190814}: on the properties of the secondary component of the binary},
  journal = {Mon. Not. Roy. Astron. Soc.},
  volume  = {505},
  pages   = {1600},
  year    = {2021},
  doi     = {10.1093/mnras/stab1383}
}

@article{Tews:2020ylw,
  author  = {Tews, Ingo and Pang, Peter T. H. and Dietrich, Tim and Coughlin, Michael W. and Antier, Sarah and Bulla, Mattia and Heinzel, Jack and Issa, Lina},
  title   = {On the Nature of {GW190814} and Its Impact on the Understanding of Supranuclear Matter},
  journal = {Astrophys. J. Lett.},
  volume  = {908},
  pages   = {L1},
  year    = {2021},
  doi     = {10.3847/2041-8213/abdaae}
}

@article{Most:2020bba,
  author  = {Most, Elias R. and Papenfort, L. Jens and Weih, Lukas R. and Rezzolla, Luciano},
  title   = {A lower bound on the maximum mass if the secondary in {GW190814} was once a rapidly spinning neutron star},
  journal = {Mon. Not. Roy. Astron. Soc.},
  volume  = {499},
  pages   = {L82},
  year    = {2020},
  doi     = {10.1093/mnrasl/slaa168}
}

@article{Nathanail:2021tay,
  author  = {Nathanail, Antonios and Most, Elias R. and Rezzolla, Luciano},
  title   = {{GW170817} and {GW190814}: tension on the maximum mass},
  journal = {Astrophys. J. Lett.},
  volume  = {908},
  pages   = {L28},
  year    = {2021},
  doi     = {10.3847/2041-8213/abdfc6}
}

@article{Lourenco:2021lpn,
  author  = {Louren{\c c}o, Odilon and Lenzi, C{\'e}sar H. and Dutra, Mariana and Ferrer, Efrain J. and de la Incera, Vivian and Paulucci, Laura and Horvath, J. E.},
  title   = {Tidal deformability of strange stars and the {GW170817} event},
  journal = {Phys. Rev. D},
  volume  = {103},
  pages   = {103010},
  year    = {2021},
  doi     = {10.1103/PhysRevD.103.103010}
}

@article{Ayriyan:2025rub,
  author  = {Ayriyan, A. and Ivanytskyi, O. and Blaschke, D.},
  title   = {Bayesian inference favors quark matter in neutron star interiors},
  journal = {Phys. Rev. D},
  volume  = {113},
  pages   = {114045},
  year    = {2026},
  doi     = {10.1103/h4bk-8nry}
}

@article{Rezzolla:2017aly,
  author  = {Rezzolla, Luciano and Most, Elias R. and Weih, Lukas R.},
  title   = {Using gravitational-wave observations and quasi-universal relations to constrain the maximum mass of neutron stars},
  journal = {Astrophys. J. Lett.},
  volume  = {852},
  pages   = {L25},
  year    = {2018},
  doi     = {10.3847/2041-8213/aaa401}
}

@misc{Gupta:2022qgg,
  author        = {Gupta, Pawan Kumar and Puecher, Anna and Pang, Peter T. H. and Janquart, Justin and Koekoek, Gideon and Van Den Broeck, Chris},
  title         = {Determining the neutron star equation of state by next-generation gravitational-wave detectors},
  year          = {2022},
  eprint        = {2205.01182},
  archivePrefix = {arXiv},
  primaryClass  = {gr-qc}
}

@article{Walker:2024loo,
  author  = {Walker, Kyle and Smith, Rory and Thrane, Eric and Reardon, Daniel J.},
  title   = {Precision constraints on the neutron star equation of state with third-generation gravitational-wave observatories},
  journal = {Phys. Rev. D},
  volume  = {110},
  pages   = {043013},
  year    = {2024},
  doi     = {10.1103/PhysRevD.110.043013}
}

@article{Ripley:2023lsq,
  author  = {Ripley, Justin L. and others},
  title   = {A constraint on the dissipative tidal deformability of neutron stars},
  journal = {Nat. Astron.},
  volume  = {8},
  pages   = {1277},
  year    = {2024},
  doi     = {10.1038/s41550-024-02323-7}
}

@article{NANOGrav:2019jur,
  author  = {Cromartie, H. T. and Fonseca, E. and Ransom, S. M. and others},
  title   = {Relativistic {S}hapiro delay measurements of an extremely massive millisecond pulsar},
  journal = {Nat. Astron.},
  volume  = {4},
  pages   = {72},
  year    = {2020},
  doi     = {10.1038/s41550-019-0880-2}
}

@article{Lattimer:2015nhk,
  author  = {Lattimer, James M. and Prakash, M.},
  title   = {The Equation of State of Hot, Dense Matter and Neutron Stars},
  journal = {Phys. Rept.},
  volume  = {621},
  pages   = {127},
  year    = {2016},
  doi     = {10.1016/j.physrep.2015.12.005}
}

@article{Raithel:2019uzi,
  author  = {Raithel, Carolyn A.},
  title   = {Constraints on the Neutron Star Equation of State from {GW170817}},
  journal = {Eur. Phys. J. A},
  volume  = {55},
  pages   = {80},
  year    = {2019},
  doi     = {10.1140/epja/i2019-12759-5}
}

@article{Raaijmakers:2019qny,
  author  = {Raaijmakers, G. and others},
  title   = {A {NICER} view of {PSR} {J0030+0451}: Implications for the dense matter equation of state},
  journal = {Astrophys. J. Lett.},
  volume  = {887},
  pages   = {L22},
  year    = {2019},
  doi     = {10.3847/2041-8213/ab451a}
}

@article{Raaijmakers:2019dks,
  author  = {Raaijmakers, G. and others},
  title   = {Constraining the dense matter equation of state with joint analysis of {NICER} and {LIGO}/{Virgo} measurements},
  journal = {Astrophys. J. Lett.},
  volume  = {893},
  pages   = {L21},
  year    = {2020},
  doi     = {10.3847/2041-8213/ab822f}
}

@article{Tews:2018kmu,
  author  = {Tews, Ingo and Carlson, Joseph and Gandolfi, Stefano and Reddy, Sanjay},
  title   = {Constraining the speed of sound inside neutron stars with chiral effective field theory interactions and observations},
  journal = {Astrophys. J.},
  volume  = {860},
  pages   = {149},
  year    = {2018},
  doi     = {10.3847/1538-4357/aac267}
}

@article{Bombaci:2020vgw,
  author  = {Bombaci, I. and Drago, A. and Logoteta, D. and Pagliara, G. and Vidana, I.},
  title   = {Was {GW190814} a black hole -- strange quark star system?},
  journal = {Phys. Rev. Lett.},
  volume  = {126},
  pages   = {162702},
  year    = {2021},
  doi     = {10.1103/PhysRevLett.126.162702}
}

@article{Dexheimer:2020rlp,
  author  = {Dexheimer, V. and Gomes, R. O. and Klähn, T. and Han, S. and Salinas, M.},
  title   = {{GW190814} as a massive rapidly-rotating neutron star with exotic degrees of freedom},
  journal = {Phys. Rev. C},
  volume  = {103},
  pages   = {025808},
  year    = {2021},
  doi     = {10.1103/PhysRevC.103.025808}
}

@article{Kurkela:2009gj,
  author  = {Kurkela, Aleksi and Romatschke, Paul and Vuorinen, Aleksi},
  title   = {Cold Quark Matter},
  journal = {Phys. Rev. D},
  volume  = {81},
  pages   = {105021},
  year    = {2010},
  doi     = {10.1103/PhysRevD.81.105021}
}

@article{Alford:2004pf,
  author  = {Alford, Mark and Braby, Matt and Paris, Mark and Reddy, Sanjay},
  title   = {Hybrid stars that masquerade as neutron stars},
  journal = {Astrophys. J.},
  volume  = {629},
  pages   = {969},
  year    = {2005},
  doi     = {10.1086/430902}
}

@article{Baym:2017whm,
  author  = {Baym, Gordon and Hatsuda, Tetsuo and Kojo, Toru and Powell, Philip D. and Song, Yifan and Takatsuka, Tatsuyuki},
  title   = {From hadrons to quarks in neutron stars: a review},
  journal = {Rept. Prog. Phys.},
  volume  = {81},
  pages   = {056902},
  year    = {2018},
  doi     = {10.1088/1361-6633/aaae14}
}

@article{Capano:2019eae,
  author  = {Capano, Collin D. and others},
  title   = {Stringent constraints on neutron-star radii from multimessenger observations and nuclear theory},
  journal = {Nature Astron.},
  volume  = {4},
  pages   = {625},
  year    = {2020},
  doi     = {10.1038/s41550-020-1014-6}
}

@article{Legred:2021hdx,
  author  = {Legred, Isaac and Chatziioannou, Katerina and Essick, Reed and Han, Sophia and Landry, Philippe},
  title   = {Impact of the {PSR} {J0740+6620} radius constraint on the properties of high-density matter},
  journal = {Phys. Rev. D},
  volume  = {104},
  pages   = {063003},
  year    = {2021},
  doi     = {10.1103/PhysRevD.104.063003}
}

@article{Zhao:2018nyf,
  author  = {Zhao, Tianqi and Lattimer, James M.},
  title   = {Tidal Deformabilities and Neutron Star Mergers},
  journal = {Phys. Rev. D},
  volume  = {98},
  pages   = {063020},
  year    = {2018},
  doi     = {10.1103/PhysRevD.98.063020}
}

@article{Ashton:2022grj,
  author  = {Ashton, Greg and others},
  title   = {Nested sampling for physical scientists},
  journal = {Nature Rev. Methods Primers},
  volume  = {2},
  pages   = {39},
  year    = {2022},
  doi     = {10.1038/s43586-022-00121-x}
}

@article{Feroz:2008xx,
  author  = {Feroz, F. and Hobson, M. P. and Bridges, M.},
  title   = {{MultiNest}: an efficient and robust {B}ayesian inference tool for cosmology and particle physics},
  journal = {Mon. Not. Roy. Astron. Soc.},
  volume  = {398},
  pages   = {1601},
  year    = {2009},
  doi     = {10.1111/j.1365-2966.2009.14548.x}
}

@article{Hogg:2017akh,
  author  = {Hogg, David W. and Foreman-Mackey, Daniel},
  title   = {Data analysis recipes: Using {M}arkov {C}hain {M}onte {C}arlo},
  journal = {Astrophys. J. Suppl.},
  volume  = {236},
  pages   = {11},
  year    = {2018},
  doi     = {10.3847/1538-4365/aab76e}
}

@article{Fujimoto:2021zas,
  author  = {Fujimoto, Yuki and Fukushima, Kenji and Murase, Koichi},
  title   = {Extensive Studies of the Neutron Star Equation of State from the Deep Learning Inference with the Observational Data Augmentation},
  journal = {J. High Energy Phys.},
  volume  = {03},
  number  = {2021},
  pages   = {273},
  year    = {2021},
  doi     = {10.1007/JHEP03(2021)273}
}

@article{Antoniadis:2013pzd,
  author  = {Antoniadis, John and others},
  title   = {A Massive Pulsar in a Compact Relativistic Binary},
  journal = {Science},
  volume  = {340},
  pages   = {448},
  year    = {2013},
  doi     = {10.1126/science.1233232}
}

@article{Weber:2004kj,
  author  = {Weber, Fridolin},
  title   = {Strange Quark Matter and Compact Stars},
  journal = {Prog. Part. Nucl. Phys.},
  volume  = {54},
  pages   = {193},
  year    = {2005},
  doi     = {10.1016/j.ppnp.2004.07.001}
}

@article{Fattoyev:2017jql,
  author  = {Fattoyev, F. J. and Piekarewicz, J. and Horowitz, C. J.},
  title   = {Neutron skins and neutron stars in the multi-messenger era},
  journal = {Phys. Rev. Lett.},
  volume  = {120},
  pages   = {172702},
  year    = {2018},
  doi     = {10.1103/PhysRevLett.120.172702}
}

@article{Fonseca:2021wxt,
  author  = {Fonseca, Emmanuel and others},
  title   = {Refined Mass and Geometric Measurements of the High-Mass {PSR} {J0740+6620}},
  journal = {Astrophys. J. Lett.},
  volume  = {915},
  pages   = {L12},
  year    = {2021},
  doi     = {10.3847/2041-8213/ac03b8}
}

@article{Fraga:2013qra,
  author  = {Fraga, Eduardo S. and Kurkela, Aleksi and Vuorinen, Aleksi},
  title   = {Interacting quark matter equation of state for compact stars},
  journal = {Astrophys. J. Lett.},
  volume  = {781},
  pages   = {L25},
  year    = {2014},
  doi     = {10.1088/2041-8205/781/2/L25}
}

@article{Demorest:2010bx,
  author  = {Demorest, Paul and Pennucci, Tim and Ransom, Scott and Roberts, Mallory and Hessels, Jason},
  title   = {Shapiro delay measurement of a two solar mass neutron star},
  journal = {Nature},
  volume  = {467},
  pages   = {1081},
  year    = {2010},
  doi     = {10.1038/nature09466}
}

@article{Miao:2021nuq,
  author  = {Miao, Zhiqiang and Jiang, Jin-Liang and Li, Ang and Chen, Lie-Wen},
  title   = {Bayesian Inference of Strange Star Equation of State Using the
             {GW170817} and {GW190425} Data},
  journal = {Astrophys. J. Lett.},
  volume  = {917},
  pages   = {L22},
  year    = {2021},
  doi     = {10.3847/2041-8213/ac194d},
  primaryClass  = {astro-ph.HE}
}

@article{Restrepo:2022wqn,
  author  = {Restrepo, Tulio E. and Provid\^encia, Constan\c{c}a and Pinto, Marcus Benghi},
  title   = {Non-strange quark stars within resummed {QCD}},
  journal = {Phys. Rev. D},
  volume  = {107},
  pages   = {114015},
  year    = {2023},
  doi     = {10.1103/PhysRevD.107.114015},
  primaryClass  = {hep-ph}
}

@article{Restrepo:2025qgp,
  author  = {Restrepo, Tulio E. and Kneur, Jean-Lo\"ic and Provid\^encia, Constan\c{c}a and Pinto, Marcus Benghi},
  title   = {Comparing strange and non-strange quark stars within resummed {QCD} at {NLO}},
  journal = {Phys. Rev. D},
  volume  = {112},
  pages   = {054013},
  year    = {2025},
  doi     = {10.1103/7x41-j7mv},
  primaryClass  = {hep-ph}
}

@article{Ma:2023stj,
  author  = {Ma, Zhi-Jun and Lu, Zhen-Yan and Xu, Jia-Feng and Peng, Guang-Xiong and Fu, Xiang and Wang, Junnian},
  title   = {Cold quark matter in a quasiparticle model: Thermodynamic consistency and stellar properties},
  journal = {Phys. Rev. D},
  volume  = {108},
  pages   = {054017},
  year    = {2023},
  doi     = {10.1103/PhysRevD.108.054017},
  primaryClass  = {hep-ph}
}

@article{Yuan:2022dxb,
  author  = {Yuan, Wen-Li and Li, Ang and Miao, Zhiqiang and Zuo, Bingjun and Bai, Zhan},
  title   = {Interacting $ud$ and $uds$ quark matter at finite densities and quark stars},
  journal = {Phys. Rev. D},
  volume  = {105},
  pages   = {123004},
  year    = {2022},
  doi     = {10.1103/PhysRevD.105.123004},
  primaryClass  = {nucl-th}
}

@article{Xu:2014zea,
    author = "Xu, JianFeng and Peng, GuangXiong and Lu, ZhenYan and Cui, ShuaiShuai",
    title = "{Two-flavor quark matter in the perturbation theory with full thermodynamic consistency}",
    doi = "10.1007/s11433-014-5599-6",
    journal = "Sci. China Phys. Mech. Astron.",
    volume = "58",
    number = "4",
    pages = "042001",
    year = "2015"
}

@article{Cao:2020zxi,
  author  = {Cao, Zheng and Chen, Lie-Wen and Chu, Peng-Cheng and Zhou, Ying},
  title   = {{GW190814}: Circumstantial evidence for up-down quark star},
  journal = {Phys. Rev. D},
  volume  = {106},
  pages   = {083007},
  year    = {2022},
  doi     = {10.1103/PhysRevD.106.083007}
}

@article{Godzieba:2020tjn,
  author  = {Godzieba, Daniel A. and Radice, David and Bernuzzi, Sebastiano},
  title   = {On the maximum mass of neutron stars and {GW190814}},
  journal = {Astrophys. J.},
  volume  = {908},
  pages   = {122},
  year    = {2021},
  doi     = {10.3847/1538-4357/abd4dd}
}

@article{Lim:2020zvx,
  author  = {Lim, Yeunhwan and Bhattacharya, Anirban and Holt, Jeremy W. and Pati, Debdeep},
  title   = {Radius and equation of state constraints from massive neutron stars and {GW190814}},
  journal = {Phys. Rev. C},
  volume  = {104},
  pages   = {L032802},
  year    = {2021},
  doi     = {10.1103/PhysRevC.104.L032802}
}

@article{Zhang:2020zsc,
  author  = {Zhang, Nai-Bo and Li, Bao-An},
  title   = {{GW190814}'s secondary component with mass 2.50--2.67 $M_\odot$ as a superfast pulsar},
  journal = {Astrophys. J.},
  volume  = {902},
  pages   = {38},
  year    = {2020},
  doi     = {10.3847/1538-4357/abb470}
}

@article{Roupas:2020nua,
  author  = {Roupas, Zacharias and Panotopoulos, Grigoris and Lopes, Il{\'\i}dio},
  title   = {{QCD} color superconductivity in compact stars: color-flavor locked quark star candidate for the gravitational-wave signal {GW190814}},
  journal = {Phys. Rev. D},
  volume  = {103},
  pages   = {083015},
  year    = {2021},
  doi     = {10.1103/PhysRevD.103.083015}
}

@article{Oikonomou:2023otn,
  author  = {Oikonomou, P. T. and Moustakidis, Ch. C.},
  title   = {Color-flavor locked quark stars in light of the compact object in {HESS J1731-347} and the {GW190814} event},
  journal = {Phys. Rev. D},
  volume  = {108},
  pages   = {063010},
  year    = {2023},
  doi     = {10.1103/PhysRevD.108.063010}
}

@article{Lopes:2021yga,
  author  = {Lopes, Luiz L. and Menezes, Debora P.},
  title   = {On the nature of the mass-gap object in the {GW190814} event},
  journal = {Astrophys. J.},
  volume  = {936},
  pages   = {41},
  year    = {2022},
  doi     = {10.3847/1538-4357/ac81c4}
}

@article{Tan:2020ics,
  author  = {Tan, Hung and Noronha-Hostler, Jacquelyn and Yunes, Nico},
  title   = {Neutron star equation of state in light of {GW190814}},
  journal = {Phys. Rev. Lett.},
  volume  = {125},
  pages   = {261104},
  year    = {2020},
  doi     = {10.1103/PhysRevLett.125.261104}
}

@article{Romani:2026j0952,
  author  = {Romani, Roger W. and Beleznay, Maya and
             Filippenko, Alexei V. and Brink, Thomas G.
             and Zheng, WeiKang},
  title   = {PSR J0952-0607: Tightening a Record-High
             Neutron Star Mass},
  journal = {The Astrophysical Journal},
  volume  = {996},
  number  = {1},
  pages   = {101},
  year    = {2026},
  doi     = {10.3847/1538-4357/ae28c5},
  eprint  = {2512.05099},
  archivePrefix = {arXiv},
  primaryClass  = {astro-ph.HE}
}

@article{Cao:2023rgh,
  author  = {Cao, Zheng and Chen, Lie-Wen},
  title   = {Neutron star versus quark star in the multimessenger era},
  journal = {Phys. Rev. D},
  volume  = {114},
  number  = {8},
  pages   = {083022},
  year    = {2026},
  doi     = {10.1103/4fr7-xqxt},
  eprint  = {2308.16783},
  archivePrefix = {arXiv},
  primaryClass  = {astro-ph.HE}
}

@article{Cao:2026dwk,
  author  = {Cao, Zheng and Chen, Lie-Wen},
  title   = {On the Possibility of a Strong First-order Phase Transition in Neutron Stars},
  journal = {Astrophys. J.},
  volume  = {1008},
  number  = {2},
  pages   = {211},
  year    = {2026},
  doi     = {10.3847/1538-4357/ae946d},
  eprint  = {2606.06378},
  archivePrefix = {arXiv},
  primaryClass  = {nucl-th}
}

@article{De:2018TidalGW170817,
  author = {De, Soumi and Finstad, Daniel and Lattimer, James M.
    and Brown, Duncan A. and Berger, Edo and Biwer, Christopher M.},
  title = {Tidal Deformabilities and Radii of Neutron Stars
    from the Observation of GW170817},
  journal = {Phys. Rev. Lett.},
  volume = {121},
  pages = {091102},
  year = {2018},
  doi = {10.1103/PhysRevLett.121.091102},
  eprint = {1804.08583},
  archivePrefix = {arXiv},
  primaryClass = {astro-ph.HE}
}

@article{Landry:2019EOSGW,
  author = {Landry, Philippe and Essick, Reed},
  title = {Nonparametric Inference of the Neutron Star Equation
    of State from Gravitational Wave Observations},
  journal = {Phys. Rev. D},
  volume = {99},
  pages = {084049},
  year = {2019},
  doi = {10.1103/PhysRevD.99.084049},
  eprint = {1811.12529},
  archivePrefix = {arXiv},
  primaryClass = {astro-ph.HE}
}

@article{Essick:2020KDEGW,
  author = {Essick, Reed and Landry, Philippe and Holz, Daniel E.},
  title = {Nonparametric Inference of Neutron Star Composition,
    Equation of State, and Maximum Mass with GW170817},
  journal = {Phys. Rev. D},
  volume = {101},
  pages = {063007},
  year = {2020},
  doi = {10.1103/PhysRevD.101.063007},
  eprint = {1910.09740},
  archivePrefix = {arXiv},
  primaryClass = {astro-ph.HE}
}

@article{Ayriyan:2021KDE,
  author = {Ayriyan, A. and Blaschke, D. and Grunfeld, A. G.
    and Alvarez-Castillo, D. and Grigorian, H. and Abgaryan, V.},
  title = {Bayesian Analysis of Multimessenger M-R Data
    with Interpolated Hybrid EoS},
  journal = {Eur. Phys. J. A},
  volume = {57},
  pages = {318},
  year = {2021},
  doi = {10.1140/epja/s10050-021-00619-0},
  eprint = {2102.13485},
  archivePrefix = {arXiv},
  primaryClass = {astro-ph.HE}
}

@article{Lewis:2025GetDist,
  author = {Lewis, Antony},
  title = {GetDist: A Python Package for Analysing Monte Carlo Samples},
  journal = {JCAP},
  volume = {2025},
  number = {08},
  pages = {025},
  year = {2025},
  doi = {10.1088/1475-7516/2025/08/025},
  eprint = {1910.13970},
  archivePrefix = {arXiv},
  primaryClass = {astro-ph.IM}
}

@article{Botev:2010KDE,
  author = {Botev, Z. I. and Grotowski, J. F. and Kroese, D. P.},
  title = {Kernel Density Estimation via Diffusion},
  journal = {The Annals of Statistics},
  volume = {38},
  number = {5},
  pages = {2916--2957},
  year = {2010},
  doi = {10.1214/10-AOS799}
}

@article{Thrane:2019BayesianGW,
  author = {Thrane, Eric and Talbot, Colm},
  title = {An Introduction to Bayesian Inference in Gravitational-Wave
    Astronomy: Parameter Estimation, Model Selection, and
    Hierarchical Models},
  journal = {Publ. Astron. Soc. Austral.},
  volume = {36},
  pages = {e010},
  year = {2019},
  doi = {10.1017/pasa.2019.2},
  eprint = {1809.02293},
  archivePrefix = {arXiv},
  primaryClass = {astro-ph.IM}
}

@article{Greif2019,
  author = {Greif, S. K. and Raaijmakers, G. and Hebeler, K.
            and Schwenk, A. and Watts, A. L.},
  title = {Equation of state sensitivities when inferring
           neutron star and dense matter properties},
  journal = {Monthly Notices of the Royal Astronomical Society},
  volume = {485},
  number = {4},
  pages = {5363--5376},
  year = {2019},
  doi = {10.1093/mnras/stz654}
}

@misc{McGinn2024,
  author = {McGinn, Jordan and Mukherjee, Arunava and
            Irwin, Jessica and Messenger, Christopher and
            Williams, Michael J. and Heng, Ik Siong},
  title = {Rapid neutron star equation of state inference
           with Normalising Flows},
  year = {2024},
  eprint = {2403.17462},
  archivePrefix = {arXiv},
  primaryClass = {astro-ph.HE}
}

@article{Lackey:2015prd,
  author  = {Lackey, Benjamin D. and Wade, Leslie},
  title   = {Reconstructing the neutron-star equation of state with gravitational-wave detectors from a realistic population of inspiralling binary neutron stars},
  journal = {Phys. Rev. D},
  volume  = {91},
  pages   = {043002},
  year    = {2015},
  doi     = {10.1103/PhysRevD.91.043002}
}

@article{Read:2009yp,
    author = "Read, Jocelyn S. and Markakis, Charalampos and Shibata, Masaru and Uryu, Koji and Creighton, Jolien D. E. and Friedman, John L.",
    title = "{Measuring the neutron star equation of state with gravitational wave observations}",
    eprint = "0901.3258",
    archivePrefix = "arXiv",
    primaryClass = "astro-ph",
    doi = "10.1103/PhysRevD.79.124033",
    journal = "Phys. Rev. D",
    volume = "79",
    number = "12",
    pages = "124033",
    year = "2009"
}

@article{Wouters:2025zju,
    author = "Wouters, Thibeau and Pang, Peter T. H. and Koehn, Hauke and Rose, Henrik and Somasundaram, Rahul and Tews, Ingo and Dietrich, Tim and Van Den Broeck, Chris",
    title = "{Leveraging differentiable programming in the inverse problem of neutron stars}",
    eprint = "2504.15893",
    archivePrefix = "arXiv",
    primaryClass = "astro-ph.HE",
    reportNumber = "LA-UR-25-23486",
    doi = "10.1103/v2y8-kxvx",
    journal = "Phys. Rev. D",
    volume = "112",
    number = "4",
    pages = "043037",
    year = "2025"
}

@article{Wouters:2025csq,
    author = "Wouters, Thibeau and Pang, Peter T. H. and Dietrich, Tim and Van Den Broeck, Chris",
    title = "{Incorporating neutron star physics into gravitational wave inference with physics-informed priors using normalizing flows}",
    eprint = "2511.22987",
    archivePrefix = "arXiv",
    primaryClass = "astro-ph.HE",
    doi = "10.1103/8t5h-y4lf",
    journal = "Phys. Rev. D",
    volume = "114",
    pages = "043005",
    year = "2026"
}

@article{Williams:2021qyt,
    author = "Williams, Michael J. and Veitch, John and Messenger, Chris",
    title = "{Nested sampling with normalizing flows for gravitational-wave inference}",
    doi = "10.1103/PhysRevD.103.103006",
    journal = "Phys. Rev. D",
    volume = "103",
    pages = "103006",
    year = "2021"
}

@book{Scott:1992kde,
    author = "Scott, David W.",
    title = "{Multivariate Density Estimation}",
    publisher = "Wiley",
    doi = "10.1002/9780470316849",
    year = "1992"
}

@article{Chatziioannou:2024jsr,
    author = "Chatziioannou, Katerina and Cromartie, H. Thankful and Gandolfi, Stefano and Tews, Ingo and Radice, David and Steiner, Andrew W. and Watts, Anna L.",
    title = "{Neutron stars and the dense matter equation of state}",
    doi = "10.1103/ymsq-cfcw",
    journal = "Rev. Mod. Phys.",
    volume = "97",
    pages = "045007",
    year = "2025"
}

@article{Lattimer:2010kpa,
    author = "Lattimer, James M.",
    title = "{Neutron stars and the dense matter equation of state}",
    doi = "10.1007/s10509-010-0529-1",
    journal = "Astrophys. Space Sci.",
    volume = "336",
    pages = "67--74",
    year = "2011"
}

@article{Annala:2021gom,
    author = "Annala, Eemeli and Gorda, Tyler and Katerini, Evangelia and Kurkela, Aleksi and N{\"a}ttil{\"a}, Joonas and Paschalidis, Vasileios and Vuorinen, Aleksi",
    title = "{Multimessenger Constraints for Ultradense Matter}",
    eprint = "2105.05132",
    archivePrefix = "arXiv",
    primaryClass = "astro-ph.HE",
    doi = "10.1103/PhysRevX.12.011058",
    journal = "Phys. Rev. X",
    volume = "12",
    pages = "011058",
    year = "2022"
}

@article{Alford:2009qj,
    author = "Alford, Mark G.",
    title = "{Quark matter in neutron stars}",
    doi = "10.1016/j.nuclphysa.2009.09.034",
    journal = "Nucl. Phys. A",
    volume = "830",
    pages = "385C--392C",
    year = "2009"
}

@article{Alford:2019oge,
    author = "Alford, Mark G. and Han, Sophia and Schwenzer, Kai",
    title = "{Signatures for quark matter from multi-messenger observations}",
    eprint = "1904.05471",
    archivePrefix = "arXiv",
    primaryClass = "astro-ph.HE",
    doi = "10.1088/1361-6471/ab337a",
    journal = "J. Phys. G",
    volume = "46",
    pages = "114001",
    year = "2019"
}

@article{Fraga:2016yxs,
    author = "Fraga, E. S. and Kurkela, A. and Schaffner-Bielich, J. and Vuorinen, A.",
    title = "{QCD constraints on the equation of state for compact stars}",
    doi = "10.1016/j.nuclphysa.2016.01.037",
    journal = "Nucl. Phys. A",
    volume = "956",
    pages = "813--816",
    year = "2016"
}

@article{Gorda:2023usm,
    author = "Gorda, Tyler and Komoltsev, Oleg and Kurkela, Aleksi and Mazeliauskas, Aleksas",
    title = "{Bayesian uncertainty quantification of perturbative QCD input to the neutron-star equation of state}",
    eprint = "2303.02175",
    archivePrefix = "arXiv",
    primaryClass = "hep-ph",
    doi = "10.1007/JHEP06(2023)002",
    journal = "JHEP",
    volume = "06",
    pages = "002",
    year = "2023"
}

@article{Gorda:2023mkk,
    author = "Gorda, Tyler and Paatelainen, Risto and S{\"a}ppi, Saga and Sepp{\"a}nen, Kaapo",
    title = "{Equation of State of Cold Quark Matter to $O(\alpha_s^3 \ln \alpha_s)$}",
    eprint = "2307.08734",
    archivePrefix = "arXiv",
    primaryClass = "hep-ph",
    doi = "10.1103/PhysRevLett.131.181902",
    journal = "Phys. Rev. Lett.",
    volume = "131",
    pages = "181902",
    year = "2023"
}

@article{Drischler:2024ebw,
    author = "Drischler, C. and Giuliani, P. G. and Bezoui, S. and Piekarewicz, J. and Viens, F.",
    title = "{Bayesian mixture model approach to quantifying the empirical nuclear saturation point}",
    eprint = "2405.02748",
    archivePrefix = "arXiv",
    primaryClass = "nucl-th",
    doi = "10.1103/PhysRevC.110.044320",
    journal = "Phys. Rev. C",
    volume = "110",
    pages = "044320",
    year = "2024"
}

@article{Most:2018hfd,
    author = "Most, Elias R. and Weih, Lukas R. and Rezzolla, Luciano and Schaffner-Bielich, J{\"u}rgen",
    title = "{New constraints on radii and tidal deformabilities of neutron stars from GW170817}",
    doi = "10.1103/PhysRevLett.120.261103",
    journal = "Phys. Rev. Lett.",
    volume = "120",
    pages = "261103",
    year = "2018"
}

@article{Carson:2019rjx,
    author = "Carson, Zack and Chatziioannou, Katerina and Haster, Carl-Johan and Yagi, Kent and Yunes, Nicol{\'a}s",
    title = "{Equation-of-state insensitive relations after GW170817}",
    eprint = "1903.03909",
    archivePrefix = "arXiv",
    primaryClass = "gr-qc",
    doi = "10.1103/PhysRevD.99.083016",
    journal = "Phys. Rev. D",
    volume = "99",
    pages = "083016",
    year = "2019"
}

@article{Finch:2025bao,
    author = "Finch, Eliot and Legred, Isaac and Chatziioannou, Katerina and Essick, Reed and Han, Sophia and Landry, Philippe",
    title = "{Unified nonparametric equation-of-state inference from the neutron-star crust to perturbative-QCD densities}",
    eprint = "2505.13691",
    archivePrefix = "arXiv",
    primaryClass = "astro-ph.HE",
    doi = "10.1103/krc7-kz2l",
    journal = "Phys. Rev. D",
    volume = "112",
    pages = "103023",
    year = "2025"
}

@article{Essick:2020ghc,
    author = "Essick, Reed and Landry, Philippe",
    title = "{Discriminating between Neutron Stars and Black Holes with Imperfect Knowledge of the Maximum Neutron Star Mass}",
    eprint = "2007.01372",
    archivePrefix = "arXiv",
    primaryClass = "astro-ph.HE",
    doi = "10.3847/1538-4357/abbd3b",
    journal = "Astrophys. J.",
    volume = "904",
    pages = "80",
    year = "2020"
}

@article{Vinciguerra:2023qxq,
    author = "Vinciguerra, Serena and others",
    title = "{An Updated Mass{\textendash}Radius Analysis of the 2017{\textendash}2018 NICER Data Set of PSR J0030+0451}",
    eprint = "2308.09469",
    archivePrefix = "arXiv",
    primaryClass = "astro-ph.HE",
    doi = "10.3847/1538-4357/acfb83",
    journal = "Astrophys. J.",
    volume = "961",
    pages = "62",
    year = "2024"
}

@article{Salmi:2022cgy,
    author = "Salmi, Tuomo and others",
    title = "{The Radius of PSR J0740+6620 from NICER with NICER Background Estimates}",
    eprint = "2209.12840",
    archivePrefix = "arXiv",
    primaryClass = "astro-ph.HE",
    doi = "10.3847/1538-4357/ac983d",
    journal = "Astrophys. J.",
    volume = "941",
    pages = "150",
    year = "2022"
}

@article{Bogdanov:2019ixe,
    author = "Bogdanov, Slavko and others",
    title = "{Constraining the Neutron Star Mass{\textendash}Radius Relation and Dense Matter Equation of State with $NICER$. I. The Millisecond Pulsar X-Ray Data Set}",
    eprint = "1912.05706",
    archivePrefix = "arXiv",
    primaryClass = "astro-ph.HE",
    doi = "10.3847/2041-8213/ab53eb",
    journal = "Astrophys. J. Lett.",
    volume = "887",
    pages = "L25",
    year = "2019"
}

@article{Raaijmakers:2021uju,
    author = "Raaijmakers, G. and Greif, S. K. and Hebeler, K. and Hinderer, T. and Nissanke, S. and Schwenk, A. and Riley, T. E. and Watts, A. L. and Lattimer, J. M. and Ho, W. C. G.",
    title = "{Constraints on the Dense Matter Equation of State and Neutron Star Properties from NICER{\textquoteright}s Mass{\textendash}Radius Estimate of PSR J0740+6620 and Multimessenger Observations}",
    doi = "10.3847/2041-8213/ac089a",
    journal = "Astrophys. J. Lett.",
    volume = "918",
    pages = "L29",
    year = "2021"
}

@article{Kini:2026rjx,
    author = "Kini, Yves and others",
    title = "{A NICER View of PSR J0030+0451: Updated Constraints from 6 yr of NICER Observations}",
    eprint = "2602.23743",
    archivePrefix = "arXiv",
    primaryClass = "astro-ph.HE",
    doi = "10.3847/1538-4357/ae733e",
    journal = "Astrophys. J.",
    volume = "1005",
    pages = "201",
    year = "2026"
}

@article{LIGOScientific:2017zic,
    author = "Abbott, B. P. and others",
    collaboration = "LIGO Scientific, Virgo, Fermi-GBM, INTEGRAL",
    title = "{Gravitational Waves and Gamma-rays from a Binary Neutron Star Merger: GW170817 and GRB 170817A}",
    eprint = "1710.05834",
    archivePrefix = "arXiv",
    primaryClass = "astro-ph.HE",
    doi = "10.3847/2041-8213/aa920c",
    journal = "Astrophys. J. Lett.",
    volume = "848",
    pages = "L13",
    year = "2017"
}

@article{Kunert:2021hgm,
    author = "Kunert, Nina and Pang, Peter T. H. and Tews, Ingo and Coughlin, Michael W. and Dietrich, Tim",
    title = "{Quantifying modeling uncertainties when combining multiple gravitational-wave detections from binary neutron star sources}",
    eprint = "2110.11835",
    archivePrefix = "arXiv",
    primaryClass = "astro-ph.HE",
    doi = "10.1103/PhysRevD.105.L061301",
    journal = "Phys. Rev. D",
    volume = "105",
    pages = "L061301",
    year = "2022"
}

@article{Weih:2019rzo,
    author = "Weih, L. R. and Most, E. R. and Rezzolla, L.",
    title = "{Optimal Neutron-star Mass Ranges to Constrain the Equation of State of Nuclear Matter with Electromagnetic and Gravitational-wave Observations}",
    eprint = "1905.04900",
    archivePrefix = "arXiv",
    primaryClass = "astro-ph.HE",
    doi = "10.3847/1538-4357/ab2edd",
    journal = "Astrophys. J.",
    volume = "881",
    pages = "73",
    year = "2019"
}

@article{Alford:2006wn,
    author = "Alford, Mark G.",
    title = "{Color superconductivity in ultra-dense quark matter}",
    eprint = "hep-lat/0610046",
    archivePrefix = "arXiv",
    primaryClass = "hep-lat",
    doi = "10.22323/1.032.0001",
    journal = "PoS",
    volume = "LAT2006",
    pages = "001",
    year = "2006"
}

\end{document}